\documentclass[letterpaper,preprintnumbers,prd,twocolumn,nofootinbib,nobibnotes,showpacs]{revtex4-2}
\usepackage[colorlinks=false, linkcolor=blue, citecolor=blue, urlcolor=blue]{hyperref}
\usepackage{graphicx}
\usepackage{subfig}
\usepackage{float}
\usepackage{color}
\usepackage{amsfonts}
\usepackage{mathrsfs}
\usepackage{epsfig}
\usepackage{dcolumn}
\usepackage{amsmath}
\usepackage{color}
\usepackage{diagbox}
\usepackage{booktabs} % 三线表
\usepackage{array}  % 表格对齐增强
\usepackage{makecell}
\usepackage{comment}

\newcommand{\blue}[1]{\textcolor{black}{{#1}}}

\begin{document}
\renewcommand{\baselinestretch}{1.3}
\newcommand\beq{\begin{equation}}
\newcommand\eeq{\end{equation}}
\newcommand\beqn{\begin{eqnarray}}
\newcommand\eeqn{\end{eqnarray}}
\newcommand\nn{\nonumber}
\newcommand\fc{\frac}
\newcommand\lt{\bigg}
\newcommand\rt{\bigg}
\newcommand\pt{\partial}

\allowdisplaybreaks

\title{Photon rings and shadows of black holes with non-minimal couplings\\
between curvature and electromagnetic field}

\author{Zhixiang Yin$^{1,2}$}
\email{yinzx@bao.ac.cn}
%\affiliation{National Astronomical Observatories, Chinese Academy of Sciences, 20A Datun Road, Beijing 100101, China}

\author{Changjun Gao$^{1,2}$}
\email{gaocj@nao.cas.cn}

\author{Yun-Long Zhang$^{1,3}$}
%\email{zhangyunlong@nao.cas.cn}

\affiliation{$^{1}$National Astronomical Observatories, Chinese Academy of Sciences, 20A Datun Road, Beijing 100101, China}
\affiliation{$^{2}$School of Astronomy and Space Sciences, University of Chinese Academy of Sciences,
19A Yuquan Road, Beijing 100049, China}
\affiliation{$^{3}$School of Fundamental Physics and Mathematical Sciences, Hangzhou Institute for Advanced Study, University of Chinese Academy of Sciences, Hangzhou 310024, China.}

\begin{abstract}

We investigate black holes with non-minimal couplings between the electromagnetic field and spacetime curvature tensors, focusing on their event horizons, shadows, and photon rings. \blue{Such couplings can naturally arise from effective field theories of gravity and quantum corrections in curved spacetime.} Starting from a general action with three independent coupling terms, we derive static and spherically symmetric black hole solutions using a series expansion method. We find that all couplings enlarge the event horizon and photon sphere, while their observational consequences differ. The coupling $F^\mu_{\ \nu}F_{\mu\rho}R^{\nu\rho}$ slightly increases the shadow radius and the separation between the zeroth- and first-order photon rings, leaving higher-order spacings nearly unchanged. The coupling $F_{\mu\nu}F_{\sigma\rho}R^{\mu\nu\sigma\rho}$ significantly enlarges the shadow and the separation between the zeroth- and first-order photon rings, but rapidly suppresses the spacing between higher-order rings. In contrast, the $F^2R$ coupling reduces the shadow radius and causes the zeroth- and first-order rings to nearly coincide, leading to an enhanced photon-ring brightness,  while increasing the separation of higher-order rings and  \blue{rendering}  them easier to resolve observationally. Backward ray-tracing simulations further confirm that these characteristic signatures persist in synthetic black-hole images. These findings provide potential observational \blue{signatures} of non-minimal couplings between the electromagnetic field and spacetime curvature tensors. 

\end{abstract}

% \Keywords{ }



%11.10.Kk Field theories in dimensions other than four (see also 04.50.-h Higher-dimensional gravity and other theories of gravity; 04.60.Kz Lower dimensional models; minisuperspace models in general relativity and gravitation)

%04.50.Kd 	Modified theories of gravity

\maketitle


%%%%%%%%%%%%%%%%%%%%%%%%%%%%%%%%%%%%%%%%%%%%%%%%%%%%%%%%%%%%%%%%%%%%%%%%%%%%%%%%%%%%%%%

\tableofcontents

\section{Introduction}

The study of black-hole shadows and photon rings has attracted considerable attention since \textit{the Event Horizon Telescope} (EHT) Collaboration produced black-hole images of M87* \cite{M87EventHorizonTelescope:2019dse,M87EventHorizonTelescope:2019uob,M87EventHorizonTelescope:2019jan,M87EventHorizonTelescope:2019ths,M87EventHorizonTelescope:2019pgp,M87EventHorizonTelescope:2019ggy,M87EventHorizonTelescope:2021bee,M87EventHorizonTelescope:2021srq,M87EventHorizonTelescope:2023gtd} and Sgr A*\cite{SGRAEventHorizonTelescope:2022wkp,SGRAEventHorizonTelescope:2022apq,SGRAEventHorizonTelescope:2022wok,SGRAEventHorizonTelescope:2022exc,SGRAEventHorizonTelescope:2022urf,SGRAEventHorizonTelescope:2022xqj,SGRAEventHorizonTelescope:2024hpu,SGRAEventHorizonTelescope:2024rju}. Black-hole shadows and photon rings encode rich information about the spacetime geometry, and they serve as crucial probes of both Einstein’s \textit{general relativity} (GR) and its possible extensions \cite{Psaltis:2018xkc,Johnson:2019ljv,gralla:2020shape,Gralla:2020pra,Volkel:2020xlc}. In particular, any deviations from Reissner–Nordstr\"{o}m or Kerr solutions may leave observable imprints both on the radius and shape of shadows and on the structure of photon rings. \blue{Such deviations can originate from several broad classes of scenarios, including generalized Kerr-like or Reissner–Nordstr\"{o}m-like spacetimes, environmental effects (e.g. plasma or surrounding matter), and black-hole solutions arising in modified theories of gravity. The observational consequences of these scenarios for black-hole shadows and photon rings have been extensively reviewed in Refs.~\cite{Perlick:2021aok,Cunha:2018acu,EventHorizonTelescope:2020qrl,Bambi:2011mj,Johannsen:2013rqa}, while representative examples can be found in Refs.~\cite{Grenzebach:2014fha,Shaikh:2019fpu,Perlick:2015vta,Meng:2023htc,Hennigar:2018hza,Kumar:2020hgm}. Overall, substantial progress has been made in understanding black-hole shadows and photon rings through both analytical and numerical investigations.}
%Such deviations can arise from various sources. These include the propagation of photons in the generalized Reissner–Nordstr\"{o}m-like or Kerr-like spacetimes \cite{Grenzebach:2014fha,Amir:2016cen,Abdujabbarov:2016hnw,Wang:2017hjl,Ovgun:2018tua,Wei:2019pjf,Shaikh:2019fpu,Roy:2020dyy}, the effects of environment due to matter fields or accretion disks surrounding black holes \cite{Perlick:2015vta,Wang:2017qhh,Bisnovatyi-Kogan:2019wdd,Wang:2019skw,Pantig:2020uhp,Jusufi:2020cpn,Meng:2023htc,Raza:2023vkn,Meng:2024puu}, and the modified geometries of black holes derived from modified or alternative theories of gravity \cite{Amarilla:2011fx,Dastan:2016vhb,Wang:2018prk,Hennigar:2018hza,Zhang:2019glo,Lu:2019zxb,Ovgun:2019jdo,Kumar:2019ohr,Contreras:2019nih,Ma:2019ybz,Kumar:2020hgm,Kumar:2020owy,Contreras:2020kgy,Chen:2020aix,EslamPanah:2020hoj,Zeng:2020dco,Peng:2020wun,Belhaj:2020okh,Guo:2021bcw,Xiong:2025wgs}. Overall, substantial progress has been made in understanding black-hole shadows and photon rings through both analytical and numerical investigations.

Among the proposed extensions, theories involving non-minimal couplings between electromagnetic fields and spacetime curvature tensors (i.e., Riemann tensor $R^{\mu\nu\sigma\rho}$, Ricci tensor $R^{\mu\nu}$ and Ricci scalar $R$) represent a particularly interesting class of theories. The most general action that describes the non-minimal couplings between curvature and electromagnetic field is given by 
 \cite{Balakin:2005cqg,Balakin:2007am}
\begin{equation}
\begin{aligned}
S&=\int d^4x\;\mathscr{L}&\\
&=\int\mathrm{d}^4x \sqrt{-g}\bigg[\frac{1}{16\pi}R-\frac{1}{16\pi}F^{2}+F_{\mu\nu}F_{\sigma\rho}\times\\
&\quad\quad\Big(\alpha_1g^{\mu\sigma}R^{\nu\rho}+\alpha_2g^{\mu\sigma}g^{\nu\rho}R+\alpha_3R^{\mu\nu\sigma\rho}\Big)\bigg]\;,
\end{aligned}
\label{eq:action}
\end{equation}
where $F_{\mu\nu}$ is the electromagnetic field strength tensor defined by the one-form $A_{\mu}$, i.e. $F_{\mu\nu}=\nabla_\mu A_\nu-\nabla_\nu A_\mu$, and $\alpha_i$ denote the coupling constants. Throughout this paper, we work in units where $G=c=1$, while the electromagnetic sector is expressed in Gaussian units. The last three terms encode the coupling between the electromagnetic field and spacetime curvature tensors. With the presence of these non-minimal couplings, the conformal invariance of electromagnetism is broken. 

Non-minimal couplings between the electromagnetic field and spacetime curvature tensors naturally arise both in effective field theories \cite{horndeski:1976jmp,Balakin:2005cqg,Balakin:2007am,BeltranJimenez:2013btb} and in \textit{quantum electrodynamics} (QED) in curved spacetime \cite{drummond:1980qed,Bastianelli:2008cu}. It is found that asymptotically safe quantum gravity can give rise to such non-minimal couplings \cite{Knorr:2024yiu}. Such non-minimal couplings also arise in the Einstein--Yang--Mills theory with $SU(2)$ symmetry under the Wu--Yang ansatz \cite{Balakin:2006gv}. In addition, a particular combination of the coupling constants can  be obtained from the Kaluza-Klein reduction of the Gauss--Bonnet terms in five dimensions \cite{HABuchdahl:1979jpa}.  

As a concrete realization within QED in curved spacetime, Drummond and Hathrell \cite{drummond:1980qed} computed the coefficients $\alpha_1$, $\alpha_2$, and $\alpha_3$ by evaluating the one-loop vacuum-polarization diagrams using the Schwinger--de Witt effective action approach. They found that $\alpha_i\propto\frac{e^2}{m_e^2}$ with $e$ and $m_e$ the charge and mass of electrons. It is important to note that Drummond and Hathrell’s analysis is suitable for electromagnetic waves with energies much less than $0.5 \,\rm MeV$. Hence, their analysis may not be reliable in the high-energy regime. Therefore, in the present work we leave $\alpha_1$, $\alpha_2$, and $\alpha_3$ as free coupling constants rather than adopting the specific values obtained in Ref.~\cite{drummond:1980qed}.  The \textit{equations of motion} (EoM) derived from the action Eq.~(\ref{eq:action}) generally contain derivatives higher than second order. Horndeski has shown that a particular combination of the coupling constants $\alpha_i$\,, i.e., $-\alpha_1/4=\alpha_2=\alpha_3$ leads to the second order EoM \cite{BeltranJimenez:2013btb,horndeski:1976jmp} such that the Ostrogradsky instability \cite{Ostrogradksi_1850} is avoided in this case. However, whether the theory generically suffers from such an instability remains an open question. Therefore, it is reasonable to treat $\alpha_i$ as free coupling constants.

Such non-minimal couplings modify both the electromagnetic field equations and the gravitational field equations. Accordingly, studies of such couplings may proceed along two complementary directions. The first is to study the modified Maxwell equations on a fixed spacetime background described by GR, as explored in, e.g., Refs.~\cite{Chen:2015cpa,Chen:2016hil,Zhang:2021hit,Chen:2023wna,ravi:2023non,jana:2024non,Carballo-Rubio:2025zwz}. The second is to investigate the impact of the non-minimal couplings on the spacetime geometry itself, such that photons propagate along null geodesics of the modified spacetime. The latter perspective constitutes the main focus of the present work. We show that these couplings can significantly modify the black-hole spacetime structure and consequently alter the observational signatures associated with light propagation.

In this work, we consider the general class of couplings between the electromagnetic field and curvature tensors. We first construct asymptotically flat, static and spherically symmetric black hole solutions using a series-expansion method, retaining terms up to $\mathcal{O}(r^{-7})$. We then analyze the photon geodesics in these space-times and apply the backward ray-tracing method to obtain shadow images. Our objective is to systematically investigate how the three independent coupling terms affect the horizon radius, shadow radius, and photon-ring structure.

The paper is organized as follows. In Sec.~\ref{sec:series solutions}, we derive the series solutions for black holes non-minimally coupled to the electromagnetic field. In Sec.~\ref{sec:photon dynamics}, we establish the equations of motion for photons. In Sec.~\ref{sec: accretion disk model}, we introduce the accretion disk model adopted in this work. Based on this setup, we theoretically analyze the effects of the three independent coupling terms on the horizon, photon sphere, shadow, and photon rings in Sec.~\ref{sec: general theoretical analysis}. The corresponding black-hole images are then presented in Sec.~\ref{sec:images of shadows and rings}, which further verify the theoretical predictions. Finally, Sec.~\ref{sec:conclusions} summarizes our main results and discusses their implications for testing modified gravity through black-hole observations. In addition, the Appendix is organized into three parts: Appendix \ref{app: EoM for the metric and the series solutions} details the derivation of the non-minimally coupled black-hole solutions, Appendix \ref{app: EoM for integration} presents the equations of motion for the numerical integration of photon trajectories, and Appendix \ref{app:backward ray-tracing approach} describes the backward ray-tracing method used to generate the images. Throughout this paper, we adopt the metric signature $(-,+,+,+)$.

\section{Series solutions for black holes non-minimally coupled to the electromagnetic field}
\label{sec:series solutions}

Varying the action in Eq.~\eqref{eq:action} with respect to $g_{\mu\nu}$ yields the modified Einstein field equations, from which the corresponding black-hole solutions can in principle be obtained. However, due to the complexity of the full field equations, we instead adopt a reduced-action approach. Specifically, we first impose a static and spherically symmetric ansatz at the level of the action and then derive the corresponding Euler--Lagrange equations for the metric functions. Without loss of generality, the spacetime metric is assumed to take the form

\begin{equation}
\mathrm{d}s^2=-U(r)\mathrm{d}t^2+\frac{N(r)^2}{U(r)}\mathrm{d}r^2+r^2\left(\mathrm{d}\theta^2+\sin^2\theta\,\mathrm{d}\varphi^2\right)\;,
\label{eq:metric}
\end{equation}
which allows for a relatively simple analysis and facilitates the investigation of the effects of the non-minimal couplings on the black-hole geometry and its observational signatures.

Since the spacetime is static and spherically symmetric, the one-form $A_\mu$ can be assumed to be
\begin{equation}
A_{\mu}=\bigg(\phi(r),\;0,\;0,\;0\bigg)\;,
\label{eq:one-form}
\end{equation}
with $\phi$ the potential function to be determined. 

Plugging Eq.~\eqref{eq:metric} and Eq.~\eqref{eq:one-form} into Eq.~\eqref{eq:action}, we obtain the Lagrangian function as follows

\begin{equation}
\begin{aligned}
\mathscr{L}&=\frac{1}{16\pi}N^{-4}\bigg\{32\pi\alpha_3r^2\phi'^2\bigg(NU''-N'U'\bigg)\\
&+32\pi\alpha_2\phi'^2\bigg[-2N^3-rN'\bigg(4U+rU'\bigg)\\
&\quad\quad+N\bigg(4rU'+r^2U''+2U\bigg)\bigg]\\
&+16\pi\alpha_1r\phi'^2\bigg[N\bigg(2U'+rU''\bigg)-N'\bigg(2U+rU'\bigg)\bigg]\\
&+N^2\bigg[2N^3+rN'\bigg(4U+rU'\bigg)\\
&\quad\quad-N\bigg(4rU'+r^2U''-2r^2\phi'^2
+2U\bigg)\bigg]\bigg\}\sin\theta\;,
\end{aligned}
\label{eq:lagrangian}
\end{equation}
where the prime notation denotes differentiation with respect to $r$. By varying the reduced Lagrangian with respect to $U(r)$, $N(r)$, and $\phi(r)$, one obtains the corresponding equations of motion. The detailed derivation of these equations, as well as their series solutions, is provided in Appendix~\ref{app: EoM for the metric and the series solutions}.

Since we seek static and spherically symmetric black-hole solutions, it is physically natural to impose asymptotic flatness on the spacetime. Under these assumptions, we obtain the series solutions truncated at $\mathcal{O}(r^{-8})$, given as follows

\begin{equation}
\begin{aligned}
\phi=&\frac{Q}{r}+\frac{16\pi\alpha_3MQ}{r^4}\\
&-\frac{8\pi\big(-\alpha_1-10\alpha_2+9\alpha_3\big)Q^3}{5r^5}\\
&+\frac{512\pi^2Q}{7r^7}\bigg[8\alpha_3^2M^2+\big(-3\alpha_1^2-25\alpha_2\alpha_1\\
&\quad\quad-36\alpha_2^2+10\alpha_3^2-26\alpha_2\alpha_3\big)Q^2\bigg]\\
&+\mathcal{O}(r^{-8})\;,
\end{aligned}
\label{eq:phi}
\end{equation}
\begin{equation}
\begin{aligned}
N=&1+\frac{8\pi\big(3\alpha_1+10\alpha_2+3\alpha_3\big)Q^2}{r^4}\\
&+\frac{2048\pi^2\big(21\alpha_1+56\alpha_2+30\alpha_3\big)\alpha_3MQ^2}{7r^7}\\
&+\mathcal{O}(r^{-8})\;,
\end{aligned}
\label{eq:N}
\end{equation}
\begin{equation}
\begin{aligned}
U=&1-\frac{2M}{r}+\frac{Q^2}{r^2}-\frac{16\pi\big(\alpha_3-2\alpha_2\big)Q^2}{r^4}\\
&+\frac{16\pi\big(-\alpha_1-6\alpha_2+\alpha_3\big)MQ^2}{r^5}\\
&-\frac{16\pi\big(-4\alpha_1-20\alpha_2+\alpha_3\big)Q^4}{5r^6}\\
&-\frac{4096\pi^2\alpha_3\big(5\alpha_3-7\alpha_2\big)MQ^2}{7r^7}\\
&+\mathcal{O}(r^{-8})\;,
\end{aligned}
\label{eq:U}
\end{equation}

The series expansion is truncated at $\mathcal{O}(r^{-8})$ for two main reasons. First, this order is sufficient to capture the leading effects induced by the non-minimal coupling constants, while higher-order corrections are expected to remain subdominant. Second, truncating the expansions at $\mathcal{O}(r^{-8})$ ensures that the highest-order contributions to $U(r)$, $N(r)$, and $\phi(r)$ appear at the same order in $1/r$, leading to a consistent asymptotic treatment of the solutions. We emphasize that the above solutions are valid only for static, spherically symmetric, and asymptotically flat spacetimes. Otherwise, the resulting geometry may develop a conical deficit at spatial infinity, signaling a nonphysical asymptotic structure within the present setup.

As shown in Eqs.~\eqref{eq:phi}, \eqref{eq:N}, and \eqref{eq:U}, the contributions induced by the non-minimal couplings are proportional to powers of the electric charge $Q$. In the neutral limit $Q=0$, the solution reduces exactly to the Schwarzschild spacetime, and all coupling effects disappear. To make the influence of the electric charge more manifest, we shall focus on the representative case $Q=M/2$ in the following analysis. 

We note that astrophysical black holes are generally expected to be nearly electrically neutral. Nevertheless, recent studies suggest that black holes may still acquire a small but non-zero electric charge under certain physical conditions. For instance, Ref.~\cite{Araya:2022few} showed that primordial black holes can possess a nonzero net charge at formation due to either Poisson fluctuations at horizon crossing or high-energy particle collisions. They further argued that the plasma within virialized dark-matter halos may endow primordial black holes with a net negative charge. Therefore, charged black-hole spacetimes can still provide a meaningful framework for exploring possible deviations from standard black-hole solutions and their potential observational signatures. 

Although the choice $Q=M/2$ appears significantly larger than the charge estimates reported in Ref.~\cite{Araya:2022few}, the two situations are not directly comparable. The analysis in Ref.~\cite{Araya:2022few} was performed within the standard Einstein--Maxwell theory, whereas the present work considers a modified Einstein--Maxwell theory with explicit non-minimal couplings between the electromagnetic field and spacetime curvature tensors. In the present study, the charge parameter $Q$ should therefore be regarded primarily as a phenomenological parameter introduced to illustrate the effects of the non-minimal couplings on the spacetime geometry and the associated observational signatures.

\section{Photon dynamics}
\label{sec:photon dynamics}

The optical structure of a black hole arises from the motion of photons in its gravitational field. Therefore, before analyzing shadows and photon rings, we first derive \blue{EoM} for photons based on the previously obtained spacetime metric.

For the spacetime under consideration, the metric components Eq.~\eqref{eq:metric} are independent of the coordinates $t$ and $\varphi$, i.e., $\partial_t g_{\mu\nu}=0$ and $\partial_\varphi g_{\mu\nu}=0$. This implies that the spacetime admits two Killing vectors, $\xi_{(t)}^\mu=(\partial_t)^\mu$ and $\xi_{(\varphi)}^\mu=(\partial_\varphi)^\mu$, associated with stationarity and axial symmetry, respectively. These vectors satisfy the Killing equation $\nabla_{(\mu}\xi_{\nu)}=0$ with $\xi_\mu=g_{\mu\nu}\xi^\nu$ \cite{wald:2010general,carroll:2019spacetime}.

For a photon moving along a geodesic with tangent vector $\dot{x}^\mu:=\frac{\mathrm{d}x^\mu}{\mathrm{d}\tau}=(\dot{t},\dot{r},\dot{\theta},\dot{\varphi})$ and canonical momentum $p_\mu=g_{\mu\nu}\dot{x}^\nu$, where $\tau$ represents the affine parameter (or the proper time for massive particles), the quantity $p_\mu \xi^\mu$ is conserved along the trajectory. This follows from
\begin{equation}
\begin{aligned}
\frac{\mathrm{d}}{\mathrm{d}\tau}(p_\mu \xi^\mu)&=\dot{x}^\nu \nabla_\nu (\dot{x}^\mu\xi_\mu)\\
&=\xi_\mu\dot{x}^\nu\nabla_\nu\dot{x}^\mu+\dot{x}^\mu\dot{x}^\nu\nabla_{(\mu}\xi_{\nu)}=0
\end{aligned}
\end{equation}
where the geodesic equation $\dot{x}^\nu\nabla_\nu\dot{x}^\mu=\ddot{x}^\mu+\Gamma^\mu_{\sigma\rho}\dot{x}^\sigma\dot{x}^\rho=0$ and the Killing equation have been used. As a result, the existence of the above symmetries leads to two conserved quantities for photon motion,
\begin{equation}
\mathcal{E}=p_\mu\xi^\mu_{(t)}=-p_t\;, \qquad \mathcal{L}_\varphi=p_\mu\xi^\mu_{(\varphi)}=p_\varphi\;,
\end{equation}
which correspond to the energy and the azimuthal angular momentum, respectively. In terms of the tangent vector and the metric Eq.~\eqref{eq:metric},  we obtain the first two EoM,
\begin{equation}
\left\{
\begin{aligned}
\dot{t}&=\frac{\mathcal{E}}{U}\;,\\
\dot{\varphi}&=\frac{\mathcal{L}_\varphi}{r^2\sin^2\theta}\;.
\end{aligned}
\right.
\label{eq:motion equation 03}
\end{equation}

Plugging the metric ansatz in Eq.~\eqref{eq:metric} into the normalization condition of tangent vector of null geodesics, i.e. $g_{\mu\nu}\dot{x}^\mu\dot{x}^\nu=0$, we derive
\begin{equation}
-\frac{\mathcal{E}^2}{U}+\frac{N^2}{U}\dot{r}^2+r^2\dot{\theta}^2+\frac{\mathcal{L}_\varphi^2}{r^2\sin^2\theta}=0\;.
\end{equation}
This equation can be separated into another two EoM \cite{Carter:1968rr}
\begin{equation}
\left\{
\begin{aligned}
r^2\dot{r}&=\pm\sqrt{\mathsf{R}(r)}\;,\\
r^2\dot{\theta}&=\pm\sqrt{\mathsf{\Theta}(\theta)}\;,
\end{aligned}
\right.
\label{eq:motion equation 12}
\end{equation}
where $\mathsf{R}(r)$ and $\mathsf{\Theta}(\theta)$ are given by
\begin{equation}
\left\{
\begin{aligned}
\mathsf{R}(r)&=\frac{\mathcal{E}^2}{N^2}r^4-\frac{(\mathcal{C}+\mathcal{L}_\varphi^2)U}{N^2}r^2\;,\\
\mathsf{\Theta}(\theta)&=\mathcal{C}-\mathcal{L}_\varphi^2\mathrm{cot}^2\theta\;.
\end{aligned}
\right.
\label{eq:separation}
\end{equation}
This separation closely resembles the Hamilton--Jacobi separability structure to geodesic motion in the Kerr spacetime \cite{Carter:1968rr}. In the present case, it reduces to the non-rotating limit, where the black hole angular momentum vanishes. The sign ``$\pm$" in Eq.~\eqref{eq:motion equation 12} arises from taking the square root and reflects the symmetry of photon trajectories when passing by the black hole. The quantity $\mathcal{C}$ is known as the Carter constant, and defined as
\begin{equation}
\mathcal{C} = p_\theta^2+\mathcal{L}_\varphi^2\cot^2\theta\;,
\end{equation}
which represents the part of the total angular momentum orthogonal to the azimuthal component $\mathcal{L}_\varphi$. 

Although all EoM are obtained, ``$\pm$'' in Eq.~\eqref{eq:motion equation 12} can lead to the accumulation of numerical errors, especially for higher-order rings. The detailed discussion of this issue is provided in Appendix~\ref{app: EoM for integration}. Thus, although these EoM provide valuable insight into the conserved quantities and analytical structure of the system, they are not the most suitable form for numerical integration. We therefore do not directly use these EoM for constructing black-hole shadows, and instead adopt an alternative set of \blue{EoM} introduced in Appendix~\ref{app: EoM for integration}, which is better suited for stable numerical integration. Based on these EoM, we numerically integrate the photon trajectories and generate the corresponding black-hole images. The detailed procedure is described in Appendix~\ref{app:backward ray-tracing approach}.

\section{Thin accretion disk model}
\label{sec: accretion disk model}
Since light serves as \blue{one of} the most direct messengers of the black hole spacetime geometry, and the radiation surrounding astrophysical black holes mainly arises from accretion disks, an appropriate disk model is required.

It should be noted that the positions and widths of photon-ring features in black-hole images discussed in Sec.~\ref{sec:images of shadows and rings} depend on the choice of accretion disk model, particularly for low-order rings. In contrast, higher-order photon rings arise from photons undergoing multiple windings around the black hole due to strong gravitational lensing, and their structure is expected to be primarily governed by the underlying spacetime geometry rather than the detailed emission model. This suggests that the results concerning higher-order rings in Secs.~\ref{sec: general theoretical analysis} and~\ref{sec:images of shadows and rings} are largely insensitive to the specific disk model.

Nevertheless, the choice of accretion disk model can affect the visibility and separability of individual photon-ring features in the simulated images. As shown in Ref.~\cite{Guerrero:2022qkh}, Disk Model I leads to a more discrete intensity distribution across different photon-ring orders, whereas the other models produce smoother radial profiles (see Figs.~8–10 therein). This difference may facilitate the identification of individual photon-ring contributions in the image. In addition, the inner edge of Disk Model I coincides with \textit{the innermost stable circular orbit} ISCO, ensuring that the emitting matter remains on physically timelike orbits. These features motivate the use of this model in the present work.

In this disk model, the emission intensity is given by
\begin{equation}
I_{em}/I_0 =
\begin{cases}
[r-(r_{isco}-1)]^{-2} & r\geq r_{isco}\;,\\
0&\text{otherwise}\;,
\end{cases}
\label{eq: emission intensity}
\end{equation}
where $r_{isco}$ is the radius of ISCO of timelike particles, and $I_0$ is the maximum of emission intensity. In a curved spacetime with metric ansatz Eq.~\eqref{eq:metric}, the radial equation of a timelike particle is \cite{Chandrasekhar:1998}
\begin{equation}
\bigg(\frac{N(r)\dot{r}}{\mathscr{E}}\bigg)^2=1-U(r)\big(\mathscr{E}^{-2}+\frac{\ell^2}{r^2}\big):=V^{isco}_{eff}(r)\;,
\end{equation}
where $\mathscr{E}=\mathcal{E}/\mathscr{M}$ and $\ell=\mathcal{L}/\mathscr{M}$ denote the specific energy and the specific angular momentum respectively, with $\mathscr{M}$ and $\mathcal{L}$ here denoting the mass and the total angular momentum of particle respectively. $V^{isco}_{\mathrm{eff}}$ is the effective potential governing radial motion. Then $r_{isco}$ is the solution of the following equations
\begin{equation}
V^{isco}_{eff}(r) =\frac{\mathrm{d}V^{isco}_{eff}(r)}{\mathrm{d}r}=
\frac{\mathrm{d}^2V^{isco}_{eff}(r)}{\mathrm{d}r^2}= 0\;.
\label{eq: isco effective potential equation}
\end{equation}
%or equivalently
%\begin{equation}
%UU^{\prime\prime}-2(U^\prime)^2+3\frac{UU^\prime}{r}=0.
%\label{eq: isco equation}
%\end{equation}

Accordingly, the observed intensity is calculated by the following formula \cite{Guerrero:2022qkh,Gralla:2019xty}
\begin{equation}
I_{obs}(b)=\sum_m U(r)^2 I_{em}(r)\bigg|_{r_m(b)}\;,
\label{eq: observed intensity}
\end{equation}
where $r_m(b)$ is known as the transfer function, representing the radial coordinate of the $m$-th intersection of the light ray with impact parameter $b$ and the accretion disk. The transfer function $r_m(b)$ is constructed by tracking the radial coordinate $r$ of a photon with impact parameter $b$ at its $m$-th intersection with the accretion disk during the backward ray-tracing procedure described in Appendix~\ref{app:backward ray-tracing approach}.

\begin{figure}
  \centering
  \includegraphics[width=\linewidth]{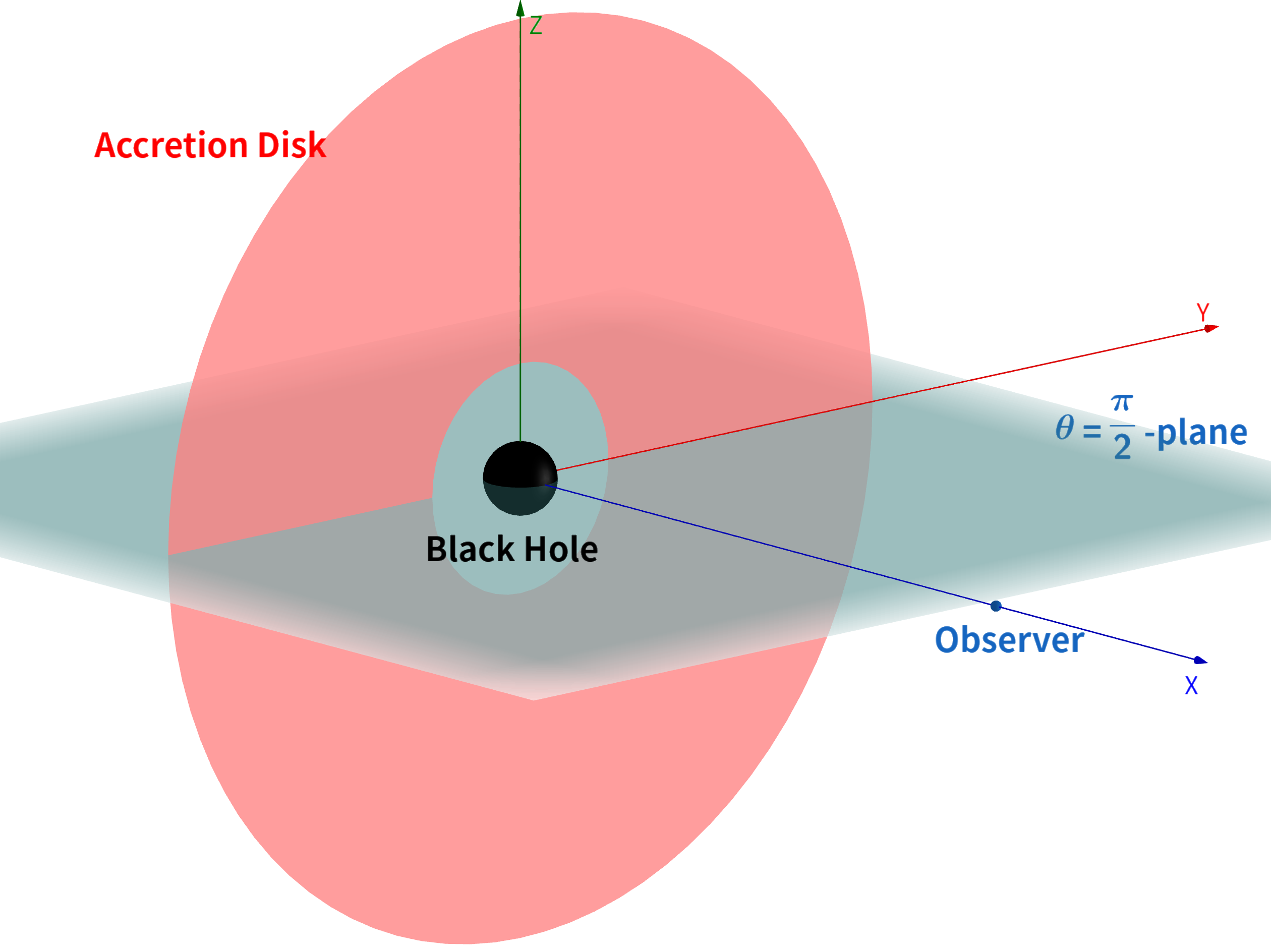}
  \caption{Schematic diagram of the relative positions of observer, black hole, and accretion disk.}
  \label{fig: disk}
\end{figure}

In addition, we suppose the disk plane is perpendicular to the line connecting the observer and the black hole, as shown in Fig.\ref{fig: disk}. For convenience in the numerical ray-tracing and visualization of the black-hole image, we adopt a Cartesian coordinate system in which the observer is located on the $X$-axis and the accretion disk lies in the $Y$--$Z$ plane, as shown in Fig.~\ref{fig: disk}. In this coordinate system, the spherical coordinates are defined as
\begin{equation}
\left\{
\begin{aligned}
r&=\sqrt{X^2+Y^2+Z^2}\,, \\
\theta&=\arccos(Z/r)\,,\\ 
\varphi&=\arctan(Y/X)\,.\\ 
\end{aligned}
\right.
\end{equation}
Hence, the $\theta=\frac{\pi}{2}$ plane (the blue plane in Fig.\ref{fig: disk}) is orthogonal to the disk plane.

\begin{figure}
  \centering
  \includegraphics[width=\linewidth]{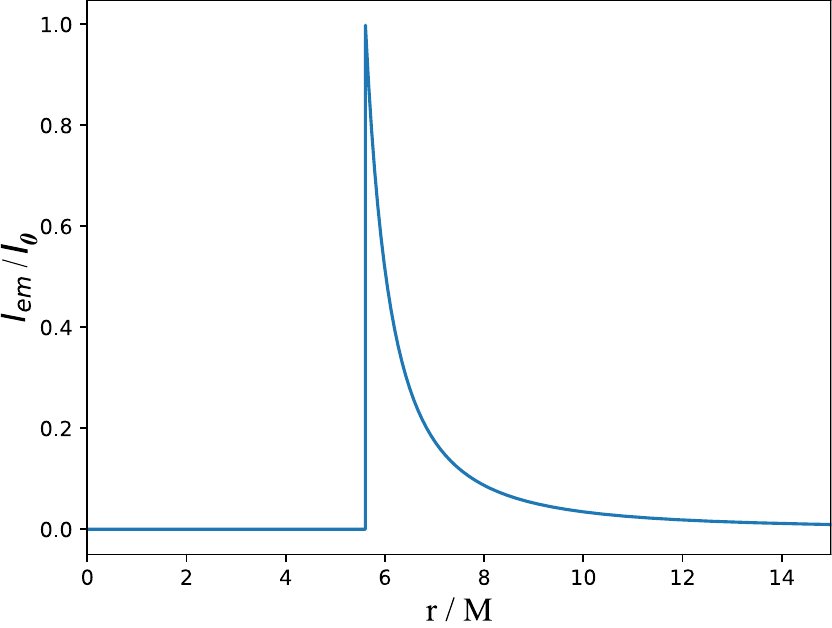}
  \caption{Function profile of emission intensity Eq.~\eqref{eq: emission intensity}.}
  \label{fig: RN Iem}
\end{figure}

\begin{figure}
  \centering
  \includegraphics[width=\linewidth]{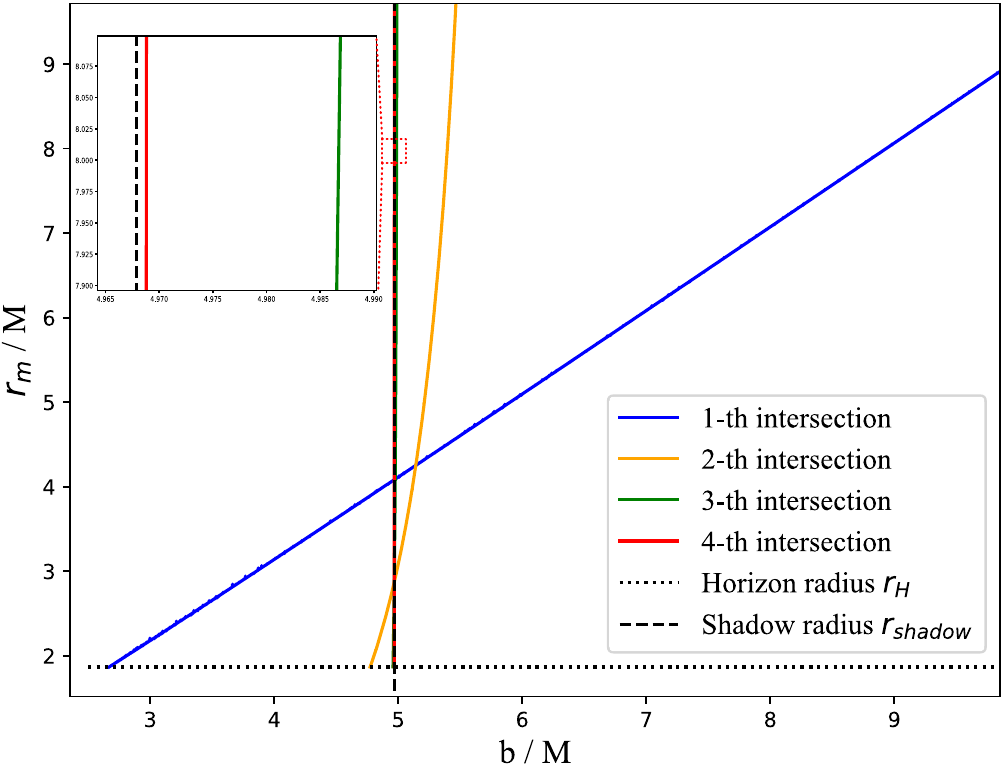}
  \caption{\blue{The first four transfer functions of the Reissner--Nordstr\"{o}m black hole, together with the horizon radius $r_m=r_H$ and the shadow radius $b=r_{shadow}$. The third transfer-function branch already approaches the curve $b=r_{shadow}$ closely, while the fourth branch nearly coincides with it, indicating that the corresponding impact parameters provide good approximations to the shadow radius. All transfer functions are defined only outside the horizon, with their domain starting from $r_m=r_H$.}}
  \label{fig: RN transfer function}
\end{figure}

\begin{figure}
  \centering
  \includegraphics[width=\linewidth]{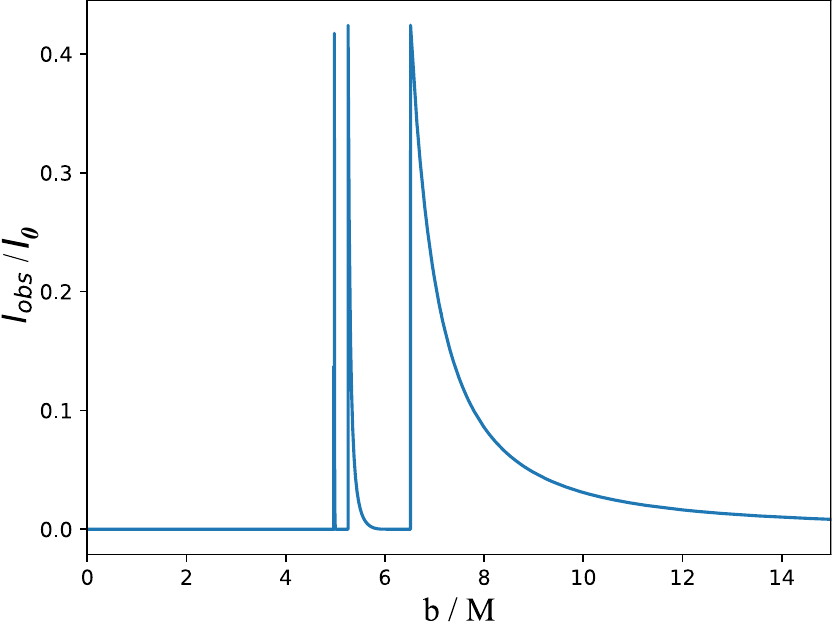}
\caption{The profile of observed intensity Eq.~\eqref{eq: observed intensity}. It is obvious that there are three peaks corresponding to the three rings in Fig.~\ref{fig: RN BH}.}
  \label{fig: RN Iobs}
\end{figure}

\begin{figure}
  \centering
  \includegraphics[width=\linewidth]{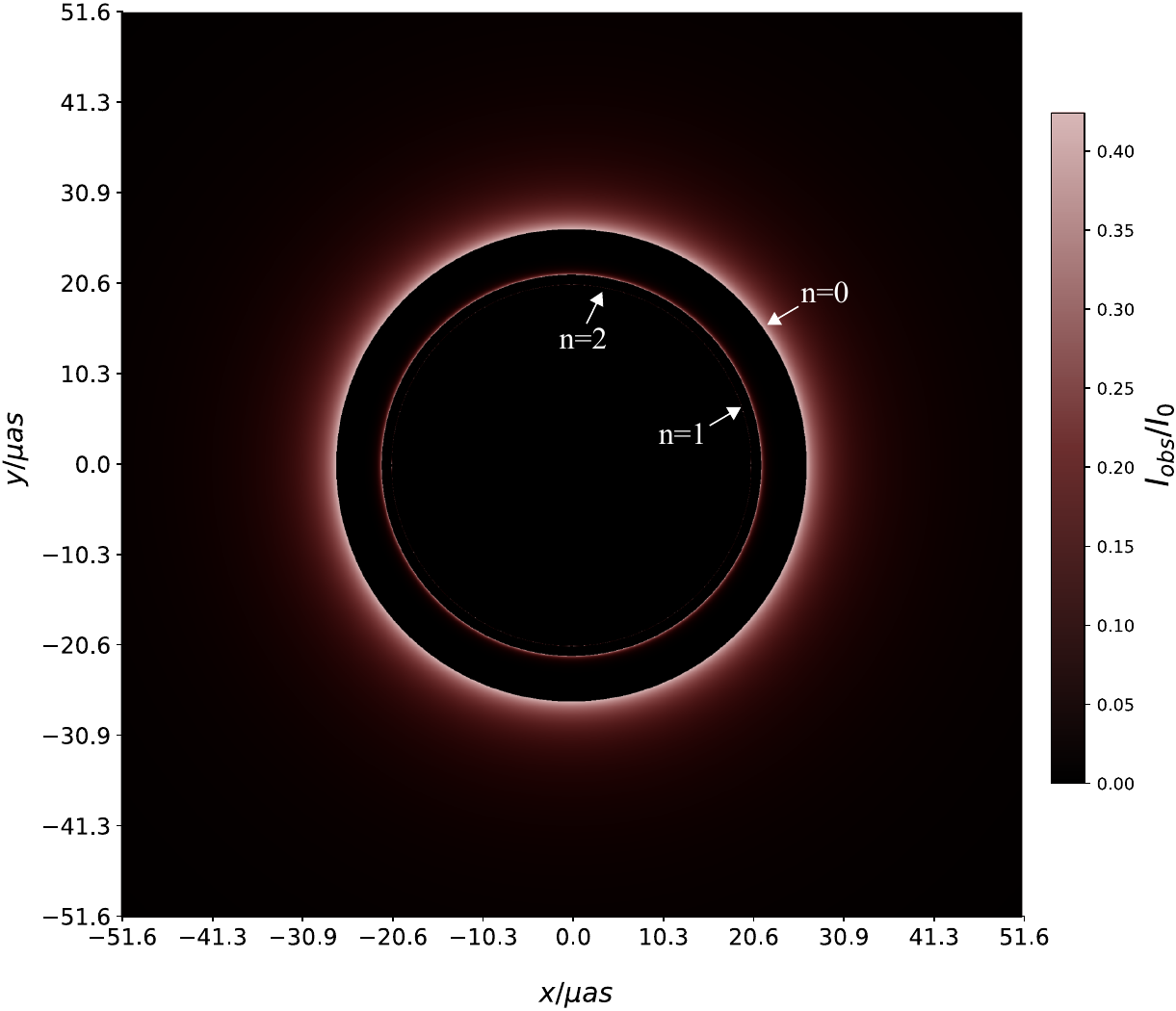}
  \caption{The image of a Reissner--Nordstr\"{o}m black hole, where the color intensity encodes the observed intensity. The outer two rings ($n=0,1$) are easily visible, whereas the third ring ($n=2$) is difficult to resolve due to its extremely narrow width.}
  \label{fig: RN BH}
\end{figure}

Taking Reissner-Nordstr\"{o}m black hole with $Q=M/2$ as an example, Fig.~\ref{fig: RN Iem} shows the emission intensity profile and Fig.~\ref{fig: RN transfer function} shows the corresponding transfer function. Based on these, Fig.~\ref{fig: RN Iobs} presents the resulting observed intensity, and Fig.~\ref{fig: RN BH} shows the final observed image of the black hole. The detailed interpretation of these figures will be discussed in later sections and is therefore not repeated here.

\section{General theoretical analysis}
\label{sec: general theoretical analysis}

In this section, we present a general theoretical analysis of the effects of the non-minimal couplings in Eq.~\eqref{eq:action} on black hole properties, based on the solutions in Eqs.~\eqref{eq:phi}, \eqref{eq:N}, and \eqref{eq:U}. In particular, we focus on the radii of the event horizon and the photon sphere, and the structure of photon rings. For simplicity, we label each coupling term by its corresponding coefficient: $\alpha_1$ for $\alpha_1 F^\sigma_{\ \nu} F_{\sigma\rho} R^{\nu\rho}$, $\alpha_2$ for $\alpha_2 F^2 R$, and $\alpha_3$ for $\alpha_3 F_{\mu\nu} F_{\sigma\rho} R^{\mu\nu\sigma\rho}$.

\subsection{Effects on the event horizon and photon sphere}

\begin{figure}
  \centering
  \includegraphics[width=\linewidth]{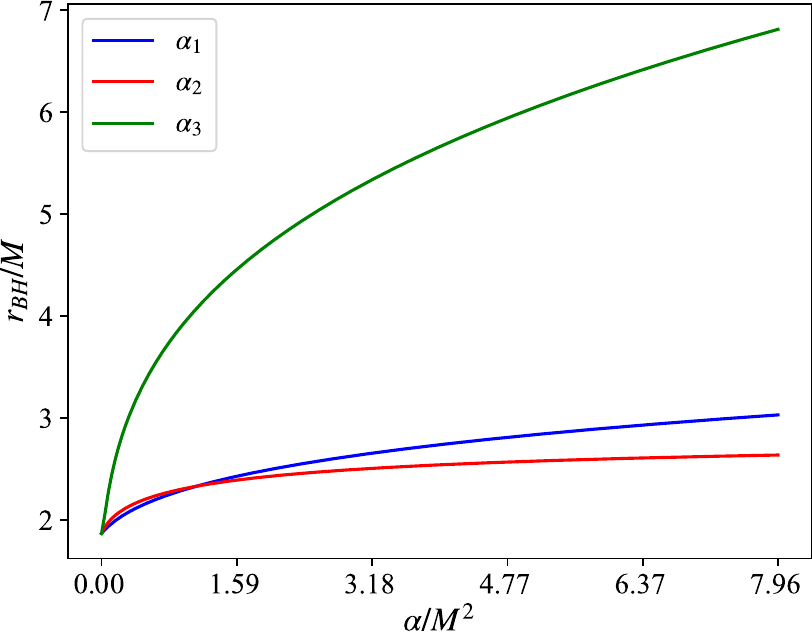}
  \caption{Effects of couplings on the radius of the horizon.}
  \label{fig: profile horizon}
\end{figure}

In this subsection, we will explore how non-minimal couplings alter the event horizon and photon sphere of black holes. Based on the metric ansatz Eq.~\eqref{eq:metric}, radius of the event horizon is the maximum root of Eq.~\eqref{eq:U}, i.e.
\begin{equation}
r_H=\max\{\, r>0 \mid U(r)=0 \,\}\;.
\end{equation}
Fig.~\ref{fig: profile horizon} illustrates the influence of the non-minimal couplings on the event horizon. From the figure, we find that all coupling terms lead to an enlargement of the horizon radius, with the differences lying primarily in the rate at which the horizon grows, i.e. $\frac{\mathrm{d}r_H}{\mathrm{d}\alpha_3}\gg \frac{\mathrm{d}r_H}{\mathrm{d}\alpha_1} \gtrsim \frac{\mathrm{d}r_H}{\mathrm{d}\alpha_2}$.

We now proceed to the photon sphere, which is closely related to the formation of black-hole shadows. Rewriting the radial equation in Eq.~\eqref{eq:motion equation 12} in terms of an effective potential, we obtain
\begin{equation}
\bigg(\frac{N(r)\dot{r}}{\mathcal{E}}\bigg)^2
= 1 - U(r)\frac{b^2}{r^2}
:= V^{ps}_{eff}(r)\;,
\end{equation}
where $b=\sqrt{\mathcal{C}+\mathcal{L}_\varphi^2}/\mathcal{E}$ is the photon impact parameter. 
The photon sphere radius $r_{ps}$ corresponds to an unstable circular null orbit and is therefore determined by the conditions
\begin{equation}
V^{ps}_{eff}(r_{ps})=0,
\qquad
\left.\frac{\mathrm{d}V^{ps}_{eff}}{\mathrm{d}r}\right|_{r=r_{ps}}=0\;.
\label{eq: ps eff}
\end{equation}
The impact parameter $b_{ps}$ corresponds to the apparent radius of the photon sphere as seen by a distant observer and therefore defines the boundary of the black-hole shadow, i.e. $r_{\rm shadow}\equiv b_{ps}$. Hence, we shall hereafter refer to $b_{ps}$ simply as the shadow radius for convenience. Any photon with an impact parameter $b<b_{ps}$ will inevitably be captured by the black hole. According to Eq.~\eqref{eq: ps eff}, $b_{ps}$ is given by
\begin{equation}
b_{ps}=\frac{r_{ps}}{\sqrt{U(r_{ps})}}\;.
\end{equation}

\begin{figure}
  \centering
  \includegraphics[width=\linewidth]{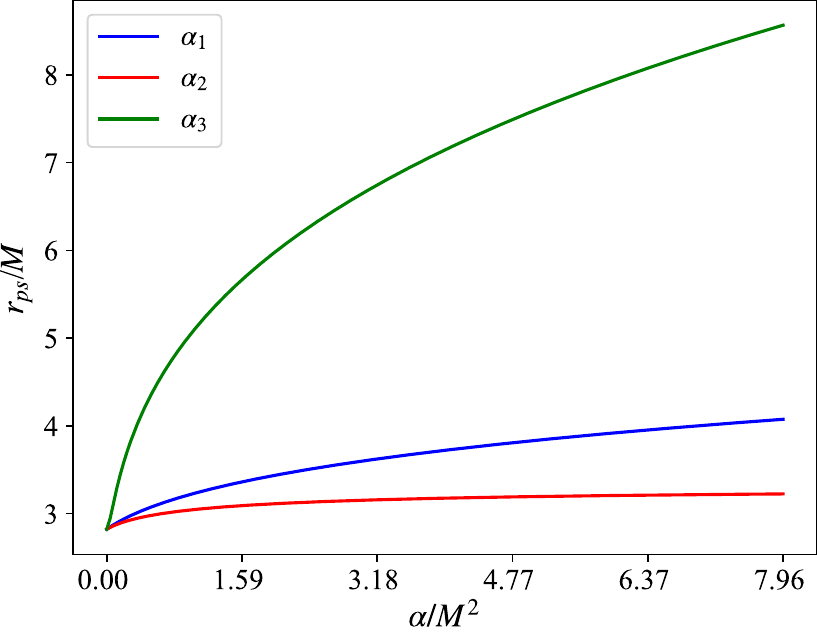}
  \caption{Effects of couplings on the photon-sphere radius.}
  \label{fig: profile rps}
\end{figure}

\begin{figure}
  \centering
  \includegraphics[width=\linewidth]{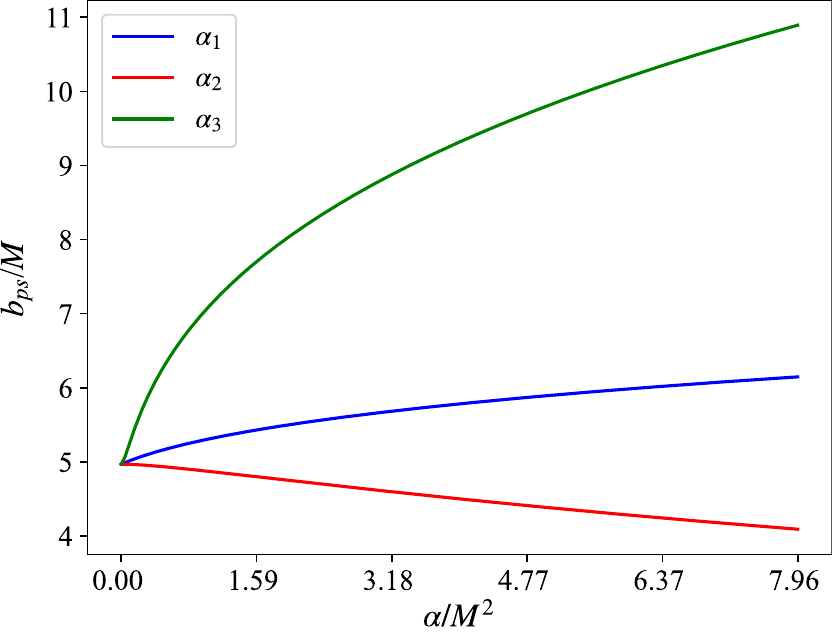}
  \caption{Effects of the couplings on the shadow radius.}
  \label{fig: profile bps}
\end{figure}

Fig.~\ref{fig: profile rps} shows the dependence of the photon sphere radius $r_{ps}$ on the coupling coefficients $\alpha_1$, $\alpha_2$, and $\alpha_3$. In all three cases, $r_{ps}$ increases with the coupling, with the growth rates following $\frac{\mathrm{d}r_{ps}}{\mathrm{d}\alpha_3} \gg \frac{\mathrm{d}r_{ps}}{\mathrm{d}\alpha_1} > \frac{\mathrm{d}r_{ps}}{\mathrm{d}\alpha_2}$. In contrast, as shown in Fig.~\ref{fig: profile bps}, the corresponding impact parameter $b_{ps}$ also increases for $\alpha_1$ and $\alpha_3$ with $\frac{\mathrm{d}b_{ps}}{\mathrm{d}\alpha_3} > \frac{\mathrm{d}b_{ps}}{\mathrm{d}\alpha_1}$, but decreases for $\alpha_2$.

\subsection{Effects on photon rings}
\begin{figure}
  \centering
  \includegraphics[width=\linewidth]{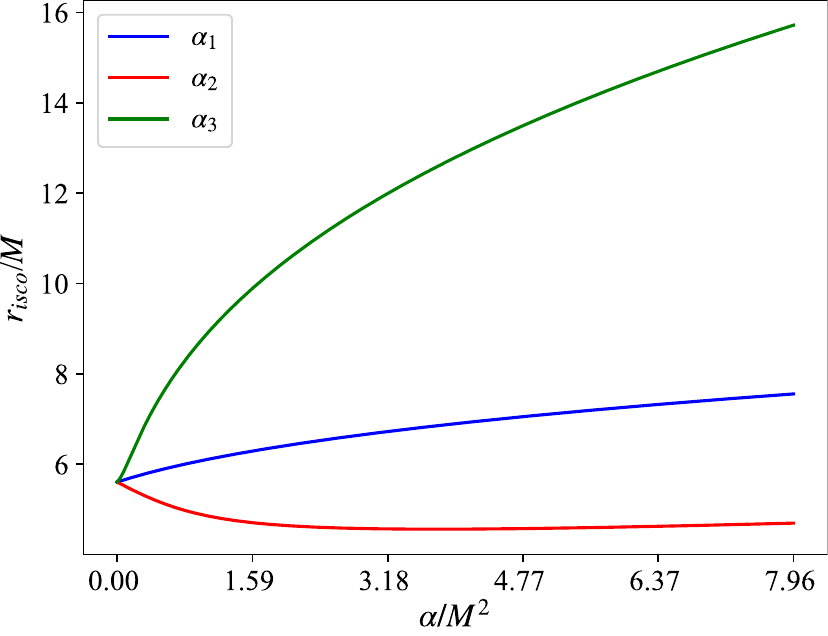}
  \caption{Effects of couplings on the radius of ISCO.}
  \label{fig: profile risco}
\end{figure}

In this subsection, we examine how each non-minimal coupling term in Eq.~\eqref{eq:action} influences the structure of photon rings produced by the thin accretion disk model introduced in Sec.~\ref{sec: accretion disk model}. As shown in Fig.~\ref{fig: RN Iem}, the dominant contribution to the emission intensity originates from photons emitted near ISCO ($r = r_{isco}$, which is the solution of Eq.~\eqref{eq: isco effective potential equation}). Although photons emitted from other regions of the accretion disk can also contribute to the formation of photon rings, their intensities are typically subdominant compared with the intensity associated with emission near ISCO. Moreover, their relative positions exhibit a qualitatively similar pattern, differing mainly by an overall rescaling. Consequently, we primarily focus on the photon rings formed by photons emitted near ISCO in this section. Fig.~\ref{fig: profile risco} presents the dependence of ISCO radius on the coupling parameters, from which we find that the $\alpha_1$ and $\alpha_3$ increase $r_{isco}$, whereas $\alpha_2$ leads to a reduction of ISCO radius.

Exploiting the spherical symmetry of the spacetime, photon trajectories can, without loss of generality, be restricted to the plane $\theta=\pi/2$, corresponding to the blue plane in Fig.~\ref{fig: disk}. Although the geodesic equations are generally valid for arbitrary values of $\theta$, this choice provides a convenient representative case for analysis. Moreover, since the spacetime is stationary and null geodesics are invariant under reversal of the affine parameter, photon trajectories connecting the observer and the emission region can be equivalently computed using a backward ray-tracing scheme. In practice, we trace photons from an observer located at $(r_O,\pi/2,0)$ to ISCO at $(r_{\mathrm{isco}},\pi/2,\tilde{\varphi}_{\tilde{n}})$ in the analysis. The trajectories of such photons are governed by
\begin{equation}
\frac{\mathrm{d}r}{\mathrm{d}\varphi}
=\frac{\dot{r}}{\dot{\varphi}}
=\pm \frac{r^2}{bN(r)}\sqrt{1-\dfrac{b^2}{r^2}U(r)}\;.
\label{eq: r-phi equation}
\end{equation}
Without loss of generality, we consider only trajectories with increasing azimuthal angle $\varphi$. Accordingly, the plus (minus) sign in Eq.~\eqref{eq: r-phi equation} corresponds to outgoing (ingoing) photon trajectories, where ``outgoing'' and ``ingoing'' are defined with respect to the black-hole radial coordinate, with $dr/d\tau>0$ and $dr/d\tau<0$, respectively.

\begin{figure}
  \centering
  \includegraphics[width=\linewidth]{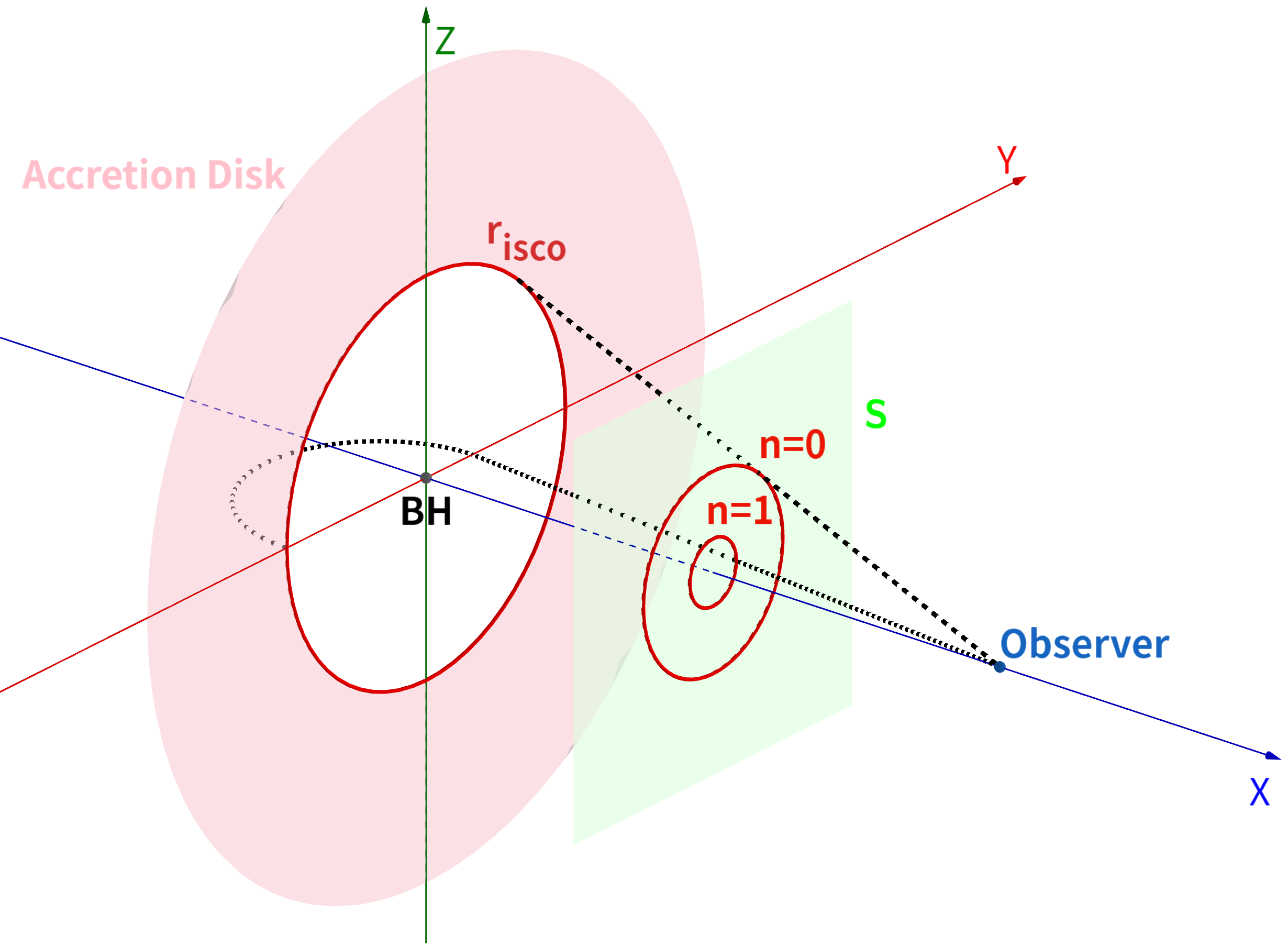}
  \caption{Schematic diagram of photon ring formation. The green S-plane is called the ``camera screen" which is introduced in Appendix~\ref{app:backward ray-tracing approach}. Two dashed lines represent the different paths of photons.}
  \label{fig: rings}
\end{figure}

To characterize different photon trajectories, we define the path order as
\begin{equation}
\tilde{n} = \left\lfloor \frac{\tilde{\varphi}_{\tilde{n}}-0}{\pi} \right\rfloor\;,
\end{equation}
where $\tilde{\varphi}_i = \frac{(2i+1)\pi}{2}$ with $i=0,1,\cdots$ represents the azimuthal coordinate at which the photon reaches the disk and $\left\lfloor \cdot \right\rfloor$ denotes the floor function. As illustrated in Fig.~\ref{fig: rings}, photon trajectories of different orders are perceived by the observer as a sequence of concentric photon rings, and the path order corresponds to the number of half-orbits completed by the photon. Specifically, the zeroth-order photon ring is formed by photons that propagate directly from the observer to ISCO corresponding to an azimuthal angle change of $\Delta\varphi=\pi/2$ (from $\varphi=0$ to $\varphi=\pi/2$), while the first-order photon ring corresponds to photons that complete one half-orbit around the black hole, corresponding to $\Delta\varphi=3\pi/2$ before reaching ISCO (from $\varphi=0$ to $\varphi=3\pi/2$). Higher-order photon rings are generated in an analogous manner, with increasing numbers of orbital windings. In the limit $\tilde{n} \to \infty$, the trajectories asymptotically approach the photon sphere, thereby forming the boundary of the black-hole shadow.

Suppose that the radius of $\tilde{n}$th-order photon ring is $b_{isco}^{\tilde{n}th}$ from the observer's perspective. Then we can calculate the radius of each photon ring via Eq.~\eqref{eq: r-phi equation}. For simplicity, we first define an auxiliary function as
\begin{equation}
\begin{aligned}
\mathcal{F}(x_1,x_2;b) &= \int_{x_1}^{x_2}\frac{-N(r)\mathrm{d}r}{r^2\sqrt{\mathcal{G}(r,b)}}\;,\\
\mathcal{G}(r,b)&=\frac{1}{b^2}-\frac{U(r)}{r^2}\;,  
\end{aligned}
\end{equation}
which comes from the integration of the negative branch of Eq.~\eqref{eq: r-phi equation}.

For the zeroth-order photon ring, as illustrated in Fig.~\ref{fig: rings}, photons propagate directly from the observer to ISCO along trajectories with a monotonically decreasing radial coordinate $r$ until reaching ISCO at $\tilde{\varphi}_0=\pi/2$. Integrating Eq.~\eqref{eq: r-phi equation} along such a trajectory yields
\begin{equation}
\frac{\pi}{2}
=\int_{0}^{\tilde{\varphi}_0} \mathrm{d}\varphi
=\mathcal{F}(r_O,r_{isco};b)\;,
\end{equation}
which implicitly determines the impact parameter $b$. The radius of the zeroth-order photon ring, denoted by $b_{isco}^{0th}$, is given by the minimal root of this equation, namely
\begin{equation}
b^{0th}_{isco}
=\min\bigg\{
b \,\bigg|\, \mathcal{F}(r_O,r_{isco};b)=\frac{\pi}{2}
\bigg\}\;.
\end{equation}
The radii of higher-order photon rings ($\tilde{n} \geq 1$) can be computed in a similar manner. It should be noted, however, that along such trajectories the photon first propagates along an ingoing branch, reaches a periastron, which is the maximal root of $\mathcal{G}(r,b)$, i.e.
\begin{equation}
r_{per}(b)=\max\bigg\{r\bigg|\mathcal{G}(r,b)=0\bigg\}\;,
\end{equation}
and then transitions to an outgoing branch before finally arriving at ISCO. Therefore the radius of $\tilde{n}$th-order photon ring ($\tilde{n}\geq1$) is 
\begin{equation}
\begin{aligned}
b^{\tilde{n}th}_{isco}=\min\bigg\{b\bigg|\mathcal{F}(r_O,r_{per};b)-\mathcal{F}(r_{per},r_{isco};b)=\frac{(\tilde{n}+1)\pi}{2}\bigg\}.
\end{aligned}
\label{eq: b_nth_isco}
\end{equation}
Notably, as $\tilde{n} \to \infty$, the radii of the photon rings converge to a critical value, which defines the boundary of the black-hole shadow, i.e. $b_{ps}\equiv b^{\infty th}_{isco}$.

\begin{figure}
  \centering
  \includegraphics[width=0.95\linewidth]{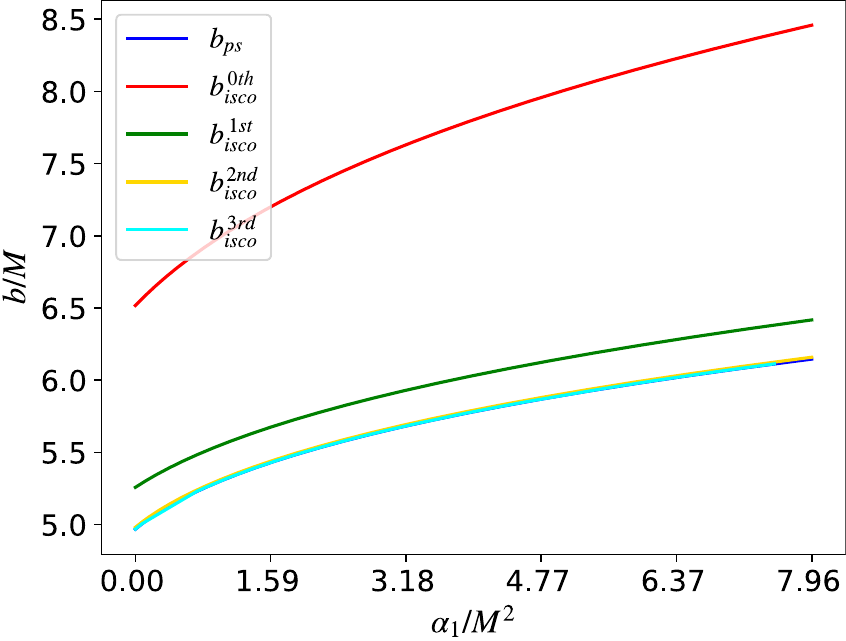}
  \caption{Effects of $\alpha_1 F^\sigma_\nu F_{\sigma\rho} R^{\nu\rho}$ on the radius of photon rings.}
  \label{fig: rings alpha1}
\end{figure}

The effects of the couplings in Eq.~\eqref{eq:action} on the photon-ring structure are illustrated in Figs.~\ref{fig: rings alpha1}, \ref{fig: rings alpha2}, and \ref{fig: rings alpha3} for $\alpha_1$, $\alpha_2$, and $\alpha_3$, respectively.

\begin{figure}
  \centering
  \includegraphics[width=0.95\linewidth]{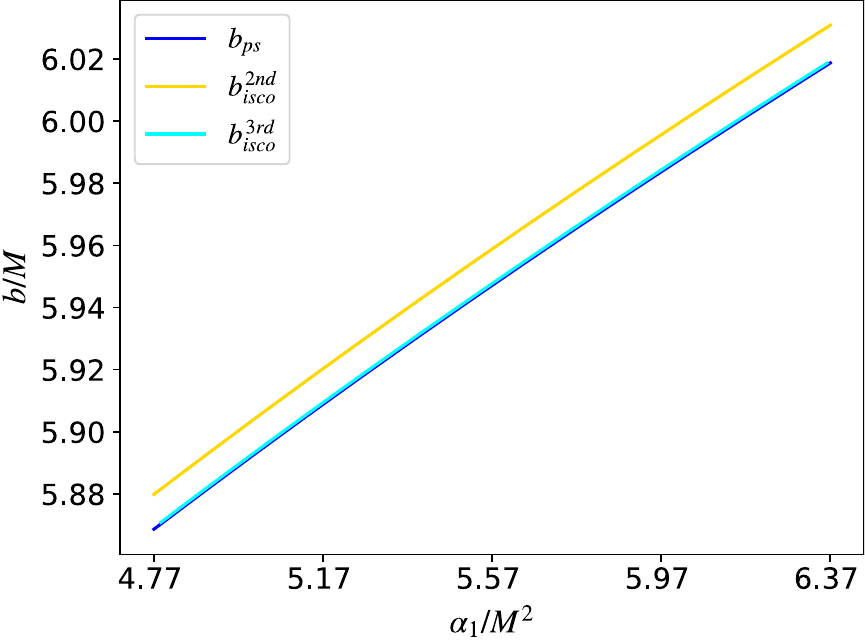}
  \caption{A zoomed-in version of Fig.\ref{fig: rings alpha1}.}
  \label{fig: rings alpha1 zoomed}
\end{figure}

\begin{figure}
  \centering
  \includegraphics[width=0.95\linewidth]{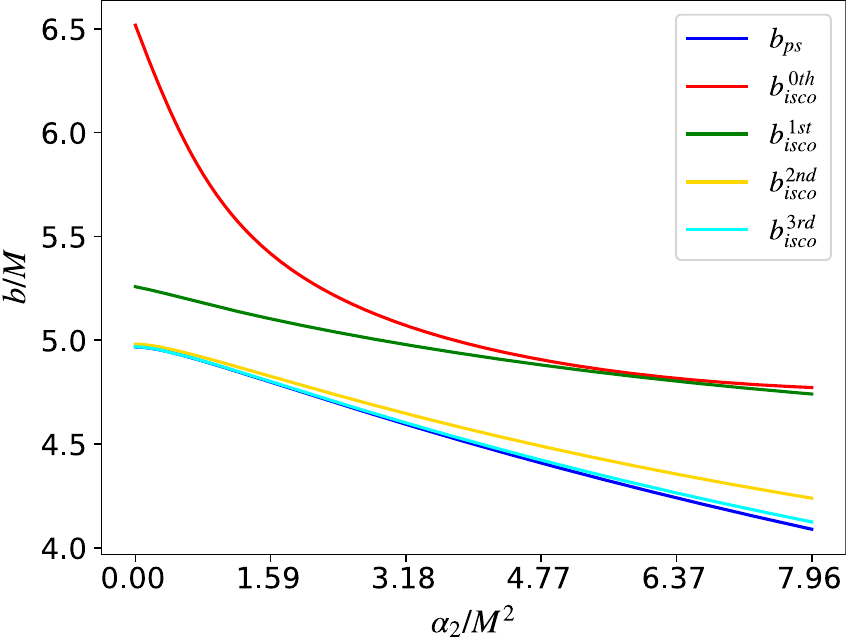}
  \caption{Effects of $\alpha_2 F^2R$ on the radius of photon rings.}
  \label{fig: rings alpha2}
\end{figure}

\begin{figure}
  \centering
  \includegraphics[width=0.95\linewidth]{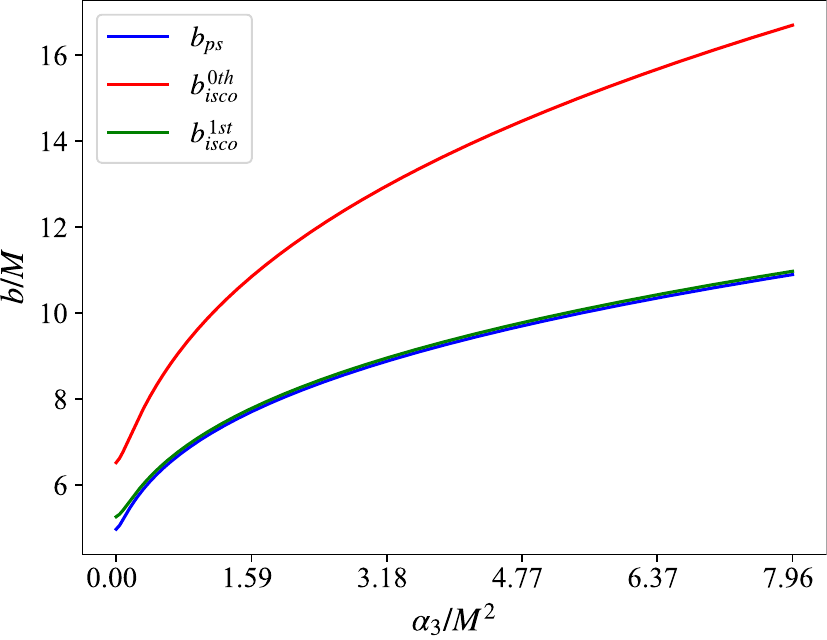}
  \caption{Effects of $\alpha_3 F_{\mu\nu}F_{\sigma\rho}R^{\mu\nu\sigma\rho}$ on the radius of photon rings.}
  \label{fig: rings alpha3}
\end{figure}

From Fig.~\ref{fig: rings alpha1}, we observe that increasing $\alpha_1$ leads to a systematic outward shift of the photon-ring radii, as well as an increase in the separation between the zeroth- and first-order rings. Although higher-order photon rings approach the black-hole shadow radius $b_{\mathrm{ps}}$ as the ring order increases, the spacing between successive higher-order rings remains approximately uniform. This behavior can be seen from the separation between the first- and second-order rings in Fig.~\ref{fig: rings alpha1}, as well as between the second- and third-order rings in the zoomed-in Fig.~\ref{fig: rings alpha1 zoomed}. Furthermore, Fig.~\ref{fig: rings alpha1 zoomed} shows that the deviation between the second-order ring $b^{2\mathrm{nd}}_{\mathrm{isco}}$ and the shadow boundary $b_{\mathrm{ps}}$ is of order $10^{-3}M$, corresponding to a relative difference below $0.4\%$. Therefore, in this coupling scenario, the second-order photon ring provides a very good approximation to the boundary of the black-hole shadow.

Fig.~\ref{fig: rings alpha2} shows that the radii of all photon rings decrease as $\alpha_2$ increases. Notably, the separation between the zeroth- and first-order photon rings shrinks continuously and becomes nearly vanishing for $\alpha_2 \gtrsim 5\,M^2$, where the two rings almost overlap. In contrast, the separations among higher-order photon rings increase with $\alpha_2$, suggesting that higher-order rings may become more distinguishable in this case.

As illustrated in Fig.~\ref{fig: rings alpha3}, all photon rings expand rapidly with increasing $\alpha_3$, accompanied by a significant growth in the separation between the zeroth- and first-order rings. Although only photon rings up to first order are shown in the figure due to numerical limitations\footnote{\blue{Since $b_{isco}^{\tilde{n}th}$ does not admit a closed-form expression, it must be obtained numerically by solving Eq.~\eqref{eq: b_nth_isco}. The required numerical precision increases rapidly with the photon-ring order. This issue is particularly pronounced for the $\alpha_3 F_{\mu\nu}F_{\sigma\rho}R^{\mu\nu\sigma\rho}$ coupling, since higher-order photon rings converge rapidly toward} \blue{the shadow boundary, making their corresponding roots increasingly difficult to resolve. However, the first-order photon-ring curve $b_{isco}^{1st}$ is already nearly indistinguishable from the shadow-boundary curve $b_{ps}$, while all higher-order photon rings must lie between them. Therefore, further increasing the numerical precision to determine higher-order rings would not affect the physical conclusions presented here.}}, it is evident that the spacing between higher-order photon rings rapidly diminishes, leading to their near coincidence at the shadow boundary. Hence, the first-order ring provides a good approximation to the black-hole shadow boundary in this case.

%As illustrated in Fig.~\ref{fig: rings alpha3}, all photon rings expand rapidly with increasing $\alpha_3$, accompanied by a significant growth in the separation between the zeroth- and first-order rings. Although only photon rings up to first order are shown in the figure due to numerical limitations, as higher-order rings require substantially higher precision{\footnote{Since $b_{\rm ISCO}^{\tilde n{\rm th}}$ does not admit a closed-form expression, it must be obtained numerically by solving Eq.~\eqref{eq: b_nth_isco}. The numerical precision required increases rapidly with the photon-ring order, making the computation of higher-order rings substantially more demanding. However, as shown in Fig.~\ref{fig: rings alpha3}, the first-order ring is already nearly indistinguishable from the shadow boundary $b_{\rm ps}$, while all higher-order rings must lie between them. Therefore, computing still higher-order rings would not provide additional physical insight for the present discussion.}}, it is evident that the spacing between higher-order photon rings rapidly diminishes, leading to their near coincidence at the shadow boundary. Hence, the first-order ring provides a good approximation to the shadow boundary of black hole in this case.

\section{Black-hole images}
\label{sec:images of shadows and rings}

After the general theoretical analysis in Sec.~\ref{sec: general theoretical analysis}, we turn to simulated observations and present the corresponding images for several representative values of the coupling parameters, namely $0.08M^2$, $0.32M^2$, $1.27M^2$, and $5.09M^2$. 

We begin by specifying the overall setup of the system. As illustrated in Fig.~\ref{fig: disk}, a black hole of mass $M$ is placed at the origin of a Cartesian coordinate system, while the observer is located at $(X_O,Y_O,Z_O)=(5\times10^{10} M,0,0)$. This choice effectively places the observer in the asymptotic region, ensuring that lensing effects are computed in the far-field limit. For comparison, the distance from the Earth to $\mathrm{Sgr\ A}^*$ (with mass $M \simeq 4.3\times10^6 M_\odot$) is approximately $8.3\,\mathrm{kpc}$ \cite{GRAVITY:2021xju}, corresponding to a distance of order $\sim 4\times10^{10} M$ in units where $G=c=1$, consistent with the scale adopted in this work.

The accretion disk is assumed to lie in the $Y$--$Z$ plane, and its emission intensity is modeled by Eq.~\eqref{eq: emission intensity}. The observer’s field of view is taken to be a circular region of radius $12.5M$ at a distance of $5\times10^{10}M$. This choice is guided by the characteristic scale of ISCO. For a Schwarzschild black hole, ISCO is located at $6M$ \cite{Chandrasekhar:1998}. Since, in the disk model introduced in Sec.~\ref{sec: accretion disk model}, the dominant contribution to photon-ring formation arises from emission near ISCO, we extend the computational window to approximately twice this scale to ensure that the relevant lensing features are fully captured.

\blue{Under this configuration, the field of view corresponds to an angular area of approximately $1.07\times10^4\mu\mathrm{as}^2$. Discretizing the image into a $2000\times2000$ pixel grid yields an angular resolution of approximately $0.05\mu\mathrm{as}$ per pixel. This high-resolution grid is adopted primarily for visualization purposes, in order to display the image structure more clearly and to ensure that the qualitative impact of the coupling parameters on the black-hole images is not obscured. For comparison, this resolution is about three orders of magnitude finer than the nominal angular resolution of EHT, which is approximately $25\mu\mathrm{as}$ at an observing wavelength of $\sim 1.3\mathrm{mm}$ ($\lambda/D$)~\cite{M87EventHorizonTelescope:2019uob}. The resolution used here is therefore close to the idealized limit and is far beyond the reach of the current EHT array.}

%Under this configuration, the field of view corresponds to an angular area of approximately $1.07\times10^4\,\mu\mathrm{as}^2$. Discretizing the image into a $2000\times2000$ pixel grid yields an angular resolution of approximately $0.05\,\mu\mathrm{as}$ per pixel. For comparison, this resolution is about \blue{three} orders of magnitude finer than the nominal angular resolution of the Event Horizon Telescope (EHT), which is approximately $25\,\mu\mathrm{as}$ at an observing wavelength of $\sim 1.3\,\mathrm{mm}$ ($\lambda/D$)~\cite{M87EventHorizonTelescope:2019uob}. \blue{Thus, the resolution used in this study is close to the idealized limit and is far beyond the reach of the current EHT array, and likely also beyond that of foreseeable Earth-based or space-based extensions.}

\subsection{Black-hole images with $\alpha_1F^\sigma_{\ \nu} F_{\sigma\rho}R^{\nu\rho}$ coupling}
\begin{figure}
  \centering
  \includegraphics[width=\linewidth]{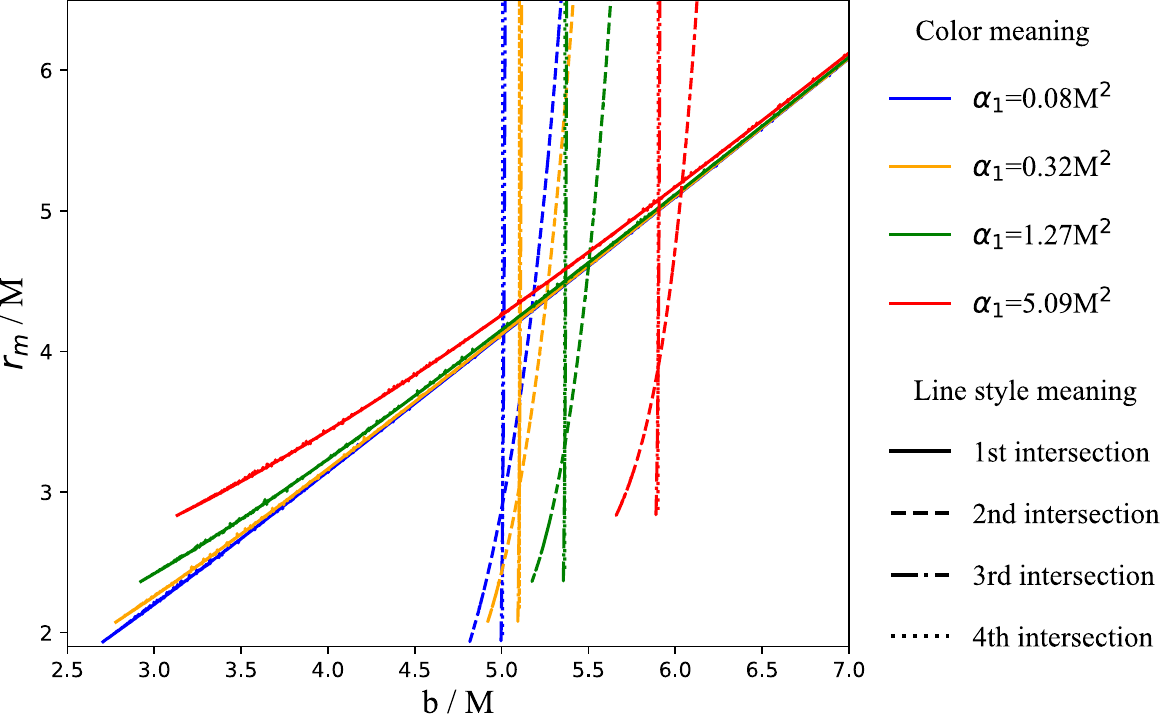}
  \caption{First four transfer functions for four different choices of coupling coefficient $\alpha_1$.}
  \label{fig: xm alpha1}
\end{figure}

\begin{figure}
  \centering
  \includegraphics[width=\linewidth]{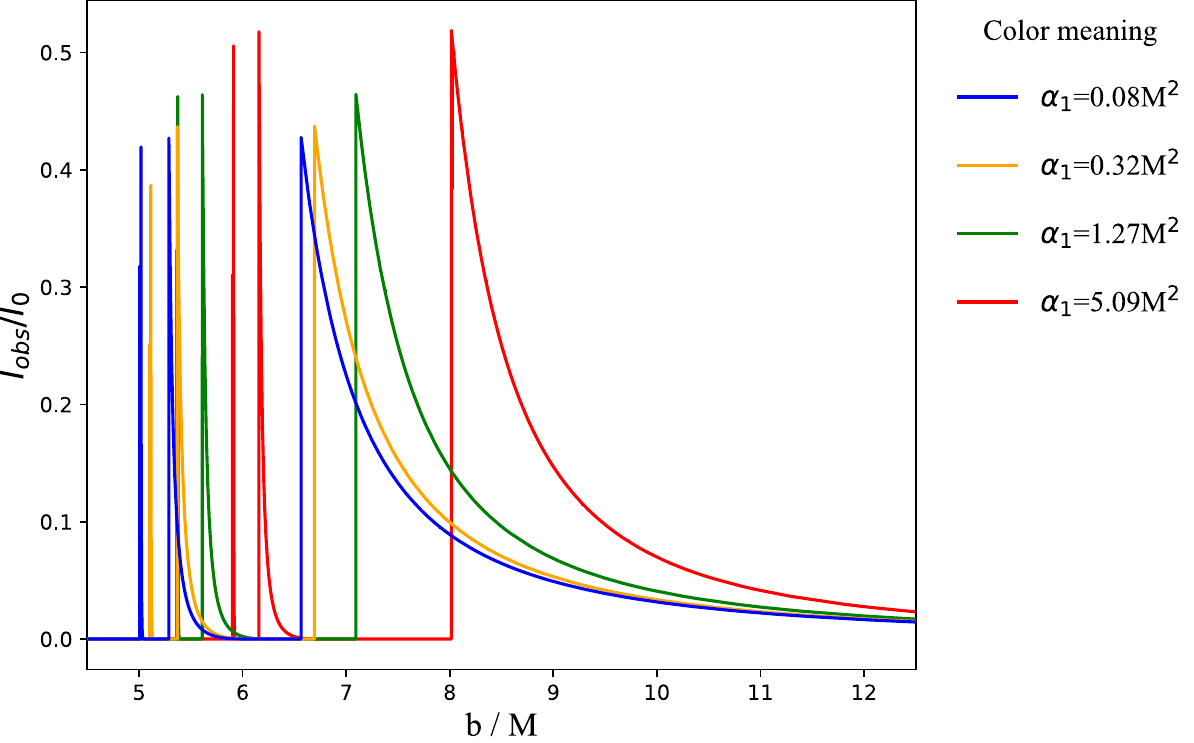}
  \caption{Observed intensity for four different choices of coupling coefficient $\alpha_1$.}
  \label{fig: Iobs alpha1}
\end{figure}

Fig.~\ref{fig: xm alpha1} and Fig.~\ref{fig: Iobs alpha1} illustrate the effects of varying $\alpha_1$ on the transfer functions and the observed intensity, respectively. 

\blue{As briefly illustrated in Fig.~\ref{fig: RN transfer function}, the transfer functions are defined for $r_m\geq r_H$. The third transfer-function branch already approaches the shadow radius closely, while the fourth branch nearly coincides with it, indicating that the corresponding impact parameters provide good approximations to the shadow radius ($r_{shadow}\equiv b_{ps}$). With this interpretation, Fig.~\ref{fig: xm alpha1} shows that increasing $\alpha_1$ leads to an overall displacement of the transfer-function curves toward larger $r_m$ and larger $b$, corresponding to the increase of both $r_H$ and $b_{ps}$. Apart from this displacement, no significant structural change is observed.}

In Fig.~\ref{fig: Iobs alpha1}, three distinct peaks can be clearly identified, corresponding (from right to left) to $b_{isco}^{0th}$ (the zeroth-order photon ring), $b_{isco}^{1st}$ (the first-order photon ring), and $b_{isco}^{2nd}$ (the second-order photon ring). As discussed in Sec.~\ref{sec: general theoretical analysis}, increasing $\alpha_1$ leads to a gradual enlargement of the separation between the zeroth- and first-order photon rings, while the spacing between the first- and second-order rings remains almost unchanged. In addition, the widths of the photon rings decrease with increasing order, which is manifested in the narrowing of the corresponding peaks in the observed intensity. In particular, the peak associated with the second-order photon ring becomes extremely narrow and nearly indistinguishable, indicating that it is hard to resolve in subsequent images.

\begin{figure*}
 \subfloat[$\alpha_1=0.08M^2$]
 {\includegraphics[width=0.48\linewidth]{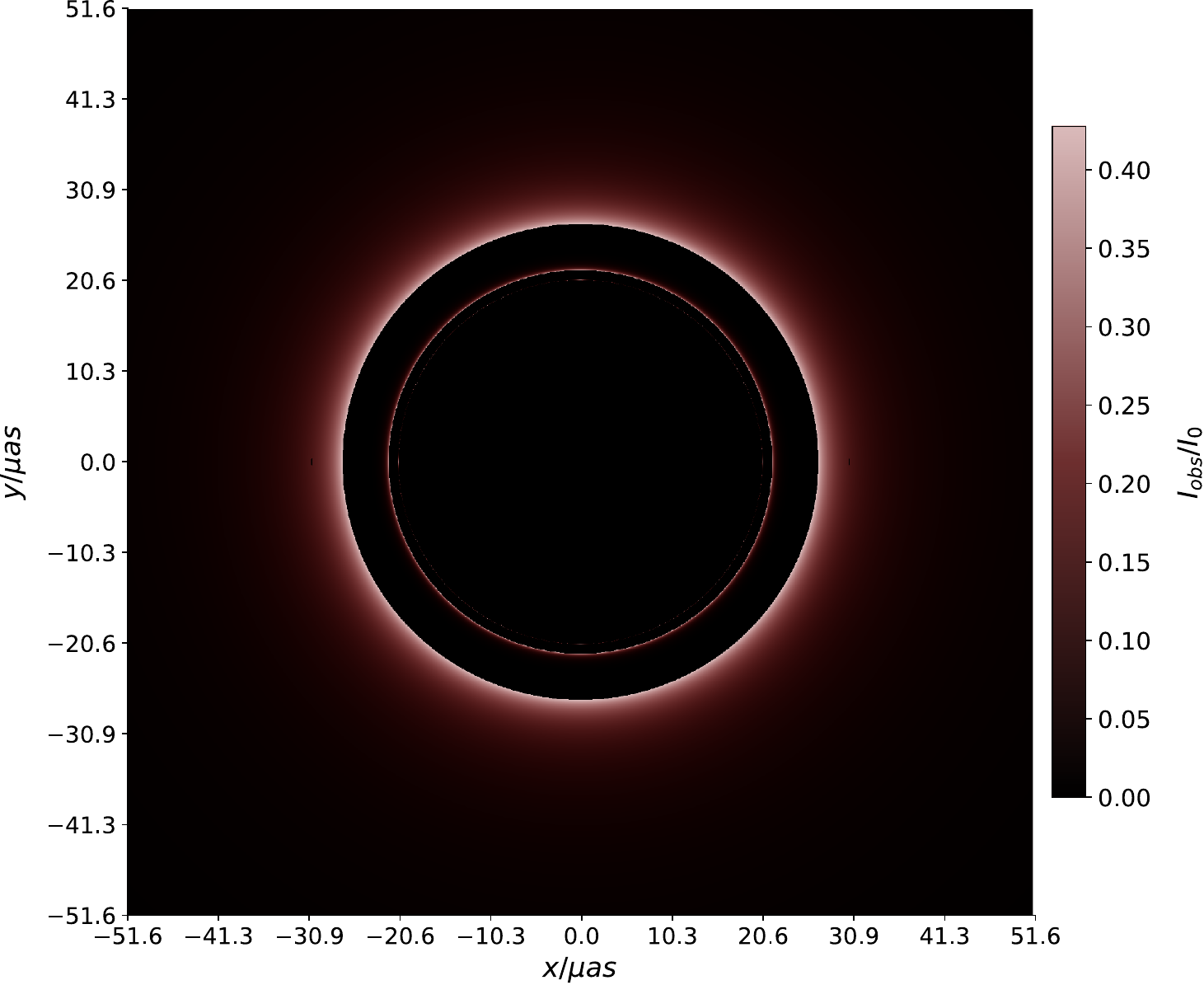}}\;
 \subfloat[$\alpha_1=0.32M^2$]
 {\includegraphics[width=0.48\linewidth]{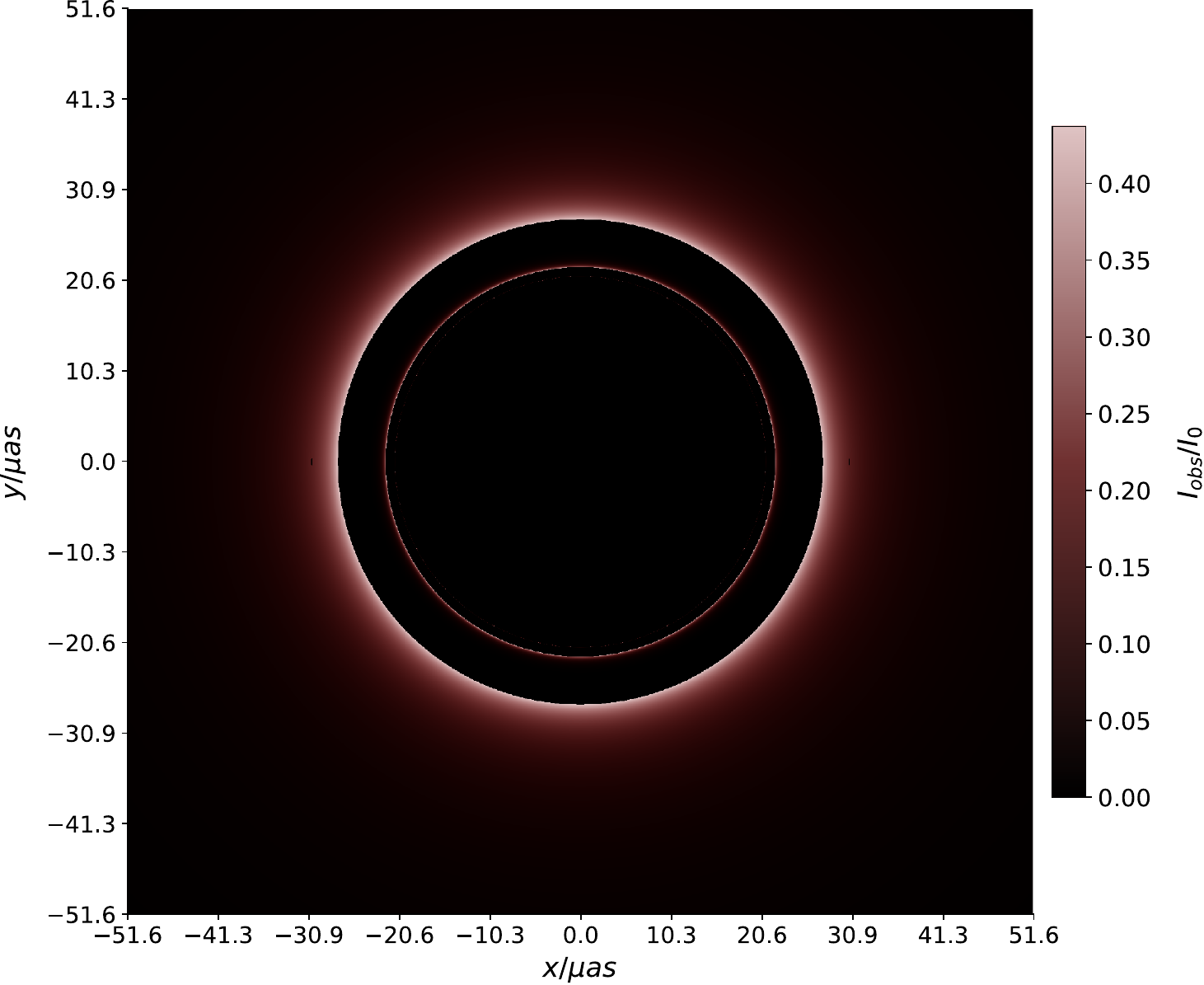}}\\
 \subfloat[$\alpha_1=1.27M^2$]
 {\includegraphics[width=0.48\linewidth]{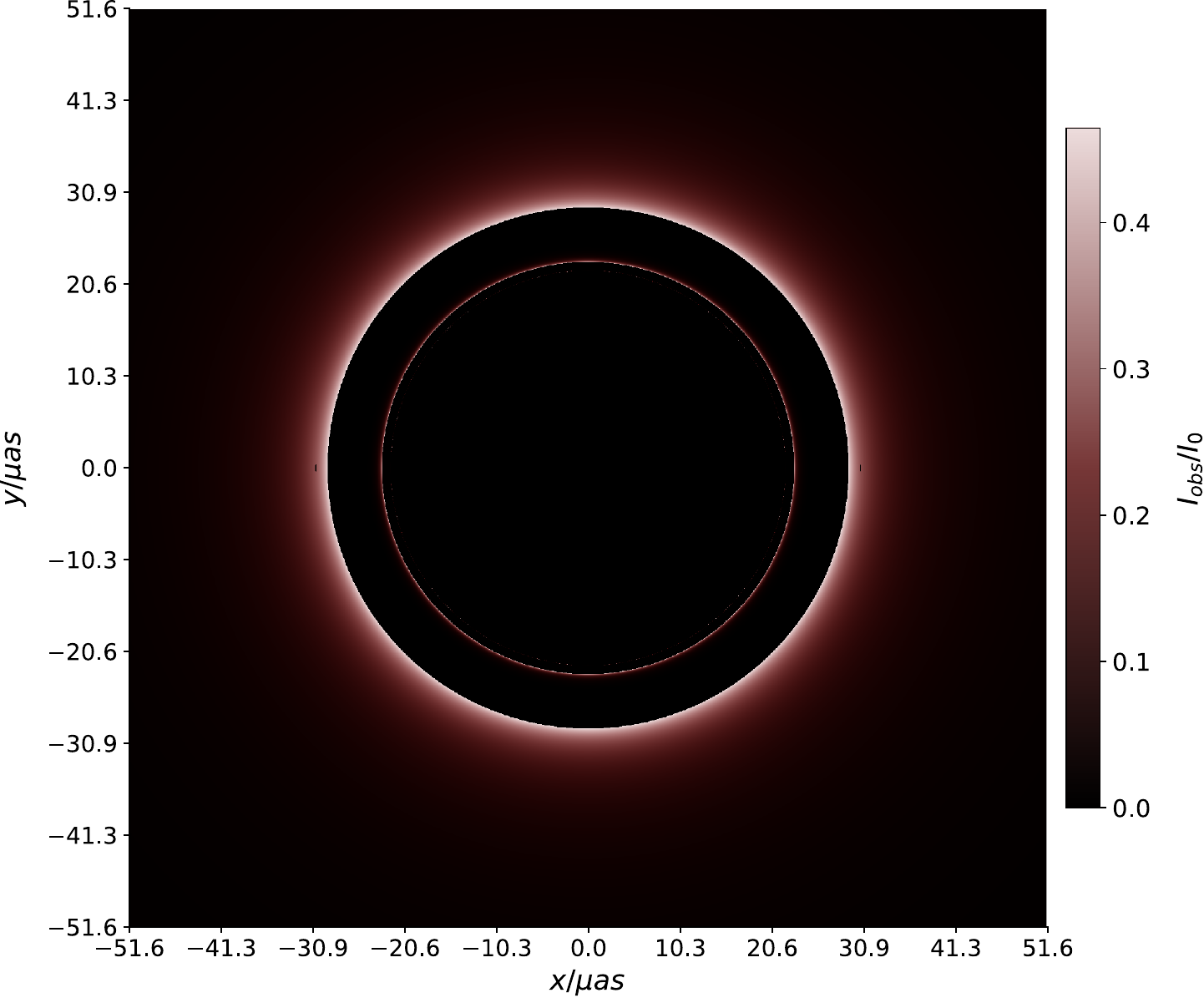}}\;
 \subfloat[$\alpha_1=5.09M^2$]
 {\includegraphics[width=0.48\linewidth]{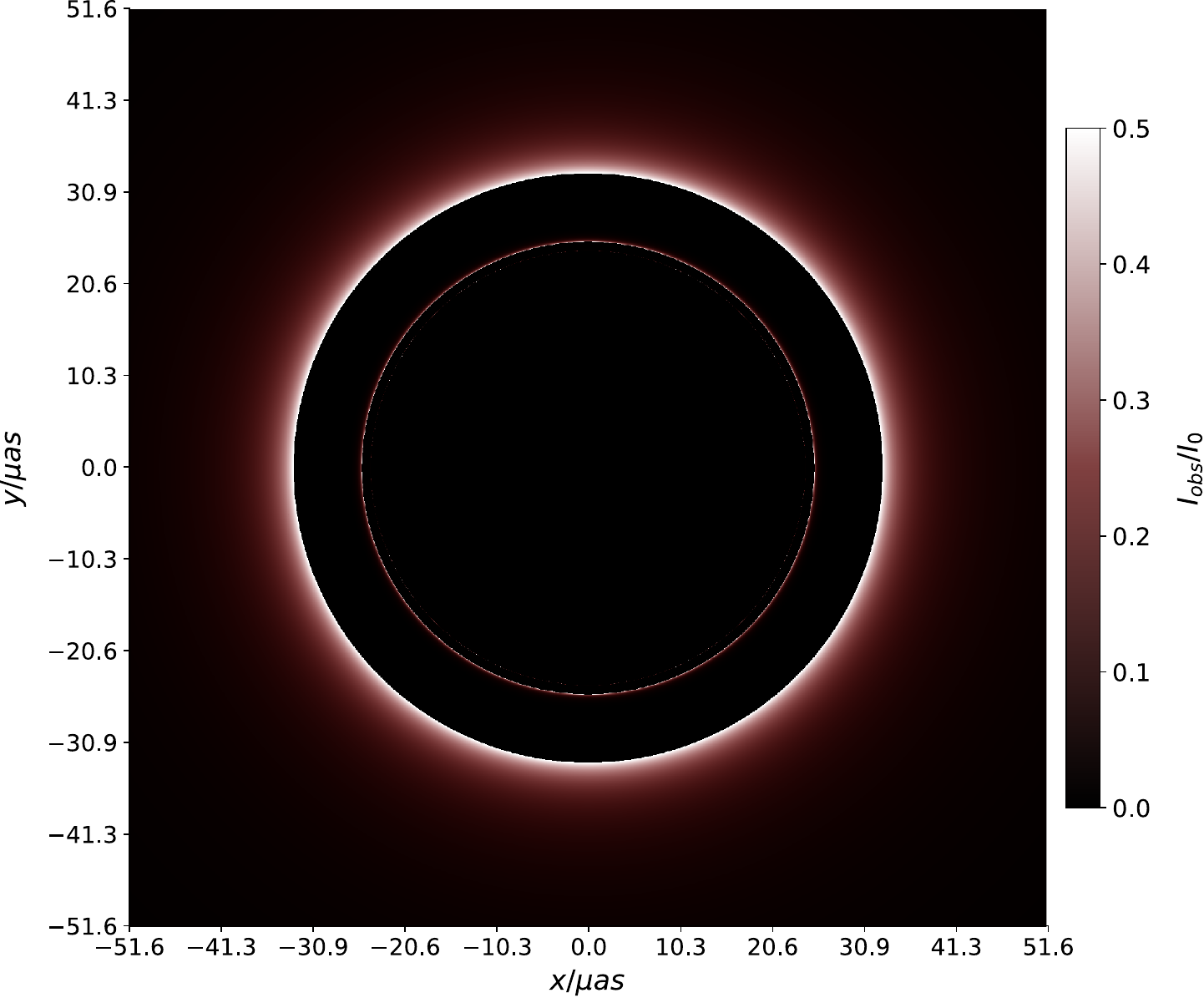}}\\
 \caption{Influence of $\alpha_1F^\sigma_{\ \nu}F_{\sigma\rho}R^{\nu\rho}$ on black-hole images (with $\alpha_2=\alpha_3=0$).}
 \label{fig:varying alpha1}
\end{figure*}

The corresponding black-hole images are displayed in Fig.~\ref{fig:varying alpha1}. From these images, we observe that the radius of the photon rings does increase mildly as $\alpha_1$ increases.

\subsection{Black-hole images with $\alpha_2F^2R$ coupling}

\begin{figure}
  \centering
  \includegraphics[width=\linewidth]{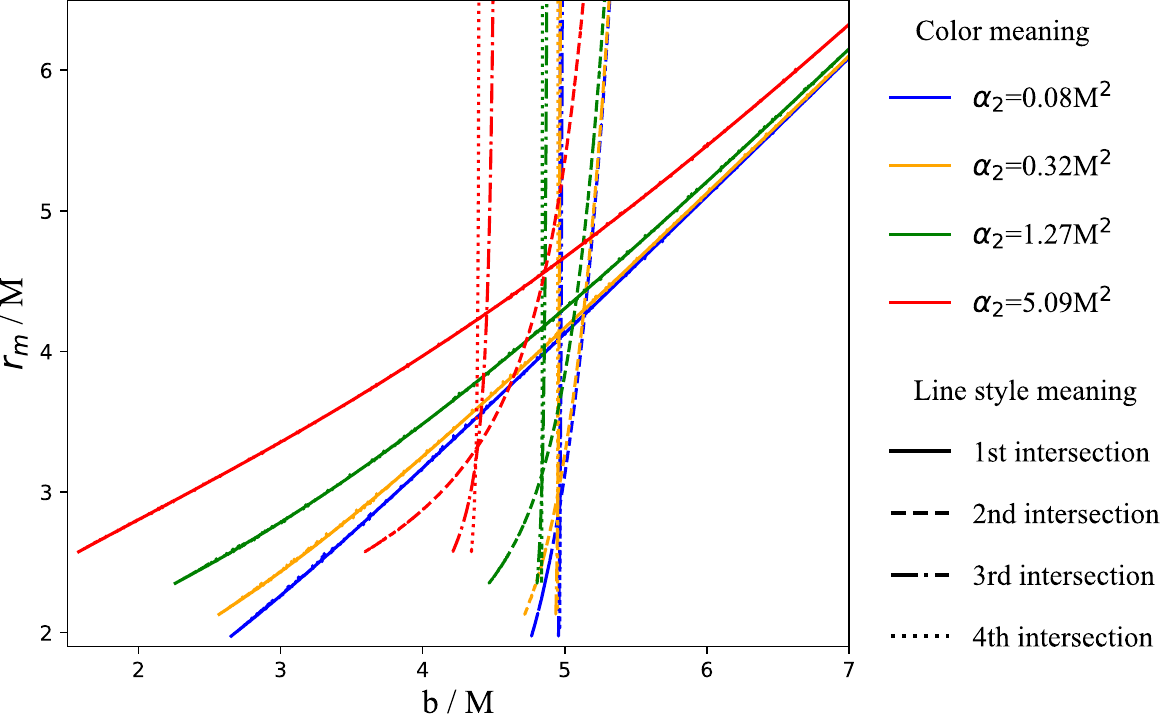}
  \caption{First four transfer functions for four different choices of coupling coefficient $\alpha_2$.}
  \label{fig: xm alpha2}
\end{figure}

\begin{figure}
  \centering
  \includegraphics[width=\linewidth]{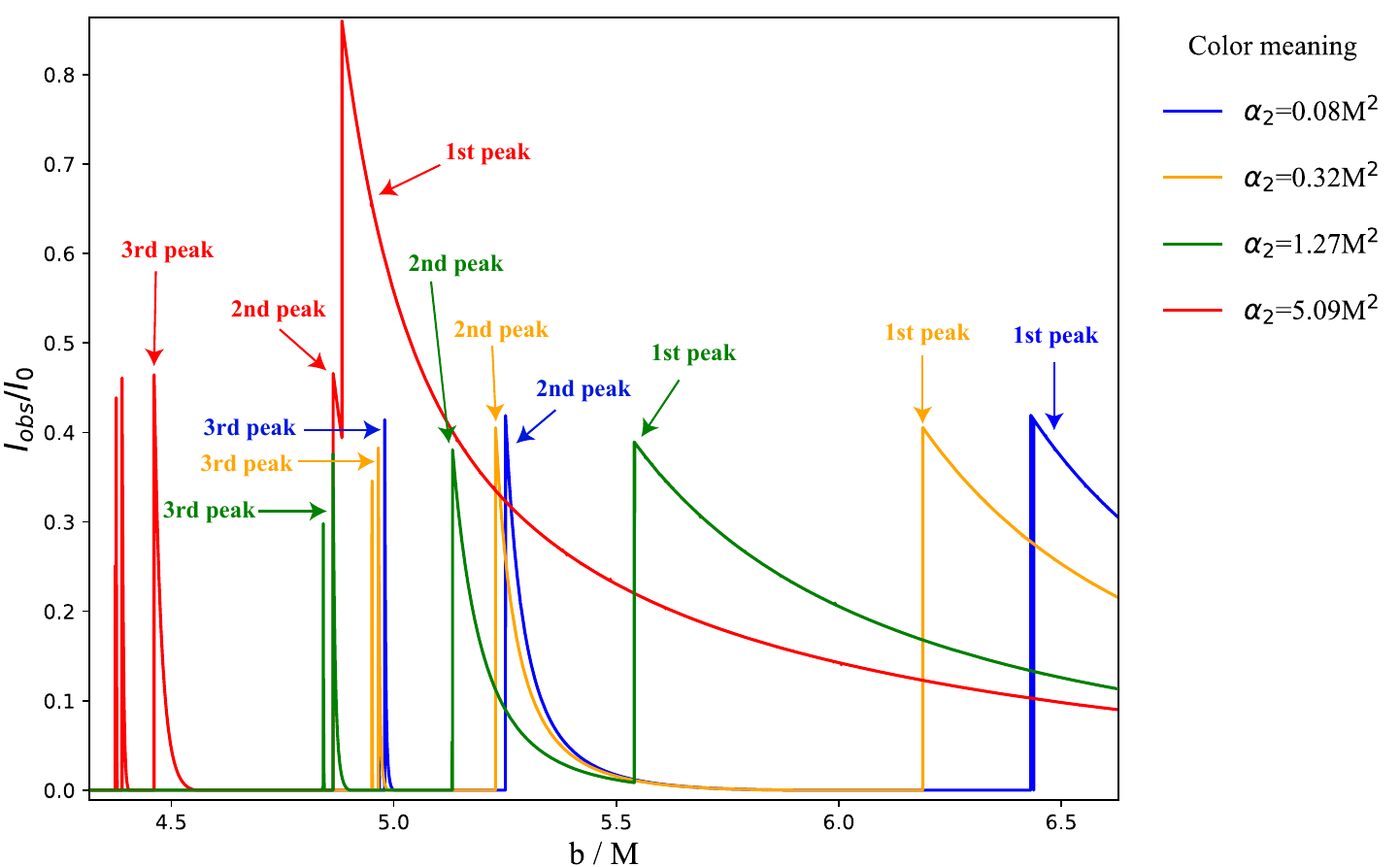}
  \caption{Observed intensity for four different choices of coupling coefficient $\alpha_2$. \blue{Under the same angular resolution, larger values of $\alpha_2$ allow higher-order photon rings to be resolved, leading to a greater number of distinguishable intensity peaks.}}
  \label{fig: Iobs alpha2}
\end{figure}

Fig.~\ref{fig: xm alpha2} and Fig.~\ref{fig: Iobs alpha2} display the effects of increasing $\alpha_2$ on the transfer functions and the observed intensity, respectively.

\blue{Unlike the case of $\alpha_1$, Fig.~\ref{fig: xm alpha2} shows that increasing $\alpha_2$ leads to an overall displacement of the transfer-function curves toward smaller $b$ and larger $r_m$, corresponding to the simultaneous enlargement of $r_H$ and reduction of $b_{ps}$. In addition, for $\alpha_2 = 5.09M^2$, the separation between the third and fourth intersection curves becomes noticeably larger than that in the other cases, indicating an enhanced possibility for the appearance of higher-order photon rings.}

In Fig.~\ref{fig: Iobs alpha2}, we observe that for $\alpha_2 = 5.09M^2$, the observed intensity of the first peak is approximately twice that in the other cases. This enhancement does not originate from an intrinsic increase in the intensity of the zeroth-order photon ring; rather, as discussed in Sec.~\ref{sec: general theoretical analysis}, it results from the coincidence of the zeroth- and first-order photon rings. Meanwhile, the width of the third peak grows as $\alpha_2$ increases, making the corresponding second-order ring more easily distinguishable, and the same trend holds for higher-order rings. \blue{As a result, under the same angular resolution, more higher-order photon rings can be resolved for larger $\alpha_2$, which explains why a larger number of intensity peaks are visible in Fig.~\ref{fig: Iobs alpha2}.}

\begin{figure*}
 \subfloat[$\alpha_2=0.08M^2$]
 {\includegraphics[width=0.48\linewidth]{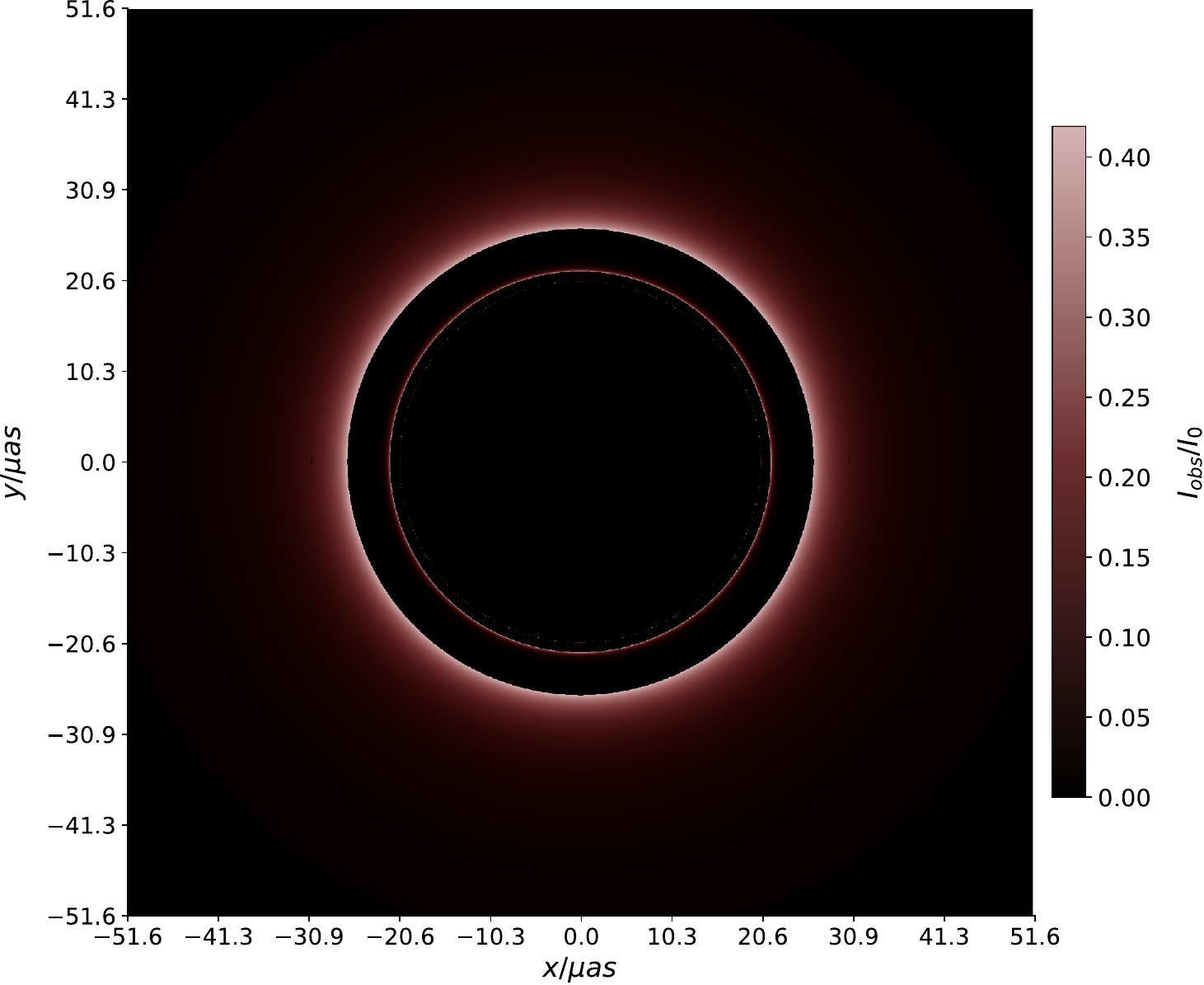}}\;
 \subfloat[$\alpha_2=0.32M^2$]
 {\includegraphics[width=0.48\linewidth]{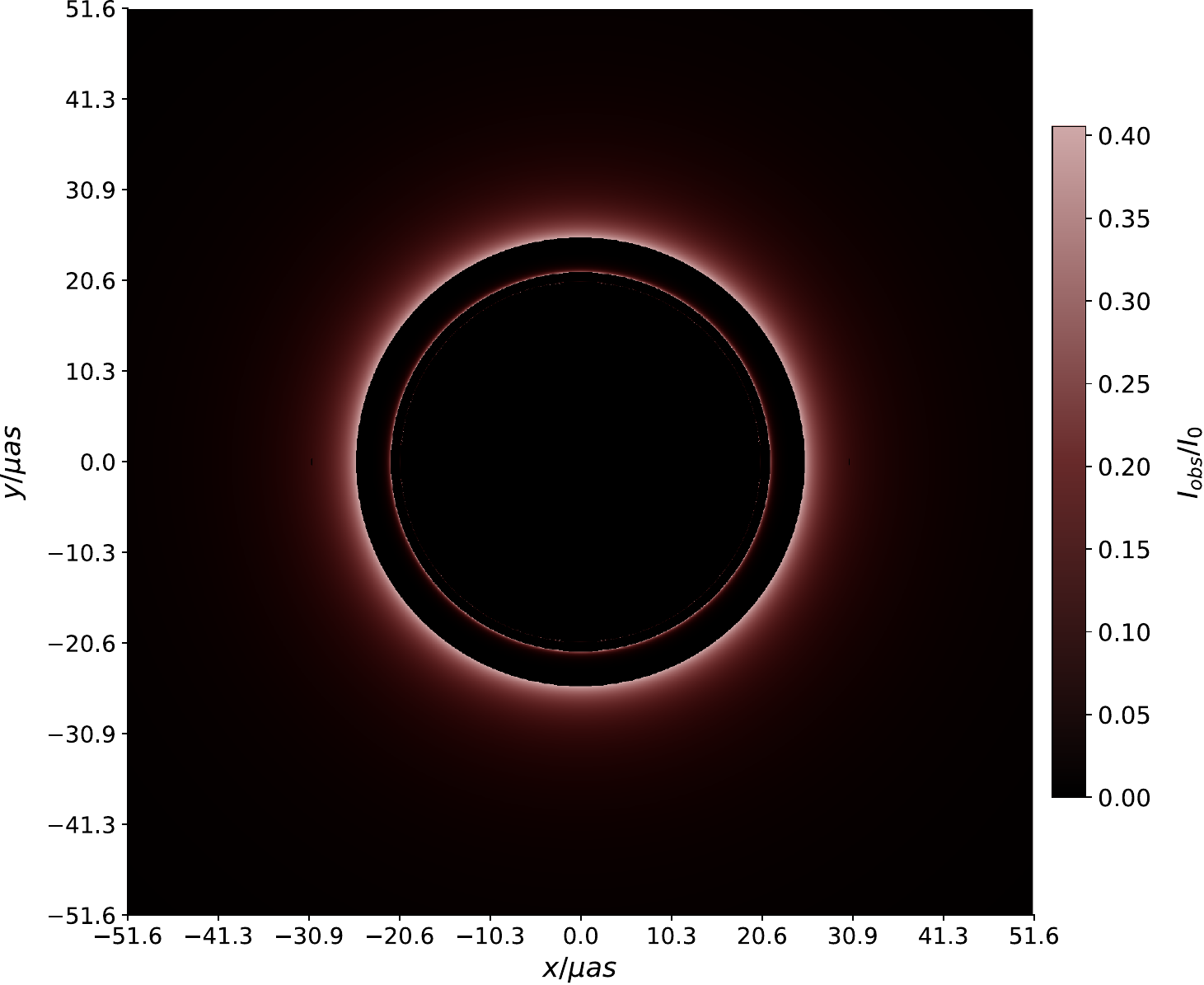}}\\
 \subfloat[$\alpha_2=1.27M^2$]
 {\includegraphics[width=0.48\linewidth]{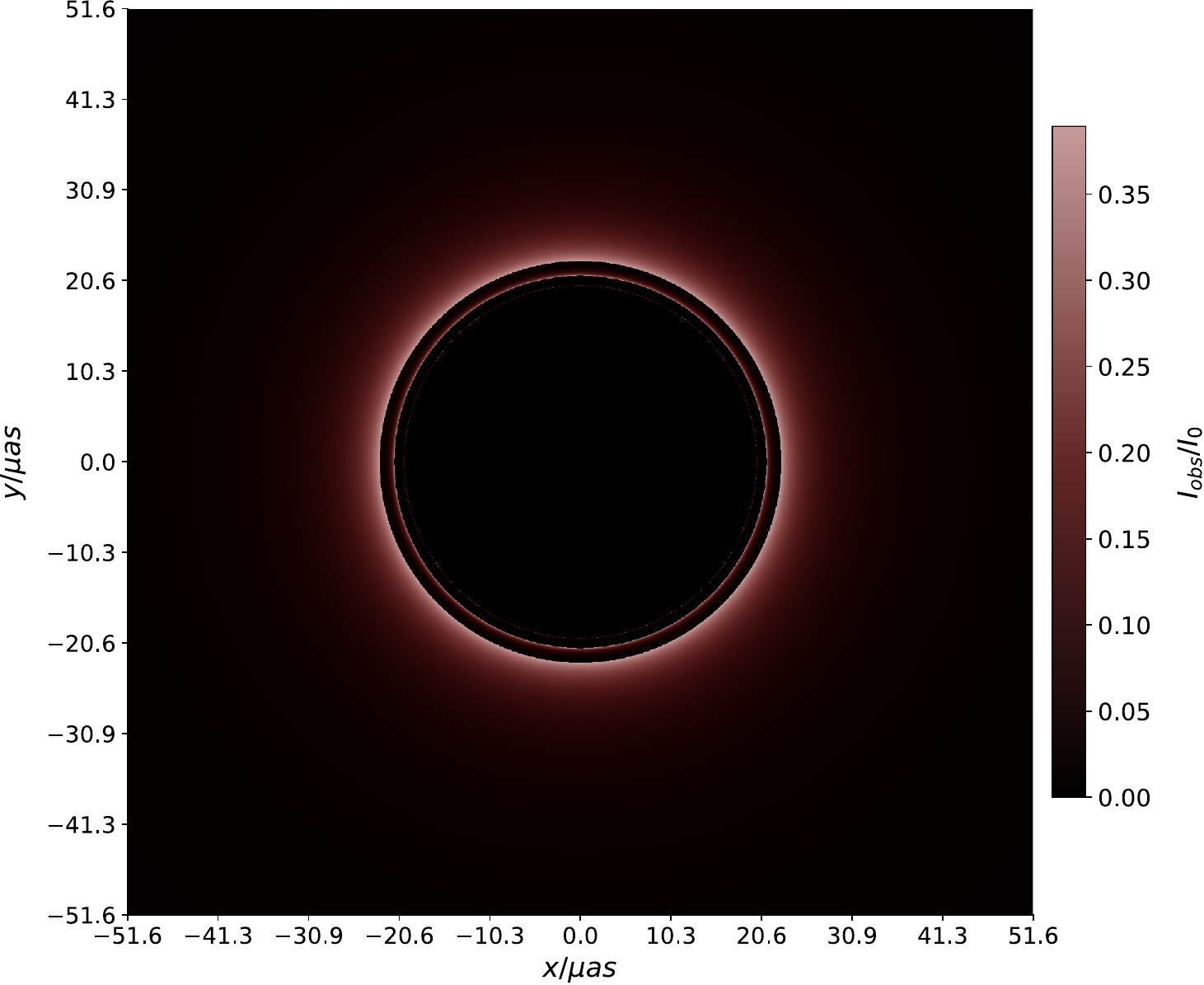}}\;
 \subfloat[$\alpha_2=5.09M^2$]
 {\includegraphics[width=0.48\linewidth]{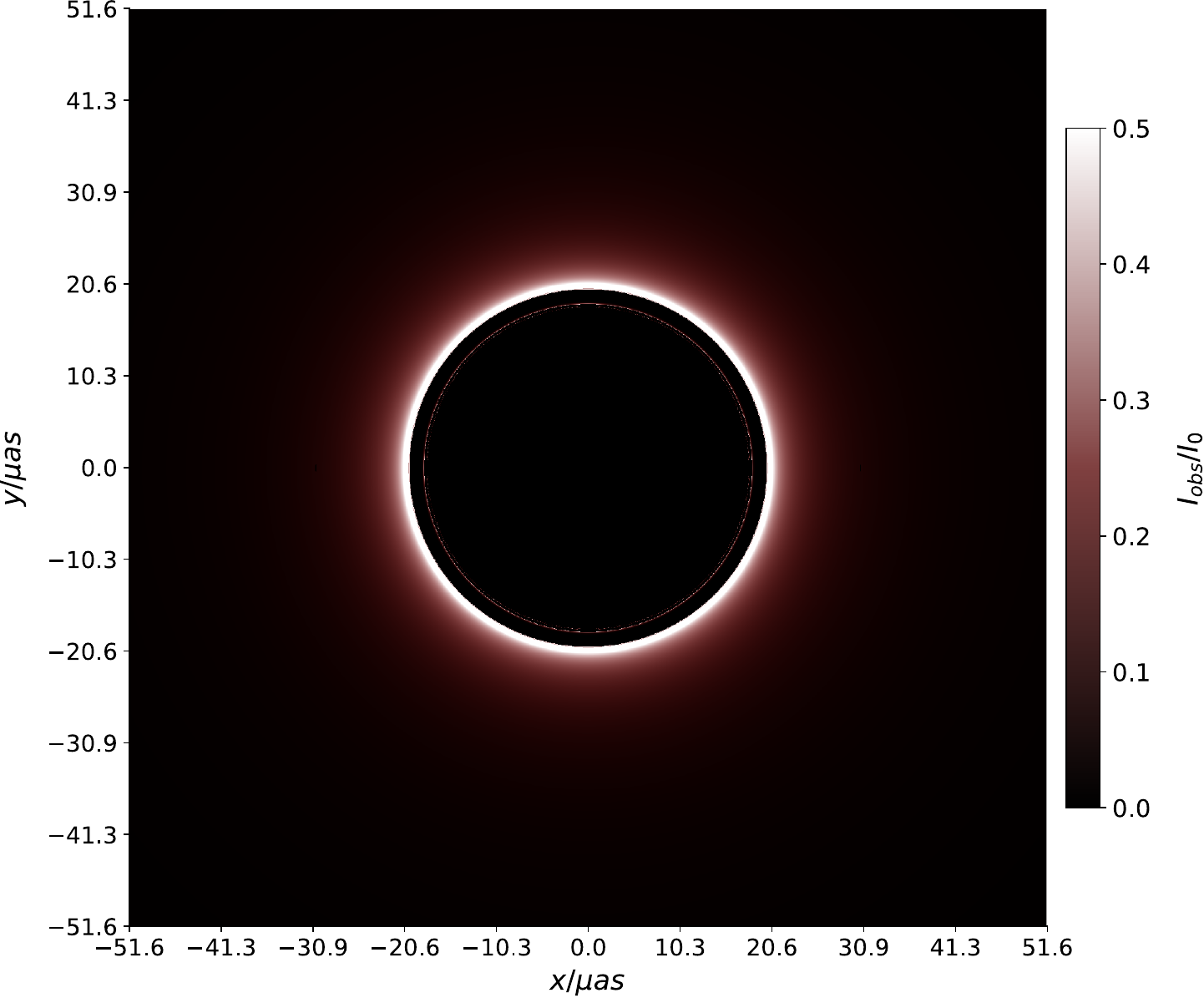}}\\
 \caption{Influence of $\alpha_2 F^2R$ on black-hole images (with $\alpha_1=\alpha_3=0$).}
 \label{fig:varying alpha2}
\end{figure*}

\begin{figure*}
 \subfloat[$\alpha_2=0.08M^2$]
 {\includegraphics[width=0.48\linewidth]{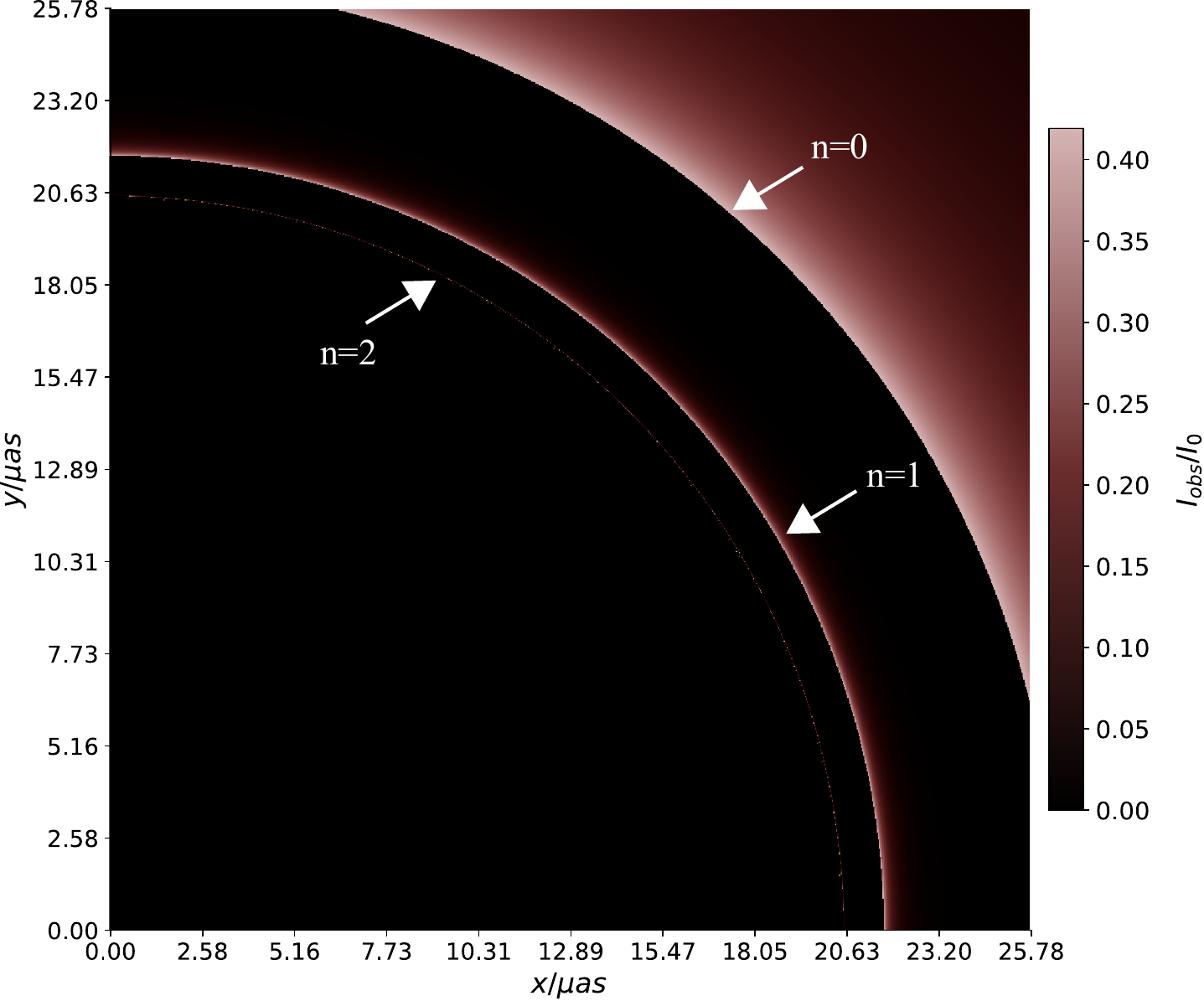}}\;
 \subfloat[$\alpha_2=0.32M^2$]
 {\includegraphics[width=0.48\linewidth]{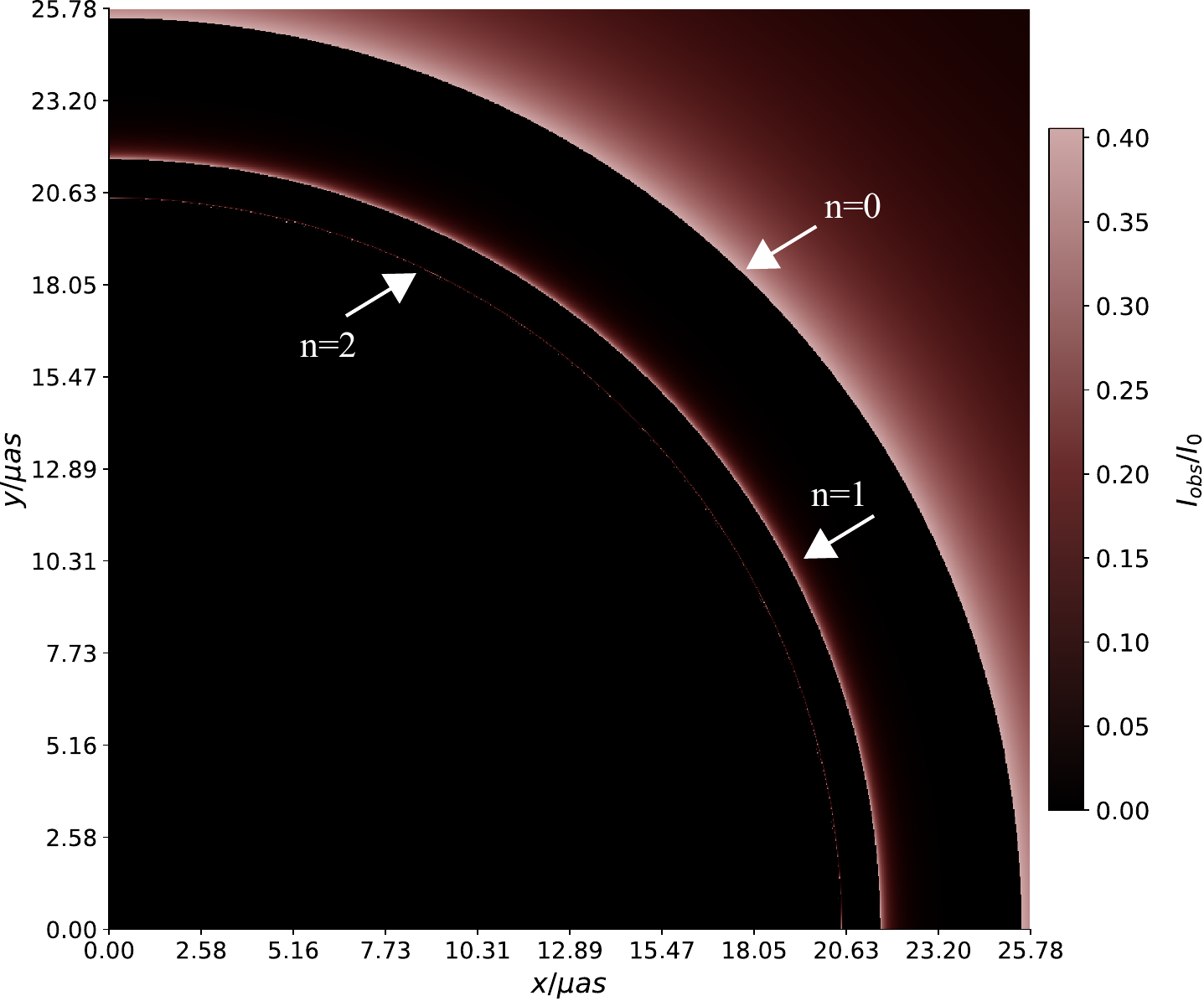}}\\
 \subfloat[$\alpha_2=1.27M^2$]
 {\includegraphics[width=0.48\linewidth]{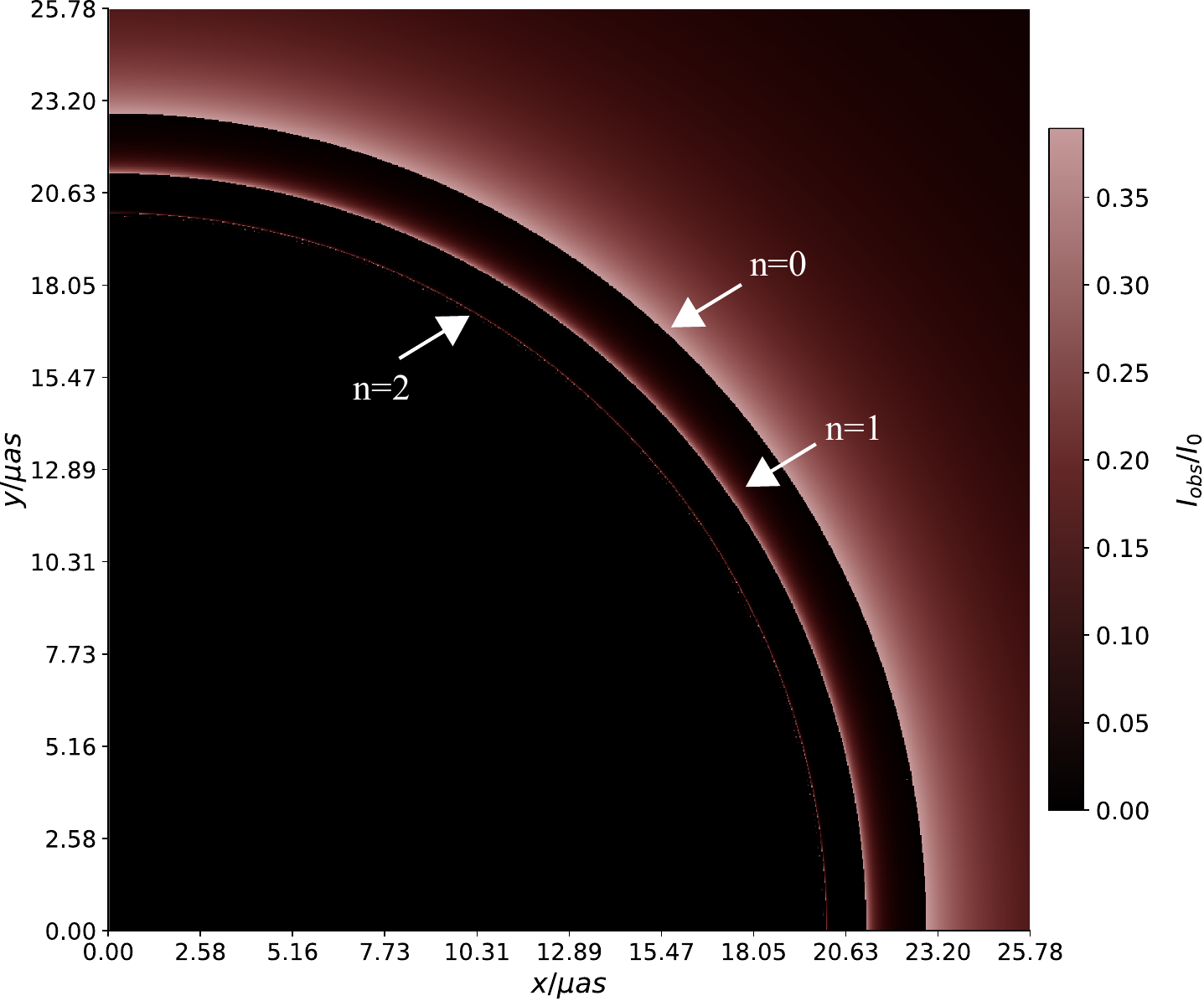}}\;
 \subfloat[$\alpha_2=5.09M^2$]
 {\includegraphics[width=0.48\linewidth]{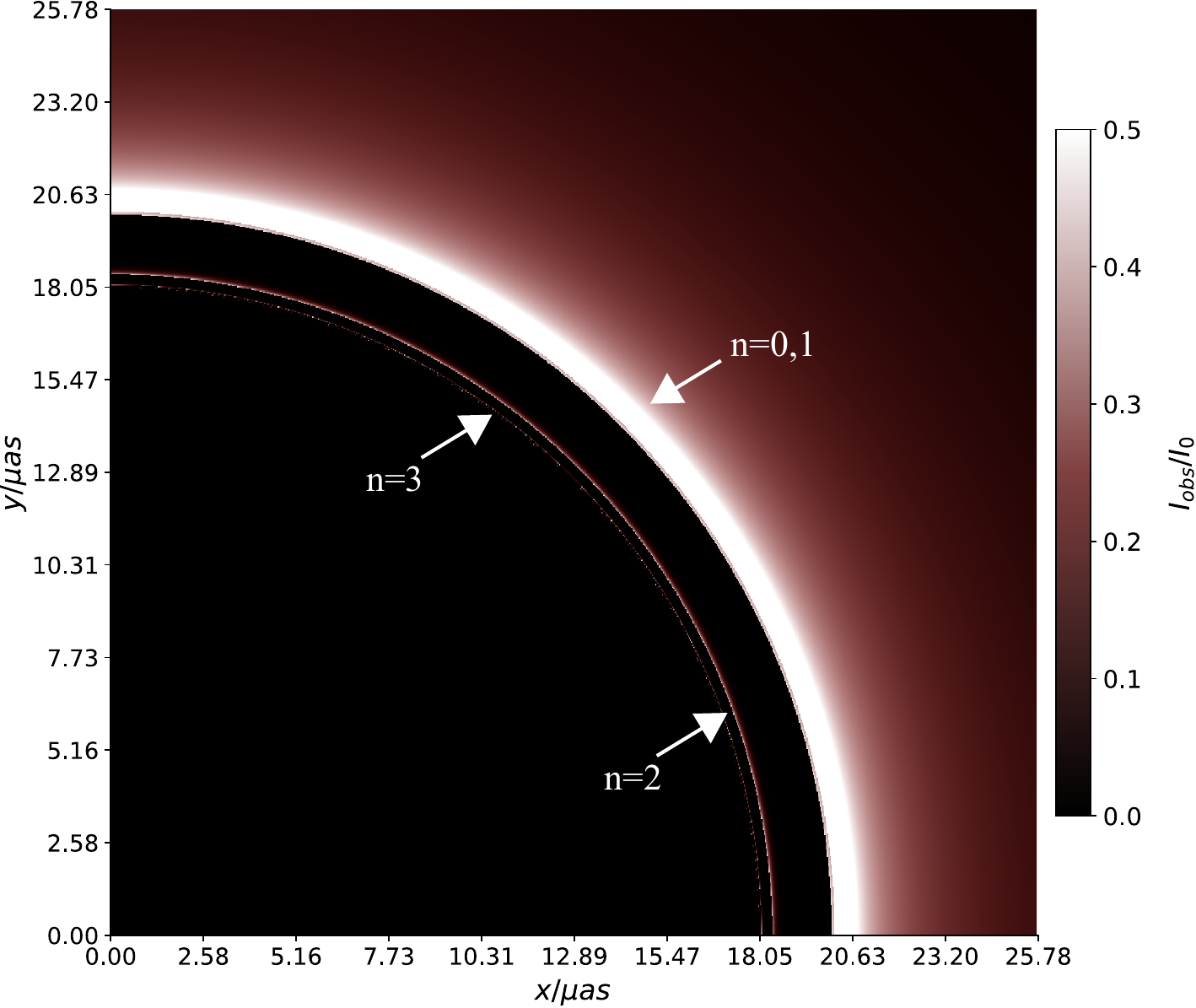}}\\
 \caption{A zoomed-in version of Fig.~\ref {fig:varying alpha2}.}
 \label{fig:zoomed-in}
\end{figure*}

The corresponding black-hole images are shown in Fig.~\ref{fig:varying alpha2}. These images demonstrate that the overall radius of the photon rings decreases gradually as $\alpha_2$ increases. In particular, for $\alpha_2 = 5.09M^2$, the merging of the zeroth- and first-order photon rings leads to a substantial enhancement of the total brightness. Moreover, a faint second-order photon ring can be discerned inside the merged ring, which is nearly unresolvable in the other cases. In the magnified view shown in Fig.~\ref{fig:zoomed-in}, these features can be seen more clearly, including the coincidence of the zeroth- and first-order photon rings, the enhanced visibility of the second-order ring, and the emergence of the third-order ring.

\subsection{Black-hole images with $\alpha_3F_{\mu\nu}F_{\sigma\rho}R^{\mu\nu\sigma\rho}$ coupling}

\begin{figure}
  \centering
  \includegraphics[width=\linewidth]{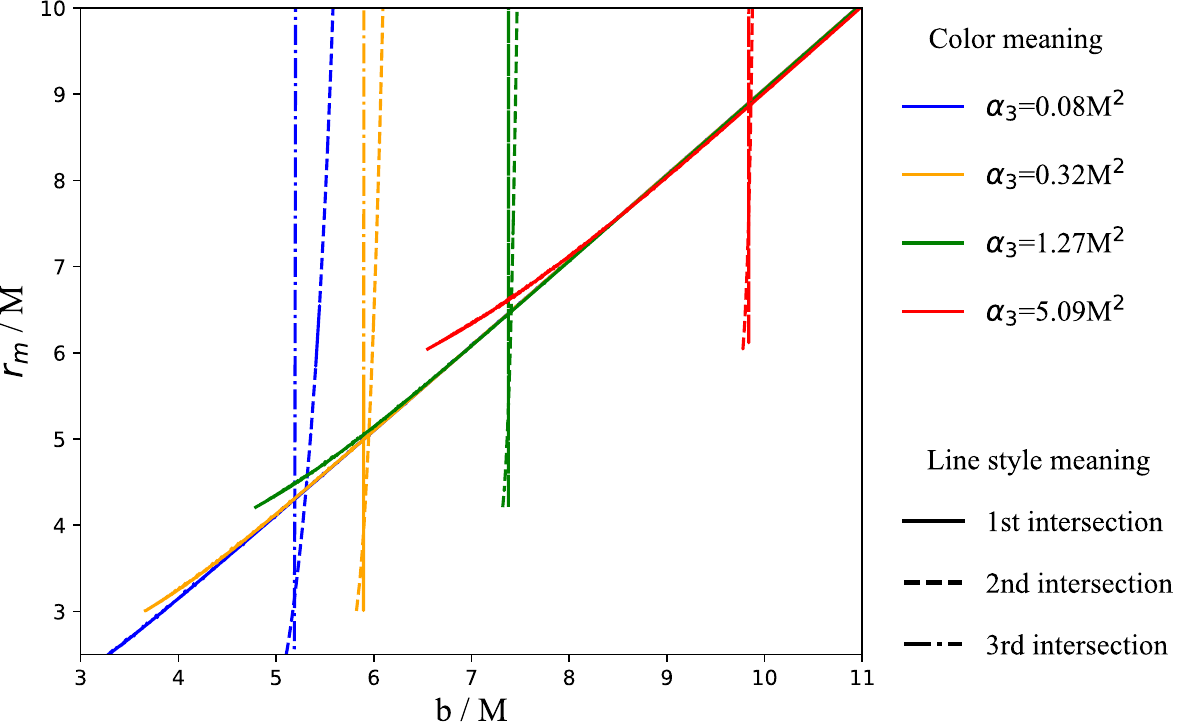}
  \caption{First three transfer functions for four different choices of coupling coefficient $\alpha_3$.}
  \label{fig: xm alpha3}
\end{figure}

\begin{figure}
  \centering
  \includegraphics[width=\linewidth]{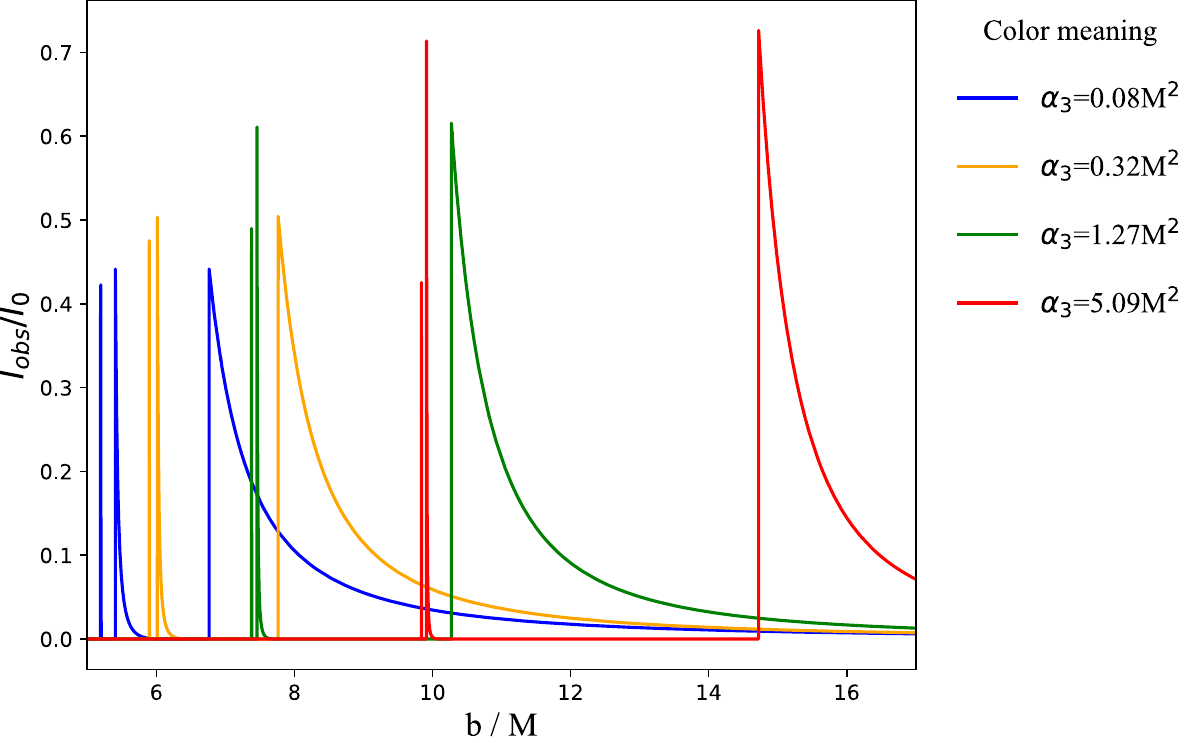}
  \caption{Observed intensity for four different choices of coupling coefficient $\alpha_3$.}
  \label{fig: Iobs alpha3}
\end{figure}

\blue{Similar to the $\alpha_1$ case, Fig.~\ref{fig: xm alpha3} shows that both the horizon radius $r_H$ and the shadow radius $b_{ps}$ increase with $\alpha_3$, but at a much faster rate.} Moreover, Fig.~\ref{fig: Iobs alpha3} indicates that, as $\alpha_3$ increases, the separation between the second and third intensity peaks gradually decreases. This behavior is consistent with the general analysis in Sec.~\ref{sec: general theoretical analysis}, where we found that the spacing between higher-order photon rings rapidly diminishes and asymptotically approaches zero, leading to their effective coincidence at the shadow boundary.

\begin{figure*}
 \subfloat[$\alpha_3=0.08M^2$]
 {\includegraphics[width=0.48\linewidth]{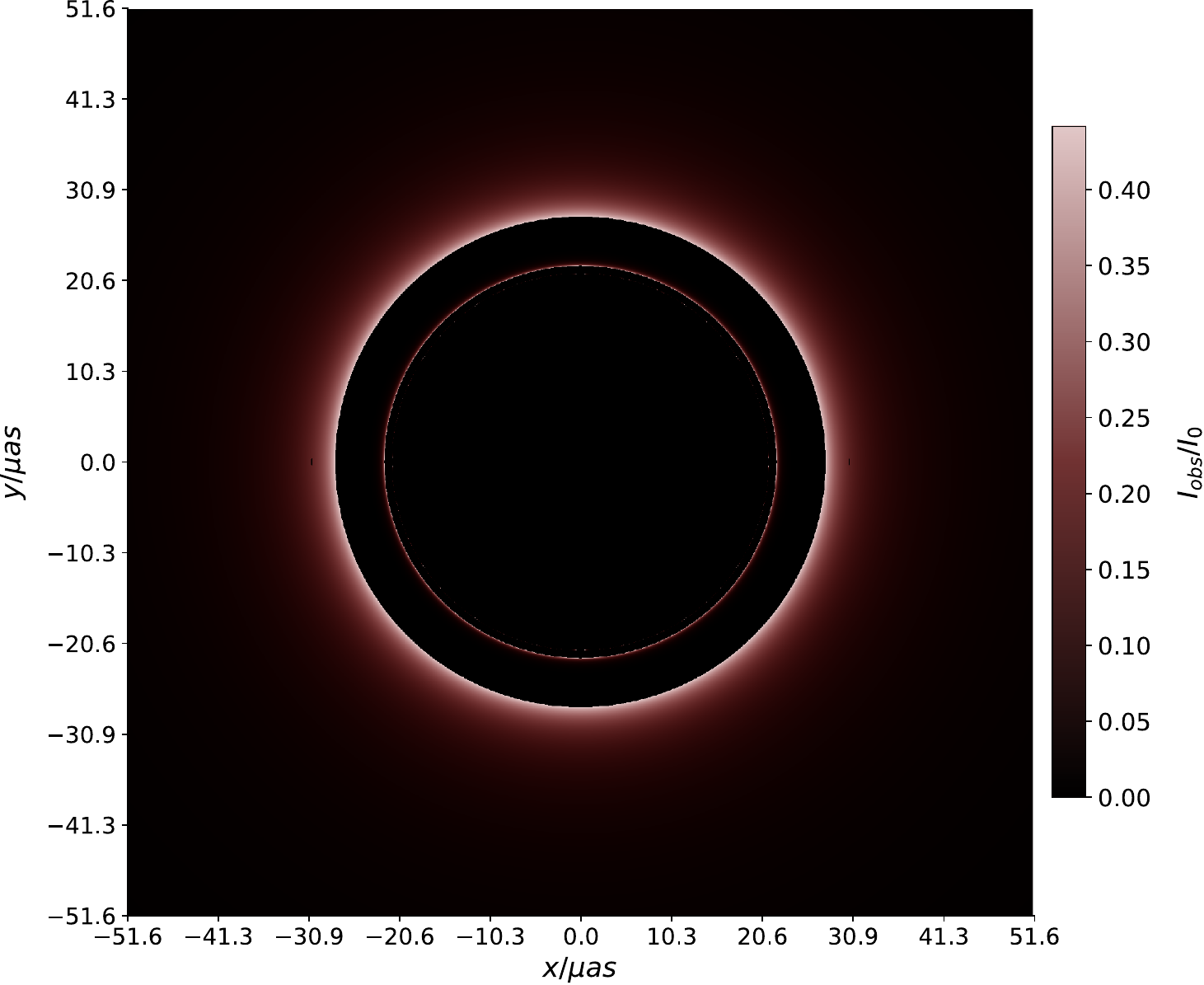}}\;
 \subfloat[$\alpha_3=0.32M^2$]
 {\includegraphics[width=0.48\linewidth]{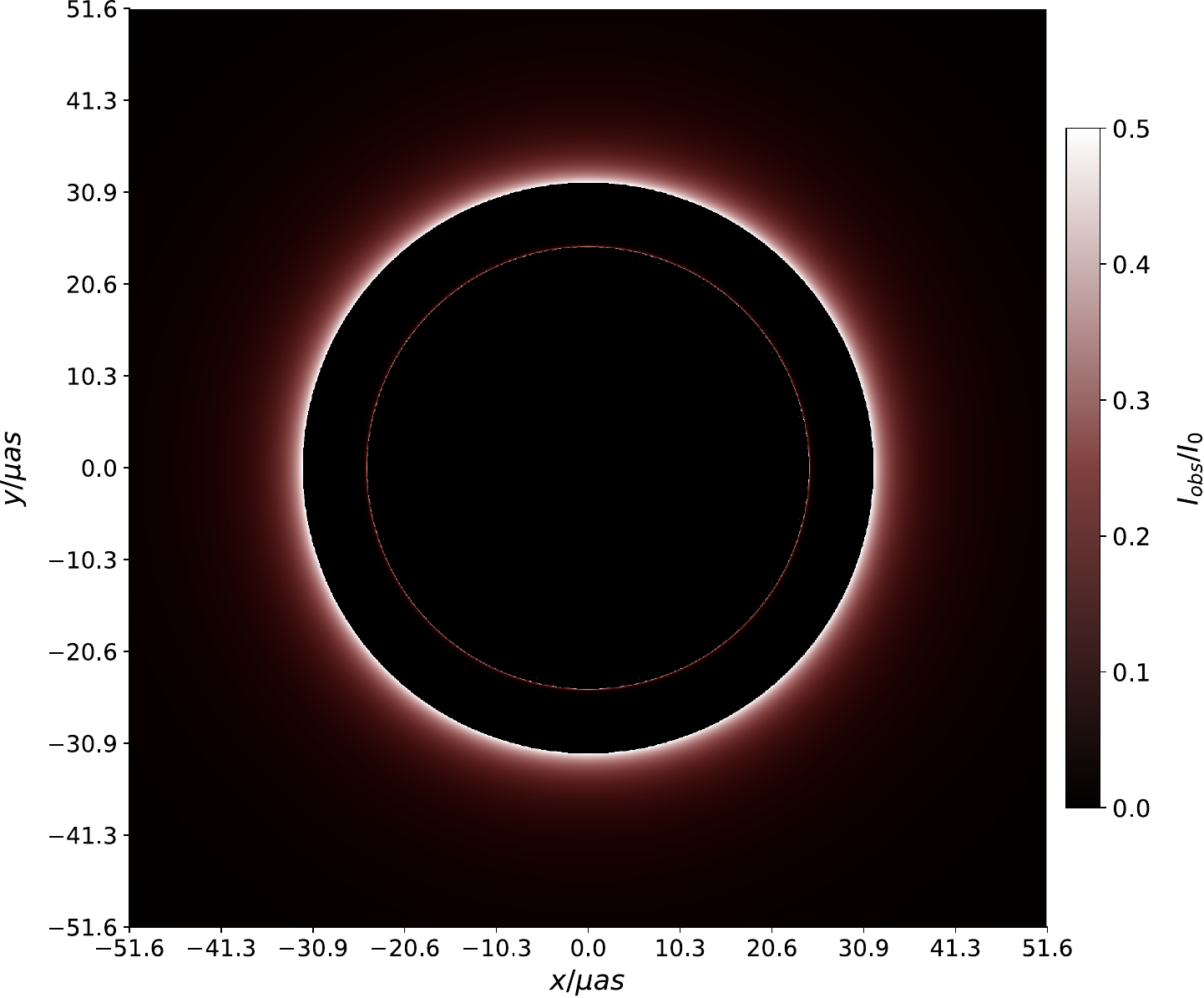}}\\
 \subfloat[$\alpha_3=1.27M^2$]
 {\includegraphics[width=0.48\linewidth]{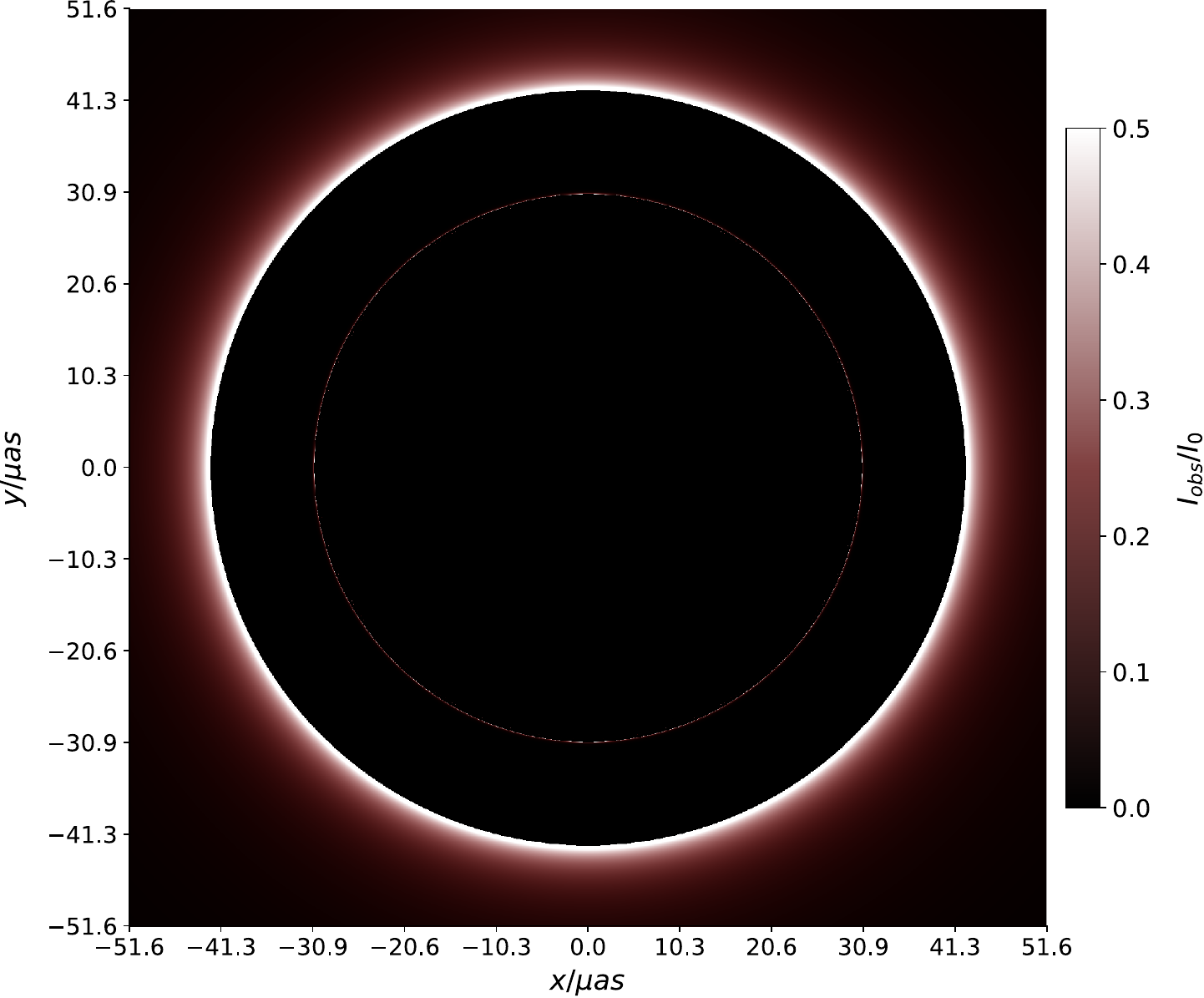}}\;
 \subfloat[$\alpha_3=5.09M^2$]
 {\includegraphics[width=0.48\linewidth]{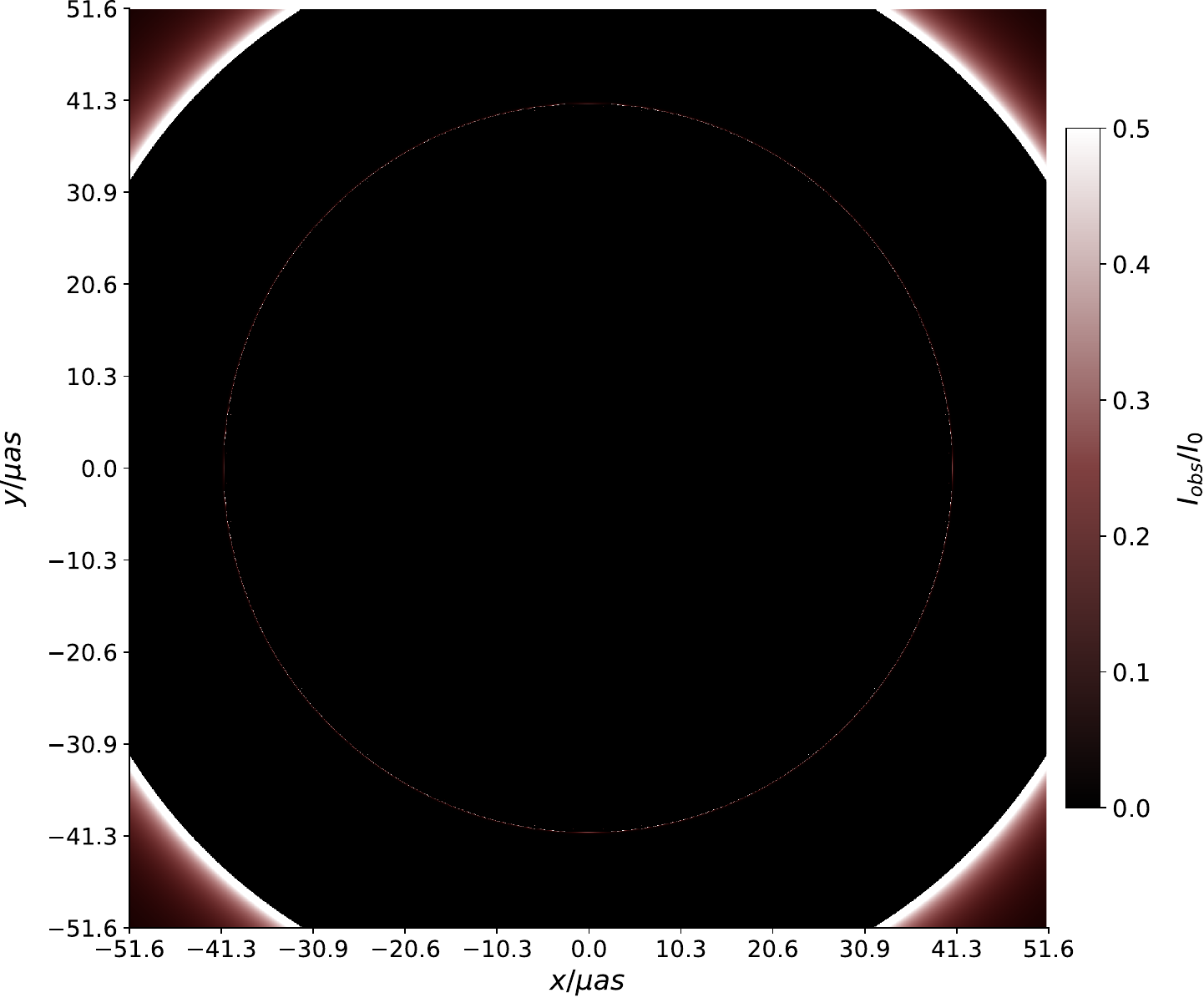}}\\
 \caption{Influence of $\alpha_3F_{\mu\nu}F_{\sigma\rho}R^{\mu\nu\sigma\rho}$ on black-hole images (with $\alpha_1=\alpha_2=0$).}
 \label{fig:varying alpha3}
\end{figure*}

Figure~\ref{fig:varying alpha3} illustrates the corresponding black-hole images for increasing $\alpha_3$. It is clear that the radius of the zeroth-order photon ring grows rapidly, accompanied by a significant increase in the separation between the zeroth- and first-order photon rings.

\section{Conclusion and discussion}
\label{sec:conclusions}

In conclusion, we have investigated static and spherically symmetric black holes with non-minimal couplings between the electromagnetic field and spacetime curvature tensors, including the Ricci scalar, Ricci tensor, and Riemann tensor. These couplings modify both the Einstein and Maxwell field equations, leading to corrections to the spacetime geometry and electromagnetic field. \blue{In the limit of asymptotically flat spacetimes and using an asymptotic expansion series in $r$ (or ``Frobenius method''), we obtain the resulting black-hole metric with power-law corrections in $r$.  These corrections give rise to noticeable modifications of the near-horizon structure and the associated observational signatures. }Black-hole shadows and photon rings may therefore provide a promising avenue for constraining such non-minimal couplings and testing possible deviations from general relativity in the strong-field regime.

We derived a series solution for a Reissner--Nordstr\"{o}m-like black hole with non-minimal couplings to the electromagnetic field. Through a detailed theoretical analysis, we found that all three types of couplings, i.e. $\alpha_1F^\sigma_{\ \nu}F_{\sigma\rho}R^{\nu\rho}$, $\alpha_2F^2R$ and $\alpha_3F_{\mu\nu}F_{\sigma\rho}R^{\mu\nu\sigma\rho}$, enlarge both the event horizon radius $r_H$ and the photon-sphere radius $r_{ps}$. However, their observational consequences are qualitatively different. In particular, the $\alpha_1F^\sigma_{\ \nu} F_{\sigma\rho} R^{\nu\rho}$ and $\alpha_3F_{\mu\nu} F_{\sigma\rho} R^{\mu\nu\sigma\rho}$ couplings increase the apparent shadow radius $b_{ps}$, whereas the $\alpha_2F^2 R$ coupling leads to a reduction of the black-hole shadow.

We then incorporated the accretion disk model I proposed in Ref.~\cite{Guerrero:2022qkh} to investigate the influence of these couplings on photon rings generated by photons emitted from ISCO. Regarding their radii, the behavior of the photon rings closely parallels that of the shadow: the $\alpha_1F^\sigma_{\ \nu} F_{\sigma\rho} R^{\nu\rho}$ and $\alpha_3F_{\mu\nu} F_{\sigma\rho} R^{\mu\nu\sigma\rho}$ couplings enlarge the radii of all photon rings, with the latter producing a more pronounced effect, while the $\alpha_2F^2 R$ coupling reduces the radii of all photon rings.

Regarding the spacing between successive photon rings, the three couplings exhibit distinct behaviors. The $\alpha_1F^\sigma_{\ \nu} F_{\sigma\rho} R^{\nu\rho}$ coupling slightly increases the separation between the zeroth- and first-order rings while leaving the spacing among higher-order rings nearly unchanged. In contrast, the $\alpha_2F^2 R$ coupling decreases the separation between the zeroth- and first-order rings, which may even become nearly coincident, resulting in an approximately doubled peak brightness. Meanwhile, it increases the spacing among higher-order rings, making them easier to resolve observationally. The $\alpha_3F_{\mu\nu} F_{\sigma\rho} R^{\mu\nu\sigma\rho}$ coupling, on the other hand, enlarges the separation between the zeroth- and first-order rings but significantly suppresses the spacing between higher-order rings.

We have selected several representative values of the coupling parameters and produced the corresponding black-hole images.  Within the observationally resolvable regime, the numerical imaging results are in good agreement with the conclusions drawn from our theoretical analysis. An important caveat is the presence of parameter degeneracies in the resulting shadow images. We treated the effect of $\alpha_1F^\sigma_{\ \nu}F_{\sigma\rho}R^{\nu\rho}$, $\alpha_2F^2R$ and $\alpha_3F_{\mu\nu}F_{\sigma\rho}R^{\mu\nu\sigma\rho}$ separately. However, these non-minimal couplings all appear at the same order and are thus expected to contribute at comparable levels. As we have shown, $\alpha_2F^2R$ and $\alpha_3F_{\mu\nu}F_{\sigma\rho}R^{\mu\nu\sigma\rho}$ have opposite effects on the radii of the shadow and photon rings. As a result, these effects may partially compensate each other, leading to shadow images similar to those predicted by GR.

\blue{Several directions for future work are worth exploring. In the first place, based on the metric we obtained, one can study the   relationship between black hole quasi-normal modes and their shadows.  Specifically, one can compute the quasinormal modes under scalar, electromagnetic, or gravitational perturbations and verify whether the correspondence between the real part of the quasinormal frequency and the shadow radius, as well as between the imaginary part and the Lyapunov exponent, remains valid.  Second, based on the metric obtained in this work, one can investigate gravitational lensing and Einstein rings. Specifically, one can calculate the strong gravitational lensing observables, such as the position and magnification of relativistic images, and explore new observable imprints of these non-minimal couplings.}

\blue{Thirdly, the current work is confined to spherically symmetric black holes. However, real black holes in the Universe (such as $\textrm{M87}$ and $\textrm{Sgr A}^{*}$ ) are expected to be rapidly rotating.  Therefore, it is necessary to extend it  to the case of rotating black holes. Rotation leads to frame dragging which causes the shadow to deviate from the circle (resulting in a ``D'' or ``gourd'' shape).  We will focus on the investigate how the coupling constants modify the critical escape parameter and whether it will result in more complex orbital bifurcations. This helps to directly compare theoretical predictions with the observed shadow asymmetry in EHT images. }  

\blue{Finally, constraints on the coupling constants based solely on the shadow size suffer from degeneracies (such as degeneracies with the black-hole mass, electric charge, and inclination angle). Therefore, there is a need to combine gravitational-wave observations and X-ray echo  data. These couplings affect the precession frequencies of bound geodesics near the ISCO. We will calculate the dephasing signatures induced by these couplings in gravitational-wave signals, establish a joint likelihood function of  ``shadow size, strong gravitational lensing time delay and gravitational wave frequency drift'' , and achieve high-precision multi-band constraints of the couplings.}

To summarize, non-minimal electromagnetic-curvature couplings arise naturally in effective field theories and from quantum corrections in curved spacetime. These couplings present both challenges and opportunities. On the one hand, the coupling parameters must remain consistent with existing observational constraints. On the other hand, they may give rise to distinctive theoretical and observational signatures, modified strong-field behavior, and new possibilities for constructing well-behaved cosmological models. Their most accessible signatures currently reside in black hole astrophysics, though the predicted effects remain subdominant and constrained by observations.  Our study reveals that different non-minimal couplings not only alter the radius of the event horizon, but also leave distinctive imprints on the morphology of black-hole shadows and photon rings. These findings provide potential observational signatures that could be used to constrain the underlying coupling parameters. Future \textit{Very Long Baseline Interferometry} (VLBI) observations, such as the \textit{Black Hole Explorer} (BHEX) mission, may be capable of detecting the first-order photon ring \cite{Johnson:2024ttr,Galison:2024bop}. Nevertheless, resolving only the first-order ring may not provide a sufficiently sensitive probe of the underlying spacetime geometry. Observations capable of resolving multiple photon rings would offer a more powerful avenue for testing non-minimal electromagnetic-curvature couplings and probing deviations from general relativity in the strong-field regime.

\section*{Acknowledgments}
We are grateful to  the referee for the professional comments which leads to significant improvements in the original manuscript.  Z. Yin acknowledges insightful and valuable discussions with Profs. Xiaomei Kuang and Minyong Guo. The work is supported by the Special Exchange Program of CAS, the China’s Space Origins Exploration Program Nos. GJ11010405 \& GJ11010401,  the National Key Research and Development Program of China (Grants No.
2022YFF0503404, No. 2022SKA0110100), the Central Guidance for Local Science and Technology Development Fund Project with Grant No.2024ZY0113,  the National Natural Science Foundation of China (No.12375059).

\appendix
\section{Equations of motion for the metric and the series solutions}
\label{app: EoM for the metric and the series solutions}

In this appendix, we present \blue{EoM} for the metric, and focus on the procedure used to solve these equations.

\subsection{Equations of motion}
We begin with the action Eq.~\eqref{eq:action} and Lagrangian Eq.\eqref{eq:lagrangian}. Utilizing the chain rule, we can obtain the equation 
\begin{equation}
  \frac{\delta S}{\delta \Xi}=\frac{\delta g_{\mu\nu}}{\delta \Xi} \frac{\delta S}{\delta g_{\mu\nu}} = 0\;,
\label{eq:EoM for undetermined function}
\end{equation}
where $\Xi = U,\;N,\;\phi$ denotes the undetermined function in the metric ansatz Eq.~\eqref{eq:metric}, and the fact that the energy-momentum tensor of metric vanishes, i.e. $T^{\mu\nu}=\frac{2}{\sqrt{-g}}\frac{\delta S}{\delta g_{\mu\nu}}=0$, is also taken into account.

Substituting  $\Xi = U,\;N,\;\phi$ into Eq.~\eqref{eq:EoM for undetermined function}, we obtain the corresponding equations for $U$,
\begin{align}
&\frac{32\pi \alpha _2 r^2}{N} \bigg[4 N'^2 \phi '^2+N^2 \bigg(\phi ''^2+\phi ^{(3)} \phi '\bigg)\bigg]\notag\\
&+4\pi\alpha _1 \bigg\{4 r N \bigg[r \phi ''^2+\phi ' \bigg(r \phi ^{(3)}+2 \phi ''\bigg)\bigg]+\frac{16 r^2 N'^2 \phi '^2}{N}\bigg\}\notag\\
&+4\pi\alpha _3 \bigg\{8 N \bigg[r^2 \phi ''^2+\phi '^2+r \phi ' \bigg(r \phi ^{(3)}+4 \phi ''\bigg)\bigg]\notag\\
&+\frac{32 r^2 N'^2 \phi '^2}{N}\bigg\} +r N^2 N'\notag\\
&=32\pi \alpha _3 r \phi ' \bigg(r N'' \phi '+5 N' \bigg(r \phi ''+\phi '\bigg)\bigg)\notag\\
&+32\pi \alpha _2 r \phi ' \bigg(r N'' \phi '+N' \bigg(5 r \phi ''+\phi '\bigg)\bigg)\notag\\
&+16\pi \alpha _1 r \phi ' \bigg(r N'' \phi '+N' \bigg(5 r \phi ''+3 \phi '\bigg)\bigg)\;,
\label{eq:EoM of U}
\end{align}
the equation for $N$,
\begin{align}
&\frac{16\pi \alpha _1 \phi '}{N} \bigg\{r \bigg(r U'+2 U\bigg) \phi ''+\phi ' \bigg[U-r \bigg(r U''+U'\bigg)\bigg]\bigg\}\notag\\
&+\frac{32\pi \alpha _2 \phi ' }{N}\bigg\{r \bigg(r U'+4 U\bigg) \phi ''-\phi ' \bigg[r \bigg(r U''+3 U'\bigg)+U\bigg]\bigg\}\notag\\
&+\frac{32\pi \alpha _3 r \phi ' }{N}\bigg[U' \bigg(r \phi ''+\phi '\bigg)-r U'' \phi '\bigg]+N^3\notag\\
&=N \bigg[r \bigg(U'+r \phi '^2\bigg)+U\bigg]-32\pi \alpha _2 N \phi '^2\;,
\label{eq:EoM of N}
\end{align}
and finally the equation for $\phi$,
\begin{align}
&4\pi\alpha _3 \bigg\{\frac{16 r^2 N'^2 U' \phi '}{N}+4 r N \bigg[r U'' \phi ''+\bigg(r U^{(3)}+2 U''\bigg) \phi '\bigg]\bigg\}\notag\\
&+4\pi\alpha _2 \bigg\{\frac{16 r N'^2}{N} \bigg(r U'+4 U\bigg) \phi '+8 N U' \bigg(2 r \phi ''+3 \phi '\bigg)\notag\\
&-8 N^3 \phi ''+4 N \bigg[\bigg(r^2 U''+2 U\bigg) \phi ''+r \bigg(r U^{(3)}+6 U''\bigg) \phi '\bigg]\bigg\}\notag\\
&+4\pi\alpha _1 \bigg\{\frac{8 r N'^2}{N} \bigg(r U'+2 U\bigg) \phi '+ 4 N U' \bigg(r \phi ''+\phi '\bigg)\notag\\
&\qquad+ 2Nr \bigg[r U^{(3)} \phi '+U'' \bigg(r \phi ''+4 \phi '\bigg)\bigg]\bigg\}\notag\\
&=r^2 N^2 N' \phi '-r N^3 \bigg(r \phi ''+2 \phi '\bigg)\notag\\
&+4\pi\alpha _1 \bigg\{2 r N' \bigg(r U'+2 U\bigg) \phi ''+4 \phi ' U \bigg(r N''+N'\bigg)\notag\\
&\qquad+ 2 \phi 'r \bigg[r N'' U'+2 N' \bigg(2 r U''+5 U'\bigg)\bigg]\bigg\}\notag\\
&+4\pi\alpha _2 \bigg\{4 r N' \bigg(r U'+4 U\bigg) \phi '' +8 \phi ' U \bigg(2 r N''+5 N'\bigg)\notag\\
&-8 \phi ' N^2 N'+4 \phi 'r \bigg[4 r N' U''+U' \bigg(r N''+18 N'\bigg)\bigg]\bigg\}\notag\\
&+16\pi \alpha _3 r \bigg\{r N'' U' \phi '+N' \bigg[r U' \phi ''+2 \phi ' \bigg(2 r U''+U'\bigg)\bigg]\bigg\} \;.
\label{eq:EoM of phi}
\end{align}
The resulting equations of motion constitute a highly coupled system of third-order differential equations, rendering exact analytic solutions difficult to obtain. We therefore construct asymptotic series solutions instead.

In addition, there exist two special choices of the coupling parameters for which the field equations reduce to second-order differential equations for $U$, $N$, and $\phi$. When $\alpha_1=-2\alpha_2=\alpha$ and $\alpha_3=0$, the equations become
\begin{align*}
-\frac{32\pi \alpha N' \phi '^2}{N}+N N'+32\pi \alpha \phi ' \phi ''=0\;,
\end{align*}

\begin{align*}
&4\pi\alpha\bigg\{\frac{8 \phi ' }{N}\bigg[\bigg(r U'+U\bigg) \phi '-r U \phi ''\bigg]-16\pi \alpha N \phi '^2\bigg\}\\
&=N \bigg[r \bigg(U'+r \phi '^2\bigg)+U\bigg]-N^3\;,
\end{align*}

\begin{align*}
&4\pi\alpha \bigg\{4 \phi '' \bigg(r U N'+N^3\bigg)+ 4 r\phi ' U N''+16 \phi ' N' \bigg(r U'+U\bigg)\\
&-\frac{4 \phi ' }{N}\bigg[4 r U N'^2+N^3 N'+N^2 \bigg(r U''+2 U'\bigg)\bigg]\\
&-4 N \bigg(r U'+U\bigg) \phi ''\bigg\}\\
&=r^2 N^2 N' \phi '-r N^3 \bigg(r \phi ''+2 \phi '\bigg)\;.
\end{align*}
For another choice, $-\alpha_1/4=\alpha_2=\alpha_3=\alpha$, the equations become
\begin{align*}
rN'+\frac{32\pi\alpha\phi'^2}{N}=0\;,
\end{align*}

\begin{align*}
32\pi\alpha\phi'^2\bigg(3U-N^2\bigg)=N^4-N^2\bigg(U+rU'+r^2\phi'^2\bigg)\;,
\end{align*}

\begin{align*}
&32\pi\alpha\bigg[N^2\phi''-\bigg(U'\phi'+U\phi''\bigg)-NN'\phi'+3\frac{N'}{N}\phi'U\bigg]\\
&=N^2\bigg(2r\phi'+r^2\phi''\bigg)-NN'r^2\phi'\;.
\end{align*}
However, even in these special cases analytic solutions remain difficult to obtain. Moreover, the resulting effects on the black-hole spacetime are qualitatively similar to those obtained by varying $\alpha_1$ or $\alpha_2$ alone. Therefore, we shall not discuss these cases further in the text.

\subsection{Series solution}
Since exact analytic solutions to Eqs.~\eqref{eq:EoM of U}--\eqref{eq:EoM of phi} are difficult to obtain, we instead construct asymptotic series solutions. Motivated by the requirement of asymptotic flatness, we consider the following large-$r$ expansions:
\begin{equation} 
\left\{
\begin{aligned}
\phi&=\frac{c_1}{r}+\frac{c_2}{r^2}+\frac{c_3}{r^3}+\frac{c_4}{r^4}+\frac{c_5}{r^5}+\cdot\cdot\cdot\;,\\
N&= b_0+\frac{b_1}{r}+\frac{b_2}{r^2}+\frac{b_3}{r^3}+\frac{b_4}{r^4}+\frac{b_5}{r^5}+\cdot\cdot\cdot\;,\\
U&= d_0+\frac{d_1}{r}+\frac{d_2}{r^2}+\frac{d_3}{r^3}+\frac{d_4}{r^4}+\frac{d_5}{r^5}+\cdot\cdot\cdot\;.
\end{aligned}
\right.
\end{equation}
Substituting these expansions into \blue{EoM}, i.e. Eqs.~\eqref{eq:EoM of U}, \eqref{eq:EoM of N}, and \eqref{eq:EoM of phi}, and solving order by order in $1/r$, we obtain
\begin{equation}
\begin{aligned}
\phi (r)&=\frac{c_1}{r}-\frac{8\pi \alpha _3 c_1 d_1}{b_0^2 r^4}\\
&-\frac{8\pi c_1^3}{5 b_0^2 r^5}\bigg(-\alpha _1-10 \alpha _2+9 \alpha _3\bigg)\\
&+\frac{512\pi^2}{7 b_0^4 r^7}\bigg(-3 \alpha _1^2 b_0^2 c_1^3-36 \alpha _2^2 b_0^2 c_1^3+10 \alpha _3^2 b_0^2 c_1^3\\
&\qquad-25 \alpha _1 \alpha _2 b_0^2 c_1^3-26 \alpha _2 \alpha _3 b_0^2 c_1^3+2 \alpha _3^2 c_1 d_1^2\bigg)\\
&\mathcal{O}(r^{-8})\;,
\end{aligned}  
\label{eq:series phi}
\end{equation}

\begin{equation}
\begin{aligned}
N(r)&=b_0+\frac{8\pi \bigg(3 \alpha _1+10 \alpha _2+3 \alpha _3\bigg) c_1^2}{b_0 r^4}\\
&-\frac{1024\pi^2 c_1^2 d_1\alpha _3}{7 b_0^3 r^7}\bigg(21 \alpha _1+56 \alpha _2+30 \alpha _3\bigg)\\
&+\mathcal{O}(r^{-8})\;,
\end{aligned}
\label{eq:series N}
\end{equation}

\begin{equation}
\begin{aligned}
U(r)&=b_0^2+\frac{d_1}{r}+\frac{c_1^2}{r^2}-\frac{16\pi c_1^2}{r^4}\bigg(\alpha _3-2 \alpha _2\bigg)\\
&+\frac{8\pi c_1^2 d_1}{b_0^2 r^5}\bigg(\alpha _1+6 \alpha _2-\alpha _3\bigg)\\
&-\frac{16\pi c_1^4}{5 b_0^2 r^6}\bigg(-4 \alpha _1-20 \alpha _2+\alpha _3\bigg)\\
&+\frac{2048\pi^2 \alpha _3 c_1^2 d_1}{7 b_0^2 r^7}\bigg(5 \alpha _3-7 \alpha _2\bigg)\\
&+\mathcal{O}(r^{-8})\;.
\end{aligned}
\label{eq:series U}
\end{equation}
As mentioned before, the main reason for truncating at $\mathcal{O}(r^{-8})$ is to ensure that all three functions have the same order in the $1/r$ expansion. The expansions contain three integration constants, $b_0$, $c_1$, and $d_1$, whose physical interpretations remain to be identified. To determine them, we consider the Einstein--Maxwell limit obtained by setting $\alpha_1=\alpha_2=\alpha_3=0$. In this case, the action Eq.~\eqref{eq:action} reduces to
\begin{equation}
S=\int\mathrm{d}^4x\sqrt{-g}\bigg[\frac{1}{16\pi}R-\frac{1}{16\pi}F^2\bigg]\;,
\end{equation}
and the corresponding solution becomes the Reissner-Nordstr\"{o}m black hole,
\begin{equation}
\left\{
\begin{aligned}
\phi &= \frac{Q}{r}\;,\\
N &= 1\;,\\
U &= 1-\frac{2M}{r}+\frac{Q^2}{r^2}\;.
\end{aligned}
\right.
\end{equation}
Comparing the above solution with the asymptotic series expansions  \blue{performed in the radial coordinate (or ``Frobenius method'')} yields
\begin{equation}
\left\{
\begin{aligned}
c_1&=Q\;,\\
b_0&=1\;,\\
d_1&=-2M\;.
\end{aligned}
\right.
\end{equation}
Substituting these constants back into the series expansions reproduces Eq.~\eqref{eq:phi}--Eq.~\eqref{eq:U}.

\section{Equations of motion for numerical integration}
\label{app: EoM for integration}
As discussed in Sec.~\ref{sec:photon dynamics}, the equations of motion in Eq.~\eqref{eq:motion equation 12} suffer from numerical instabilities associated with the explicit ``$\pm$'' prescription, particularly in the computation of higher-order photon rings. This is mainly due to the difficulty of consistently handling sign changes near turning points in numerical implementations. In this appendix, we provide a detailed explanation of this issue and present an alternative formulation suitable for numerical integration.

To make this issue more explicit, we first focus on the radial equation in Eq.~\eqref{eq:motion equation 12}. For a photon transitioning from an ingoing trajectory to an outgoing one (usually seen in a scattering process), the evolution of the radial equation changes from $r^2\dot{r}=-\sqrt{\mathsf{R}(r)}$ ($t<t_{tp}$), to $r^2\dot{r}=0$ ($t=t_{tp}$), and then to $r^2\dot{r}=+\sqrt{\mathsf{R}(r)}$ ($t>t_{tp}$), where $t_{tp}$ denotes the time at which the photon reaches its periastron (turning point), i.e., $r=r_{tp}$ with $\mathsf{R}(r_{tp})=0$. However, due to finite floating-point precision in numerical computations, it is difficult for the algorithm to accurately capture this moment. As a result, the integration may evolve from $r_{tp}+\epsilon$ at one step, where $\epsilon$ denotes a small quantity, to $r_{tp}-\epsilon$ at the next, where $\mathsf{R}(r_{tp}-\epsilon)<0$, leading to numerical errors arising from the evaluation of the square root of a negative quantity. To address this issue, one is forced to modify the equation to $r^2\dot{r}=\pm\sqrt{\mathsf{R}(r)+\epsilon}$, thereby introducing an additional numerical error in an ad hoc manner. Moreover, it should be emphasized that the equation $r^2\dot{\theta}=\pm\sqrt{\mathsf{\Theta}(\theta)}$ undergoes frequent sign changes in the computation of higher-order photon rings, leading to increasingly significant accumulation of numerical errors.

These instabilities make the direct use of Eq.~\eqref{eq:motion equation 12} unsuitable for stable numerical ray-tracing. Therefore, instead of using these equations in their original form, we adopt an alternative formulation of the equations of motion that avoids explicit sign tracking and is more robust for constructing black-hole shadow images.

To construct a formulation suitable for numerical integration, we recast the equations of motion in Hamiltonian form. Starting from the Hamiltonian
\begin{equation}
\mathcal{H}=\frac{1}{2}g^{\mu\nu}(r,\theta)p_\mu p_\nu\;,
\label{eq: Hamiltonian}
\end{equation}
the Hamiltonian equations can be derived as follows
\begin{equation}
\left\{
\begin{aligned}
\dot{x}^\mu &=\frac{\partial \mathcal{H}}{\partial p_\mu}= g^{\mu\nu}p_\nu\;,\\
\dot{p}_\mu &=-\frac{\partial\mathcal{H}}{\partial x^\mu}= -\frac{1}{2}(\partial_\mu g^{\sigma\rho}) p_\sigma p_\rho\;.
\end{aligned}
\right.
\label{eq: hamiltonian equations}
\end{equation}
For a static and spherically symmetric black hole spacetime, the metric components depend only on $r$ and $\theta$, and all off-diagonal components vanish. Expanding Eq.~\eqref{eq: hamiltonian equations}, the nontrivial equations reduce to
\begin{equation}
\left\{
\begin{aligned}
\dot{t} &= - g^{tt}\mathcal{E}\;,\\
\dot{r} &=g^{rr}p_r,\\
\dot{\theta} &= g^{\theta\theta}p_\theta,\\
\dot{\varphi} &= g^{\varphi\varphi}\mathcal{L}_\varphi\;,\\
\dot{p}_r &= -\frac{1}{2}\partial_rg^{\sigma\rho}p_\sigma p_\rho\;,\\
\dot{p}_\theta &= -\frac{1}{2}\partial_\theta g^{\sigma\rho}p_\sigma p_\rho\;,
\end{aligned}
\right.
\label{eq:components of hamitonian equations}
\end{equation}
where $\mathcal{E}=-p_t$ and $\mathcal{L}_\varphi=p_\varphi$ are conserved quantities. Aside from Hamiltonian equations, photons are subject to the null constraint on the four-momentum,
\begin{equation}
g^{\mu\nu}p_\mu p_\nu
= g^{tt} \mathcal{E}^2
+ g^{rr}p_r^2
+ g^{\theta\theta}p_\theta^2
+ g^{\varphi\varphi}p_\varphi^2
= 0\;.
\label{eq: 4-momentum constraint}
\end{equation}

Together, these equations, i.e. Eqs.~\eqref{eq:components of hamitonian equations} and~\eqref{eq: 4-momentum constraint}, form a system of \textit{differential-algebraic equations} (DAEs), which is generally challenging for direct numerical integration. To eliminate the algebraic constraint and transform the system into a set of ordinary differential equations, we rewrite Eq.~\eqref{eq: 4-momentum constraint} as
\begin{equation}
\begin{aligned}
-g^{tt}\mathcal{E}^2 - g^{\varphi\varphi}\mathcal{L}_\varphi^2&=g^{\theta\theta}p_\theta^2 + g^{rr}p_r^2 \\
&:=\rho^2(\sin^2\chi+\cos^2\chi),
\end{aligned}
\label{eq: rewrite of 4-momentum constraint}
\end{equation}
which motivates the introduction of two auxiliary functions, $\rho(r,\theta)$ and $\chi$, defined by

\begin{equation}
\left\{
\begin{aligned}
\rho &=\sqrt{g^{\theta\theta}p_\theta^2 + g^{rr}p_r^2}= \sqrt{-g^{tt}\mathcal{E}^2 - g^{\varphi\varphi}\mathcal{L}_\varphi^2}\;,\\
\chi &= \arctan \dfrac{\sqrt{g^{\theta\theta}}\,p_\theta}{\sqrt{g^{rr}}\,p_r}\;,
\end{aligned}
\right.
\label{eq: auxiliary functions}
\end{equation}

The inverse transformation is then given by
\begin{equation}
\left\{
\begin{aligned}
p_r &= -\sqrt{g_{rr}}\rho\cos\chi\;,\\
p_\theta &= -\sqrt{g_{\theta\theta}}\rho\sin\chi\;.
\end{aligned}
\right.
\label{eq: transformation from momentum to auxiliary functions}
\end{equation}
Here we have used the identities $g^{rr}g_{rr}=g^{\theta\theta}g_{\theta\theta}=1$. The minus signs are chosen such that the initial photon momentum is directed from the observer toward the black hole.

Substituting Eq.~\eqref{eq: transformation from momentum to auxiliary functions} into Eq.~\eqref{eq:components of hamitonian equations}, we finally obtain the following set of equations of motion,
\begin{equation}
\left\{
\begin{aligned}
\dot{t} &= -g^{tt}\mathcal{E}\;,\\
\dot{\varphi} &= g^{\varphi\varphi} \mathcal{L}_\varphi\;,\\
\dot{r} &= -\sqrt{g^{rr}}\,\rho \cos \chi\;,\\
\dot{\theta} &= -\sqrt{g^{\theta\theta}}\, \rho \sin \chi\;,\\
\dot{\chi}
&= \sqrt{g^{rr}}\partial_r\rho \sin\chi - \sqrt{g^{\theta\theta}}\partial_\theta\rho \cos\chi\\
 &\quad- \frac{1}{2}\rho\sqrt{g^{rr}}\partial_r\ln g^{\theta\theta}\sin\chi\;.
\end{aligned}
\right.
\label{eq: EoM for plotting}
\end{equation}

Since $\mathcal{E}$ and $\mathcal{L}_\varphi$ are fixed by the initial conditions and the metric is known, the auxiliary function $\rho(r,\theta)$ is fully determined. Consequently, Eq.~\eqref{eq: EoM for plotting} constitutes a closed and numerically solvable system. Compared with Eq.~\eqref{eq:motion equation 12} and Eq.~\eqref{eq:motion equation 03}, this formulation solves the ``$\pm$'' problem through the introduction of the auxiliary variable $\chi$, which not only facilitates numerical implementation but also significantly suppresses the accumulation of numerical errors. However, since the auxiliary functions lack a clear physical interpretation, Eq.~\eqref{eq: EoM for plotting}, despite its numerical advantages, is not well suited for analytical investigations. Therefore, we employ this formulation only for constructing black-hole images, while retaining Eqs.~\eqref{eq:motion equation 12} and~\eqref{eq:motion equation 03} for theoretical analysis of photon rings.

%In addition, for axially symmetric spacetimes, i.e. rotating black holes, a similar procedure yields
%\begin{equation}
%\left\{
%\begin{aligned}
%\dot{t} &= -\dfrac{g_{t\varphi}\mathcal{L}_\varphi + %g_{\varphi\varphi}\mathcal{E}}{g_{tt}g_{\varphi\varphi}-g_{t\varphi}^2}\;,\\
%\dot{\varphi} &= \dfrac{g_{tt}\mathcal{L}_\varphi + g_{t\varphi}\mathcal{E}}{g_{tt}g_{\varphi\varphi}-g_{t\varphi}^2}\;,\\
%\dot{r} &= -\sqrt{g^{rr}}\,\rho \cos \chi\;,\\
%\dot{\theta} &= -\sqrt{g^{\theta\theta}}\, \rho \sin \chi\;,\\
%\dot{\chi} &= \sqrt{g^{rr}}\partial_r\rho\sin\chi - \sqrt{g^{\theta\theta}}\partial_\theta\rho\cos\chi \\
%&\quad
% - \frac{1}{2}\rho\left(
% \sqrt{g^{rr}}\partial_r\ln g^{\theta\theta}\sin\chi
% - \sqrt{g^{\theta\theta}}\partial_\theta\ln g^{rr}\cos\chi
% \right)\;,
%\end{aligned}
%\right.
%\end{equation}
%where the two auxiliary functions are defined as $\rho=\sqrt{-g{tt}\mathcal{E}^2-2g^{t\varphi}\mathcal{EL}-g^{\varphi\varphi}\mathcal{L}_\varphi^2}$ and $\chi = \arctan \frac{\sqrt{g^{\theta\theta}}p_\theta}{\sqrt{g^{rr}}p_r}$.

\section{Backward ray-tracing approach}
\label{app:backward ray-tracing approach}
In this section, we provide a brief overview of the backward ray-tracing approach \cite{Luminet:1979nyg,Bardeen:1972fi}. For comprehensive descriptions of publicly available general relativistic ray-tracing and radiative-transfer codes, we refer the reader to established implementations such as RAPTOR \cite{Bronzwaer:2018lde,Bronzwaer:2020kle}, ipole \cite{Moscibrodzka:2017lcu}, Gyoto \cite{Vincent:2011wz,Aimar:2023vcs}, and related references.

\subsection{Coordinate transformation}
\label{subapp:coordinate transformation}

\begin{figure}[ht]
  \centering    
\includegraphics[width=0.7\linewidth]{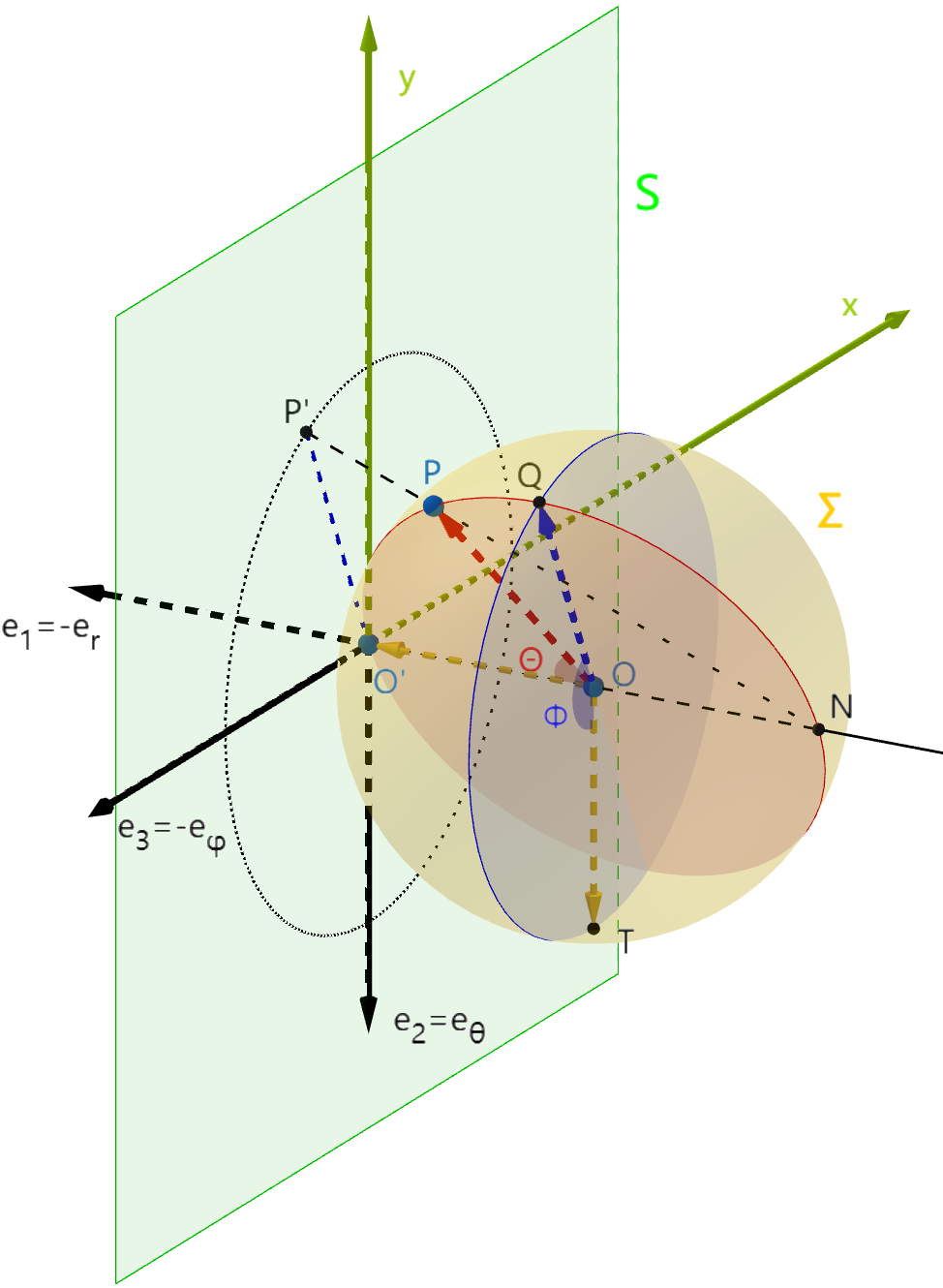}
  \caption{Geometric relations between the Boyer--Lindquist coordinates $(t,r,\theta,\varphi)$, the celestial coordinates $(\Theta,\Phi)$, and the camera screen coordinates $(x,y)$.}
  \label{fig:geometric relation}
\end{figure}

The geodesic of photons $s(\lambda)$ obtained from \blue{EoM}, Eq.~\eqref{eq:motion equation 03} and Eq.~\eqref{eq:motion equation 12}, is expressed in the Boyer--Lindquist coordinates with tangent vector 
\begin{equation}
\dot{s}=\dot{t}\partial_t+\dot{r}\partial_r+\dot{\theta}\partial_\theta+\dot{\varphi}\partial_\varphi\;,
\label{eq:initial tangent vector}
\end{equation}
which do not correspond to the coordinates perceived by the observer. Therefore, a series of coordinate transformations, as illustrated in Fig.~\ref{fig:geometric relation}, are required to represent the black hole as seen by the observer.

In Fig.~\ref{fig:geometric relation}, the semi-transparent golden sphere $\Sigma$ is centered at point $O$ with coordinate $(r_O,\theta_O,\varphi_O)$ representing the location of the observer. The red vector $\overrightarrow{OP}$, which we will discuss later, represents the radius of the sphere. For better description, we adopt a local rest frame in the vicinity of the observer with unit basis defined as:
\begin{equation}
\left\{
\begin{aligned}
\mathbf{e}_0&=\mathbf{e}_t=\frac{\partial_t}{\sqrt{-g_{tt}}}\big|_O\;,\\
\mathbf{e}_1&=-\mathbf{e}_r=-\frac{\partial_r}{\sqrt{g_{rr}}}\big|_O\;,\\
\mathbf{e}_2&=\mathbf{e}_\theta=\frac{\partial_\theta}{\sqrt{g_{\theta\theta}}}\big|_O\;,\\
\mathbf{e}_3&=-\mathbf{e}_\varphi=-\frac{\partial_\varphi}{\sqrt{g_{\varphi\varphi}}}\big|_O\;,
\end{aligned}
\right.
\label{eq:local frame}
\end{equation}
where $\mathbf{e}_0$ is parallel to the four-momentum of the observer and $\mathbf{e}_1$ points from the observer towards the center of the black hole.

Along the line parallel to $\mathbf{e}_1$, point $N$ is the farthest point away from black hole on the sphere $\Sigma$ while point $O^\prime$ is the closest one. The green plane $S$ referred as ``the camera screen" is orthogonal to the line $O^\prime N$ and tangent to the sphere $\Sigma$ at $O^\prime$. The red great circle passing through $O^\prime$ and $P$ intersects the blue great circle, which is parallel to the $S$-plane, at point $Q$, the closest intersection point to $P$. The angle $\Theta$ measured from $\overrightarrow{OO^\prime}$ to $\overrightarrow{OP}$ is the polar angle, while $\Phi$ between $\overrightarrow{OT}$ (which is parallel to $\mathbf{e}_2$) and $\overrightarrow{OQ}$, is the azimuthal angle. 

Suppose the observer detects a photon with $\overrightarrow{OP}$ being the tangent vector of its geodesic in the three-dimensional space. Then in the local frame of the observer, the tangent vector Eq.~\eqref{eq:initial tangent vector} can be expressed as
\begin{equation}
\dot{s}=\big|\overrightarrow{OP}\big|\big(\mathbf{e}_0+\cos\Theta \mathbf{e}_1+\sin\Theta\cos\Phi \mathbf{e}_2+\sin\Theta\sin\Phi \mathbf{e}_3\big)\;,
\label{eq:tangent vector in celestial coordinate}
\end{equation}
where $\big|\overrightarrow{OP}\big|$ can be set to one due to the scalability of null vectors. 

Utilizing point $N$, we can stereographically project $P$ onto the $S$-plane as $P^\prime$. In $S$-plane, we can construct a Cartesian coordinate system centered at point $O^\prime$, where the coordinate of $P^\prime$ is given by
\begin{equation}
\left\{
\begin{aligned}
x_{P^\prime}&=-2\big|\overrightarrow{OP}\big|\tan\frac{\Theta}{2}\sin\Phi\;,\\
y_{P^\prime}&=-2\big|\overrightarrow{OP}\big|\tan\frac{\Theta}{2}\cos\Phi\;.
\end{aligned}
\right.
\label{eq:celestial to cartesian}
\end{equation}

\begin{figure}
  \centering  
  \includegraphics[width=0.8\linewidth]{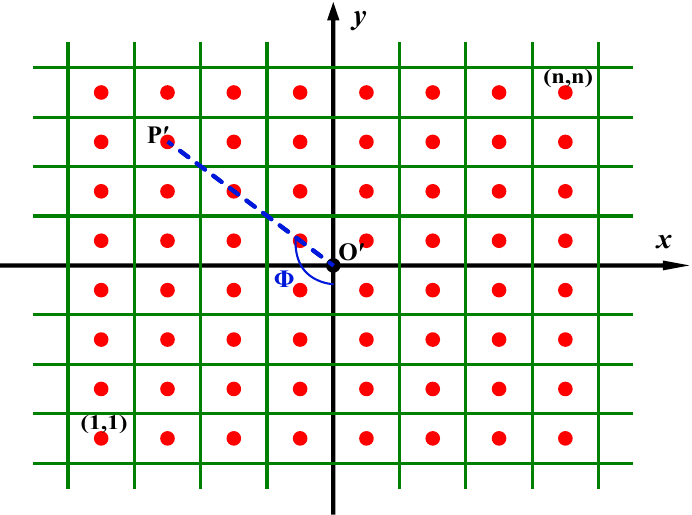}
  \caption{The camera screen $S$ is divided into $n\times n$ pixels, each labeled by $(i,j)$, with the Cartesian coordinate system originating at $O^\prime$.}
  \label{fig:camera screen}
\end{figure}

\begin{figure}[ht]
  \centering    \includegraphics[width=0.6\linewidth]{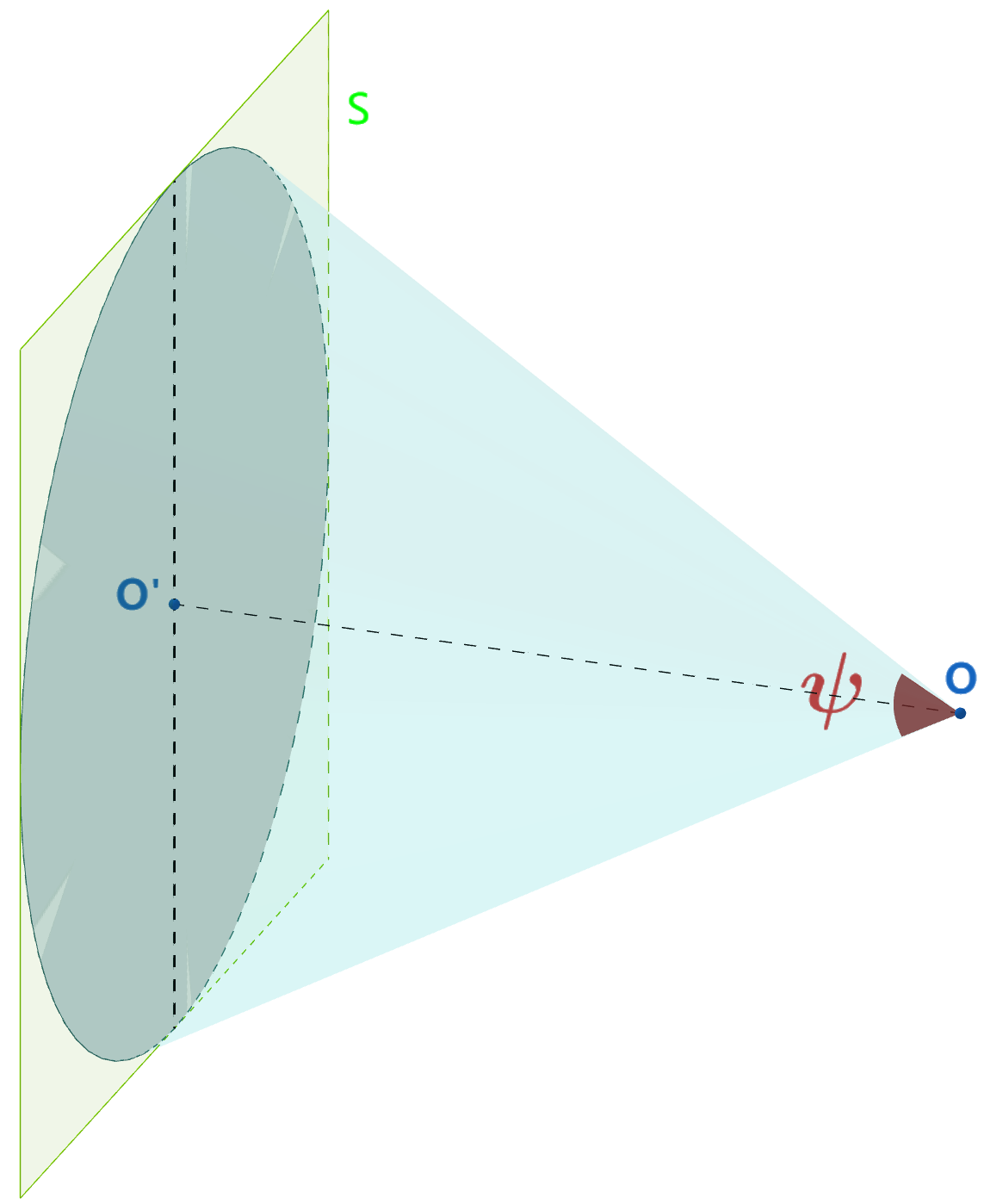}
  \caption{Observer's field of view.}
  \label{fig:field of view}
\end{figure}

In order to draw the picture of black-hole shadow and display the distortion of light, we separate $S$ into $n\times n$ pixels (we can set $n=2000$). As shown in Fig.~\ref{fig:camera screen}, each pixel is labeled by $(i,j)$ ranging from $(1,1)$ (left bottom corner) to $(n,n)$ (right top corner). We suppose the angular diameter of the $S$-plane relative to the observer $O$, or say observer's field of view, is $\psi$ as shown in Fig.~\ref{fig:field of view}. Hence, the length $\mathsf{L}$ of the screen $S$ is
\begin{equation}
\mathsf{L}=2\big|\overrightarrow{OP}\big|\tan\frac{\psi}{2}\;.
\end{equation}
Therefore, the relation between the Cartesian coordinate $(x_{P^\prime},y_{P^\prime})$ and the index $(i,j)$ of point $P^\prime$ is given by
\begin{equation}
\left\{
\begin{aligned}
x_{P^\prime}&=\iota (i-\frac{n+1}{2})\;,\\
y_{P^\prime}&=\iota (j-\frac{n+1}{2})\;,\\
\end{aligned}
\right.
\label{eq:index to cartesian}
\end{equation}
where $\iota$ is the length of each pixel given by $\iota=\frac{\mathsf{L}}{n}$.

Combing Eq.~\eqref{eq:celestial to cartesian} with Eq.~\eqref{eq:index to cartesian}, we can derive the direct transformation from the index $(i,j)$ to celestial coordinate $(\Theta,\Phi)$
\begin{equation}
\left\{
\begin{aligned}
\tan\Phi&=\frac{(n+1)/2-i}{(n+1)/2-j}\;,\\
\tan\frac{\Theta}{2}&=\frac{\tan(\psi/2)}{n}\sqrt{(\frac{n+1}{2}-i)^2+(\frac{n+1}{2}-j)^2}\;.
\end{aligned}
\right.
\label{eq:index to celestial}
\end{equation}
The polar angle $\Theta$ takes values in the range $0\leq\Theta\lesssim\psi/2$, while the azimuthal angle $\Phi$ spans the range $0\leq\Phi<2\pi$.

\subsection{Backward ray-tracing approach}
\label{subapp:backward ray-tracing approach}

In principle, the construction of the black-hole shadow requires tracing photons from the light source, after their deflection by the black hole, to their eventual detection by the observer. Since the gravitational influence of the black hole is confined to a finite region, the vast majority of photons from the source either experience negligible deflection or fail to reach the observer. To circumvent this inefficiency, we employ the backward ray-tracing approach, which exploits the reversibility of null geodesics by initiating photon trajectories at the observer and propagating them backward toward the light source or the black hole. Here, the term ``light source'' is used in a general sense and is not necessarily restricted to an accretion disk; it may also represent other emitting configurations, such as a spherical source at infinity, as considered in Refs.~\cite{Liu:2024soc,Liu:2024iec,Hu:2020usx}.

In this approach, we first select a pixel on the camera screen $S$, labeled by $(i,j)$. Then via Eq.~\eqref{eq:index to celestial}, we can determine the direction, denoted by $(\Theta,\Phi)$, in which the corresponding photon is emitted from the observer. Before numerically solving Eq.~\eqref{eq: EoM for plotting}, initial conditions are required. Plugging \blue{EoM}, Eq.~\eqref{eq:motion equation 03} and Eq.~\eqref{eq:motion equation 12}, into the tangent vector in Eq.~\eqref{eq:initial tangent vector} at the observer’s position $r_O$, and equating the result with Eq.~\eqref{eq:tangent vector in celestial coordinate}, we obtain the conserved quantities associated with each photon emitted by the observer:
%Combining the tangent vector Eq.~\eqref{eq:initial tangent vector} or Eq.~\eqref{eq:tangent vector in celestial coordinate} with these motion equations, we can deduce that each photon the observer emits has the following conserved quantities:
\begin{equation}
\left\{
\begin{aligned}
\mathcal{E}&=\sqrt{U(r_O)}\;,\\
\mathcal{L}_\varphi&=-r_O\sin\Theta\sin\Phi\;,\\
\mathcal{C}&=r_O^2\sin^2\Theta\cos^2\Phi\;,
\label{eq:contants}
\end{aligned}
\right.
\end{equation}
Given these initial conditions, we numerically integrate the trajectory of each photon and simultaneously record the number of times it intersects the accretion disk in order to compute the corresponding observed intensity via Eq.~\eqref{eq: observed intensity}. The integration is terminated if the number of intersections exceeds a specified threshold (a threshold of 4 is typically sufficient, and higher values may be used for increased precision), if the photon is captured by the black hole ($r \leq r_H$), or if it reaches a predefined large radial distance that approximates infinity (e.g., $r \geq 1.5\, r_O$). The resulting color (e.g., red) and grayscale value (corresponding to the observed intensity) are then assigned to the initial pixel $(i,j)$ on the $S$-plane. If the photon is captured by the black hole without intersecting the accretion disk, the corresponding pixel is assigned black. This procedure is repeated for all initial directions until the $S$-plane is fully mapped.
\bibliography{reference.bib} 
\end{document}